\documentclass[]{jfm}

\usepackage{graphicx}
\usepackage{epstopdf,epsfig}
\usepackage{newtxtext}
\usepackage{newtxmath}
\usepackage{natbib}
\usepackage{hyperref}
\hypersetup{
    colorlinks = true,
    urlcolor   = blue,
    citecolor  = black,
}

\newcommand{\RomanNumeralCaps}[1]

\usepackage{latexsym}
\usepackage{amsmath}
\usepackage{amssymb}
\usepackage{tabularx}
\usepackage{multirow}
\usepackage{subfigure}
\usepackage{calligra}
\usepackage{calrsfs}
\usepackage{mathrsfs}
\usepackage{extarrows}
\usepackage{booktabs}
\usepackage{bm}
\usepackage{color}

\title{Large-scale influence of numerical noises as artificial stochastic disturbances on a sustained turbulence}

\author{Shijie Qin\aff{2}
\and Shijun Liao\aff{1,2}
\corresp{\email{sjliao@sjtu.edu.cn}}
}

\affiliation{\aff{1}State Key Laboratory of Ocean Engineering, Shanghai 200240, China
\aff{2} Center of Marine Numerical Experiment, School of Naval Architecture, Ocean and Civil Engineering, Shanghai Jiao Tong University, Shanghai 200240, China
}

\begin{document}
\maketitle

\begin{abstract}
%%%
We investigate the large-scale influence of numerical noises as tiny artificial stochastic disturbances on a sustained turbulence. Using the two-dimensional (2D) turbulent Rayleigh-B{\'e}nard (RB) convection as an example, we numerically solve the NS equations, separately, by means of a traditional algorithm with double precision (marked by RKwD) and the so-called clean numerical simulation (CNS). The numerical simulation given by the RKwD is a mixture of the ``true'' physical solution and the ``false'' numerical noises that is random and can be regarded as a kind of artificial stochastic disturbances:  unfortunately, the ``true'' physical solution is  mostly at the same level as the ``false'' numerical noises. By contrast, the CNS can greatly reduce the background numerical noise to any a required level so that the ``false'' numerical noises are negligible compared with the ``true'' physical solution and thus the CNS solution can be used as a ``clean'' benchmark solution for comparison.
It is found that the numerical noises as tiny artificial stochastic disturbances could indeed lead to large-scale deviations of simulations not only in spatio-temporal trajectories but also even in statistics. Especially, these numerical noises (as artificial stochastic disturbances) even lead to different types of flows: the shearing convection occurs for the RKwD simulations, and its corresponding flow field turns to a kind of zonal flow thereafter, however the CNS benchmark solution always sustains the non-shearing vortical/roll-like convection during the whole process of simulation. Thus, we provide a rigorous evidence that numerical noises as a kind of small-scale artificial stochastic disturbances have quantitatively and qualitatively large-scale influences on a sustained turbulence, i.e. the 2D turbulent RB convection considered in this paper.
%%%%
\end{abstract}

%\begin{keyword}
\hspace{-0.4cm}{\bf Keyword} Rayleigh-B{\'e}nard convection, stochastic disturbances, NS equations, numerical noises
%\end{keyword}

\section{Introduction}

Turbulent flows of a compressible viscous fluid are mostly described by the Navier-Stokes (NS) equations in the conservation form
\begin{equation}
\frac{\partial\mathbf{U}}{\partial t}+\nabla\cdot\mathbf{F}_a=\nabla\cdot\mathbf{F}_d,
\label{NS-eq}
\end{equation}
subject to a given proper boundary condition and the initial condition
\begin{equation}
\mathbf{U}|_{t=0}=\mathbf{U}_0(\mathbf{r}), \hspace{1.0cm} \mathbf{r}\in\Omega\hspace{0.1mm},
\label{NS-ini}
\end{equation}
where $\mathbf{U}(\mathbf{r},t)=(\rho,\,\rho\hspace{0.3mm}\mathbf{u}\hspace{0.1mm},\,E)^\mathbf{T}$ is a vector of conserved variables, $\nabla$ is the Hamiltonian/gradient operator,
\begin{equation}
\mathbf{F}_a=
\begin{pmatrix}
    \rho\hspace{0.3mm}\mathbf{u} \\
    \rho\hspace{0.3mm}\mathbf{u}\otimes\mathbf{u}+P\hspace{0.3mm}\mathbf{I} \\
    (E+P)\hspace{0.3mm}\mathbf{u}
\end{pmatrix},
\hspace{1.0cm}
\mathbf{F}_b=
\begin{pmatrix}
    0 \\
    \bm{\tau} \\
    \mathbf{u}\cdot\bm{\tau}+\mathbf{q}
\end{pmatrix}
\label{NS-ad}
\end{equation}
are the advection/hyperbolic and diffusion fluxes, $\mathbf{r}\in\Omega$ is the vector of spatial position, $t$ denotes the time, $\mathbf{U}_0(\mathbf{r})$ is a given vector, $\rho$ is the density of mass, $\mathbf{u}$ is the velocity vector, $\mathbf{u}\otimes\mathbf{u}$ denotes the tensor product of $\mathbf{u}$ and $\mathbf{u}$, $P$ is the pressure, $\mathbf{I}$ denotes the unit tensor, $E$ is the total energy per unit mass, $\bm{\tau}$ denotes the tensor of viscous stress and $\mathbf{q}$ is the vector of heat diffusion flux, respectively.

Note that tiny stochastic disturbances in fluid flows resulting from either small-scale thermal fluctuations or environmental noises are unavoidable in practice and the triggered deviations might exponentially increase \citep{leith1972predictability, boffetta2001predictability, boffetta2017chaos}. However, these tiny stochastic disturbances are {\em not} considered in the above-mentioned NS equations (\ref{NS-eq})-(\ref{NS-ad}) that are deterministic. \citet{LLNS1959} proposed a stochastic form of the NS equations that models the effect of thermal fluctuations via an additional stochastic stress tensor \citep{graham1974hydrodynamic, swift1977hydrodynamic, bell2007numerical, donev2010accuracy, donev2014low}. However, the direct numerical simulation (DNS) of this stochastic model is rather difficult due to the extremely fine resolution that is required to accurately measure the velocities at dissipation-range length scales. By contrast, molecular dynamics (MD) provides molecular-level simulation techniques \citep{bird1998molecular, donev2011enhancement, smith2015molecular, mcmullen2022navier} for directly investigating the role of thermal fluctuations in turbulent flows.

Currently, \citet{mcmullen2022navier} investigated a decaying turbulent flow and found that, due to thermal fluctuations, the molecular-gas-dynamics spectra grow quadratically with wave number in the dissipation range, while the Navier-Stokes spectra decay exponentially.
Furthermore, the transition to quadratic growth occurs at a length scale much larger than the gas molecular mean free path, namely in a regime that the NS equations are widely believed to describe. Thus, they provided the first direct evidence
that ``the Navier-Stokes equations do not describe turbulent gas flows in the dissipation range because they neglect thermal fluctuations'' \citep{mcmullen2022navier}, which is in agreement with the results given by \citet{bandak2022dissipation}, \citet{bell2022thermal}, \citet{Eyink2022} and so on.
\citet{Gallis2021-PRF} separately used the direct simulation Monte Carlo (DSMC) method (molecular gas dynamics) and direct numerical simulation (DNS) of the NS equations to simulate a freely decaying turbulent flow, i.e. the compressible Taylor-Green vortex flow, and found that both methods produce basically the same energy decay for the Mach and Reynolds numbers they examined, but the molecular fluctuations in DSMC (and in experiments) can break symmetries, which in turn can cause the flows different from but basically similar to those given by the DNS.
Note that the above-mentioned investigations mainly focus on the influence of tiny stochastic disturbances resulting from thermal fluctuations on {\em small-scale} properties of freely {\em decaying} turbulence. Might the micro-level stochastic disturbances have huge influences on {\em large-scale} properties of some {\em sustained} turbulent flows? This is the primary motivation for our investigation of this paper.

Some turbulent flows might have multiple states according to \citet{frisch1986fully} and his well-known monograph on turbulence \citep{frisch1995turbulence} that ``it is typical for dissipative dynamical systems to have more than one attractor'' and ``each attractor has an associated basin'', and thus ``the statistical properties of the solution will then depend on to which basin the initial condition belongs''. The first evidence of this phenomenon in a turbulent flow is given by \citet{huisman2014multiple} for their study on the highly turbulent Taylor-Couette flow. Note that the above-mentioned work mainly focuses on the influence of initial data on the flow state of turbulence as well as its statistical properties. Considering that there are some previous investigations \citep{knobloch1989effect, kraut1999preference, masoller2002noise, de2007noise} in the effects of noises (in general whose levels are much larger than the stochastic disturbances considered later in this paper) on the dynamical systems with multiple attractors, in this paper we aim to demonstrate that not only the initial conditions but also the weak, small-scale stochastic noises/disturbances can determine in which basin the solution of a sustained turbulent flow will reside for a long time.

In theory, the influence of {\em tiny} stochastic disturbances on turbulent flows is fundamental, since this kind of stochastic disturbances is often in a micro-level of magnitude and is {\em unavoidable} in practice.
There are two kinds of stochastic disturbances: natural and artificial. The thermal fluctuation belongs to the former, whereas the latter can be caused by many sources. Note that the background numerical noises, i.e. truncation errors and round-off errors, always exist for all numerical algorithms. Besides, it is widely believed that turbulence has a close relationship to chaos and thus the background numerical noises should increase exponentially (and quickly) until to the same level of ``true'' physical solution. Therefore, a computer-generated simulation of the NS equations is a kind of mixture of the ``true'' physical solution and ``false'' numerical noises. Note that the background numerical noise is tiny and random, dependent on different numerical algorithms and data accuracy, which itself is a kind of artificial stochastic disturbance. Therefore, in a natural way, the background numerical noises can be regarded as the sum of all kinds of tiny artificial stochastic disturbances.
In this paper, we investigate the {\em large-scale} influences of numerical noises as tiny artificial stochastic disturbances on a {\em sustained} turbulence by means of numerically solving the above-mentioned NS equations via a traditional numerical algorithm (Runge-Kutta's method with double precision, marked by RKwD) and the so-called ``Clean Numerical Simulation'' (CNS) \citep{Liao2009, Liao2013A, Liao2013B, hu2020risks, qin2020influence, xu2021accurate, Li2021SCPMA, Liao2022NA, AAMM-14-799}, separately, with the same initial/boundary conditions using the same values of physical parameters. The result given by the traditional numerical algorithm RKwD is a mixture of the ``true'' physical solution and ``false'' numerical noises, which are mostly at the same order of magnitude, where the background numerical noise is regarded as the sum of all tiny artificial stochastic disturbances. By contrast, the CNS can greatly reduce the background numerical noise to any a required level so that the numerical noises are {\em negligible} compared with the ``true'' physical solution, and thus the corresponding numerical result is convergent (reproducible) in an interval of time long enough for statistics, as described below. In other words, results given by the CNS can be regarded as a ``clean'' benchmark solution. Thus, we can investigate the influences of tiny artificial stochastic disturbances on turbulent flows by means of comparing the RKwD simulations with the CNS benchmark solution.

Then, let us briefly talk about the motivation and basic ideas of the CNS. Strictly speaking, all numerical simulations are not ``clean'', since background numerical noises (i.e. truncation errors and round-off errors) always exist there. Indeed, for non-chaotic systems, the magnitude of numerical noises can be controlled on a tiny level much smaller than ``true'' physical solution so that the influences of the numerical noises can be neglected.  However, for chaotic dynamical systems \citep{Li1975Period, Parker1989Practical, Lorenz1993The, PeterSmith1998Explaining, sprott2010}, numerical noises increase exponentially due to the ``sensitivity dependence on initial condition'' (SDIC), which was first discovered by \citet{poincare1890probleme} and later rediscovered by \citet{lorenz1963deterministic} with a more famous name ``butterfly-effect'': a hurricane happening in North America might be created by a flapping of the wings of a distant butterfly in South America several weeks earlier. More importantly, it was further found by \citet{lorenz1989computational, lorenz2006computational} that a chaotic dynamical system has the sensitivity dependence {\em not only} on initial condition (SDIC) {\em but also} on numerical algorithms (SDNA) in single/double precision. Such kind of uncertainty certainly raises serious doubt about the reliability of numerical simulations of chaotic systems. For example, \citet{Teixeira2007Time} carefully investigated the time-step sensitivity of three nonlinear atmospheric models by means of some traditional numerical algorithms (in single/double precision), and made a rather pessimistic conclusion that ``for chaotic systems, numerical convergence {\em cannot} be guaranteed {\em forever}''.

To overcome the above-mentioned limitations/restrictions of traditional numerical algorithms (in single/double precision) for chaotic dynamical systems, \citet{Liao2009} suggested a numerical strategy, namely the ``Clean Numerical Simulation'' (CNS). The basic idea of the CNS \citep{Liao2013A, Liao2013B, hu2020risks, qin2020influence, Li2021SCPMA, Liao2022NA, AAMM-14-799} is to greatly decrease the background numerical noises, i.e. truncation errors and round-off errors, to such a tiny level that the influence of numerical noises can be neglected in an interval of time $0\leq t \leq T_{c}$ that is long enough for statistics, where $T_{c}$ is the so-called ``critical predictable time''.
The CNS is based on such a well-known phenomenon: for a numerical/computer-generated simulation of chaotic dynamical system, the level of simulation-deviation (in an average meaning) from its (``true'')  physical  solution increases {\em exponentially} to a macroscopic one (at $t = T_{c}$), i.e.
\begin{equation}
{\cal E}(t) =  {\cal E}_0 \exp(K\hspace{0.3mm}t), \hspace{1.0cm}  t\in[0, T_{c}],  \label{def:delta-t}
\end{equation}
where $K > 0$ is the so-called ``noise-growing exponent'', ${\cal E}_0$ denotes the level of background numerical noise, ${\cal E}(t)$ is the level of simulation-deviation (in an average meaning) from the physical solution, respectively.  In theory, the critical predictable time $T_c$ is determined by a critical level ${\cal E}_c$ of simulation-deviation from its physical solution, say,
\begin{equation}
T_{c} = \frac{1}{K} \ln \left( \frac{{\cal E}_c}{ {\cal E}_0 }\right).    \label{def:delta_c}
\end{equation}
Obviously, for a given critical level ${\cal E}_c$, the smaller the level of the background numerical noise ${\cal E}_0$, the larger the critical predictable time $T_c$. This is the reason why in the frame of the CNS we have to greatly decrease the background numerical noises, i.e. truncation errors and round-off errors, to a tiny enough level. 
So, different from the Taylor series method,  the key point of the CNS is the so-called  ``critical predictable time''    $T_c$  that determines a temporal interval  $[0,T_c]$ in which the numerical simulations are ``reliable''  and  ``clean'',  since their ``false'' numerical noises are much smaller than the ``true'' physical solution and thus are negligible.  
For more details about the CNS, please refer to \citet{Liao2009, Liao2013A, Liao2013B} and his coauthors \citep{hu2020risks, qin2020influence, xu2021accurate, Li2021SCPMA, Liao2022NA}.

The CNS has been successfully applied to many chaotic dynamical systems. For example, by means of traditional numerical algorithms (in double precision), one can get convergent (i.e. reproducible) numerical simulations of the famous Lorenz equations just in a rather short interval of time, say, approximately $t\in [0,32]$. However, using the CNS, a convergent numerical simulation of the Lorenz equations was obtained first by \citet{Liao2009} in $t\in [0,1000]$ and then by \cite{LIAO2014On} in a much longer interval of time, i.e. $t\in[0, 10000]$. Besides, since the background numerical noises of the CNS can be much smaller even than the micro-level physical uncertainty, \cite{lin2017origin} successfully applied the CNS to provide a direct rigorous evidence that the micro-level  thermal fluctuation is the origin of macroscopic randomness of the turbulent Rayleigh-B{\'e}nard convection. Especially, it is worth noting that the CNS has been successfully applied to find more than $2000$ new families of periodic orbits of three-body system \citep{Li2017More, li2018over, li2019collisionless, Li2021SCPMA, Liao2022NA}. The discovery of these new periodic orbits was reported twice in the famous popular magazine {\em New Scientist} \citep{NewScientist2017, NewScientist2018}, because only three families of periodic orbits of the three-body problem had been reported in three hundred years after Newton mentioned this problem in 1687. All of these illustrate the validity, novelty, and great potential of the CNS.

Recently, an efficient CNS algorithm has been proposed to solve the spatio-temporal chaotic systems, i.e. the complex Ginzburg-Landau equation \citep{hu2020risks} and the damped driven sine-Gordon equation \citep{qin2020influence}, respectively. Using the CNS result as a benchmark solution, one can investigate the influence of numerical noises on the computer-generated simulation of a spatio-temporal chaotic system. It was found \citep{hu2020risks, qin2020influence} that numerical noises might lead to huge deviations of computer-generated simulations of some spatio-temporal chaotic systems not only in trajectories but also even in {\em statistics}.

In this paper, we apply the CNS and a traditional algorithm (based on the 4th-order Runge-Kutta's method with double precision, marked by RKwD), {\em separately}, to solve a sustained turbulence, i.e. the two-dimensional (2D) turbulent Rayleigh-B{\'e}nard convection (RBC). Note that, using the CNS, the background numerical noises can be decreased to such a tiny level that the numerical noises are negligible in a long enough interval of time, so that the CNS result can be regarded as a benchmark solution of the ``true'' physical result to investigate the influence of tiny artificial stochastic disturbances by comparing the RKwD simulations with the CNS benchmark solution.
In this way, we provide a rigorous evidence that tiny artificial stochastic disturbances have huge influences on large-scale properties of the turbulent Rayleigh-B{\'e}nard convection not only in statistics but also even in flow types: the CNS benchmark solution always keeps the vortical/roll-like turbulent convection, however, for the RKwD simulations, the shearing convection occurs and its corresponding flow field turns to zonal flow thereafter. This phenomenon is reasonable if the boundaries of different attractor basins (mentioned above) in this multistable system are intricately interwoven as has been observed in other cases \citep{shrimali2008nature}.

\section{Mathematical model for 2D turbulent RBC}    \label{sec2}

The buoyancy-driven convection in a fluid layer between two horizontal parallel plates heated from below and cooled from above, known as the Rayleigh-B{\'e}nard convection (RBC) for compressible viscous fluids, is one of the most fundamental and classic paradigms of nonlinear dynamics in fluid mechanics. It was first investigated by \cite{Rayleigh1916On} and the continuous efforts devoted to the study of this problem have greatly enriched our understanding \citep{Chandrasekhar1961Hydrodynamic, Schluter1965On, Heslot1987Transitions, Kadanoff2001Turbulent, Niemela2006Turbulent, Ahlers2009Heat, lohse2010small, Zhou2013Thermal, Goluskin2014Convectively, wang2020From}.

\begin{figure}
    \begin{center}
        \begin{tabular}{cc}
             \subfigure[]{\includegraphics[width=2.5in]{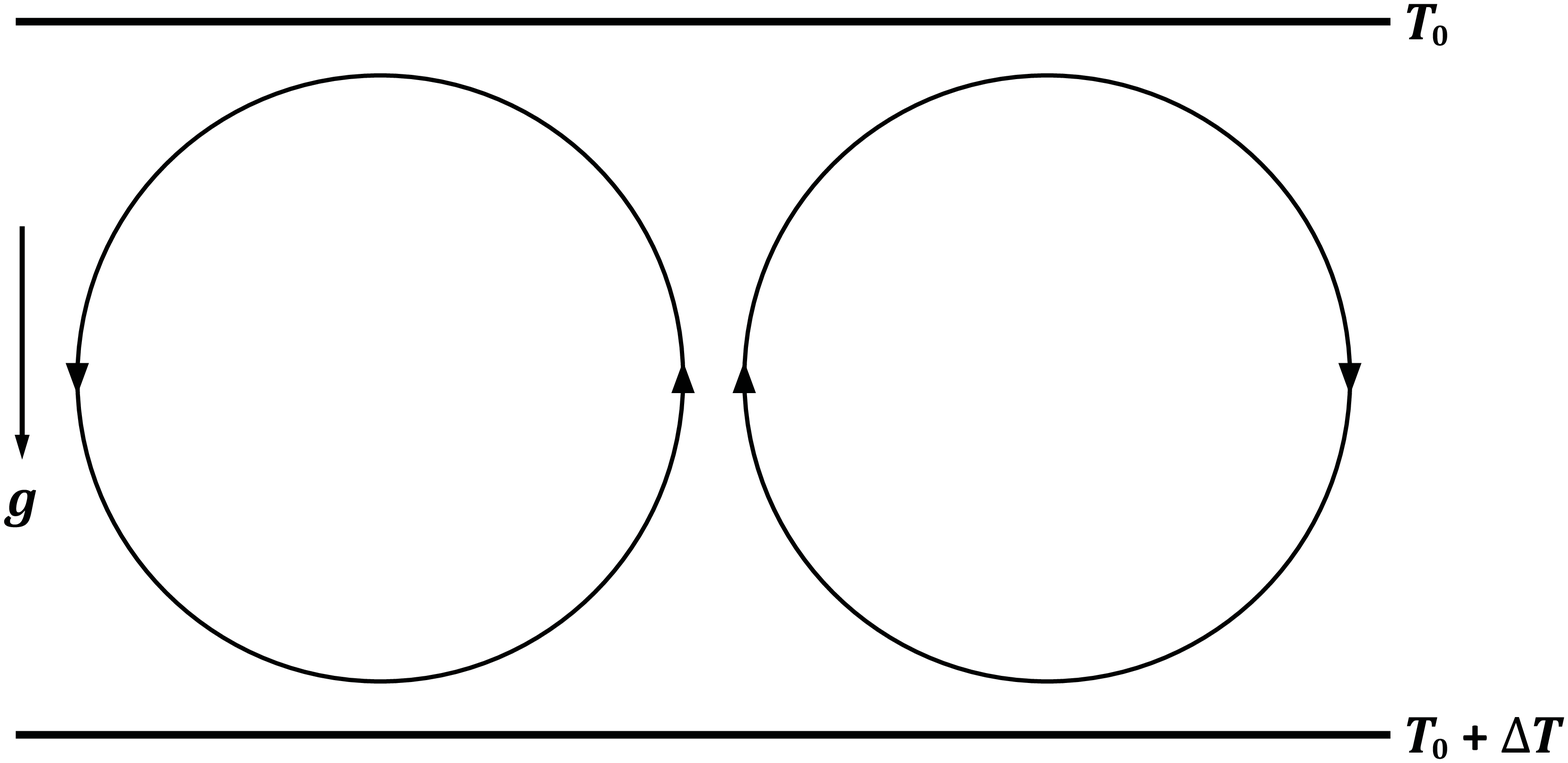}}
             \subfigure[]{\includegraphics[width=2.5in]{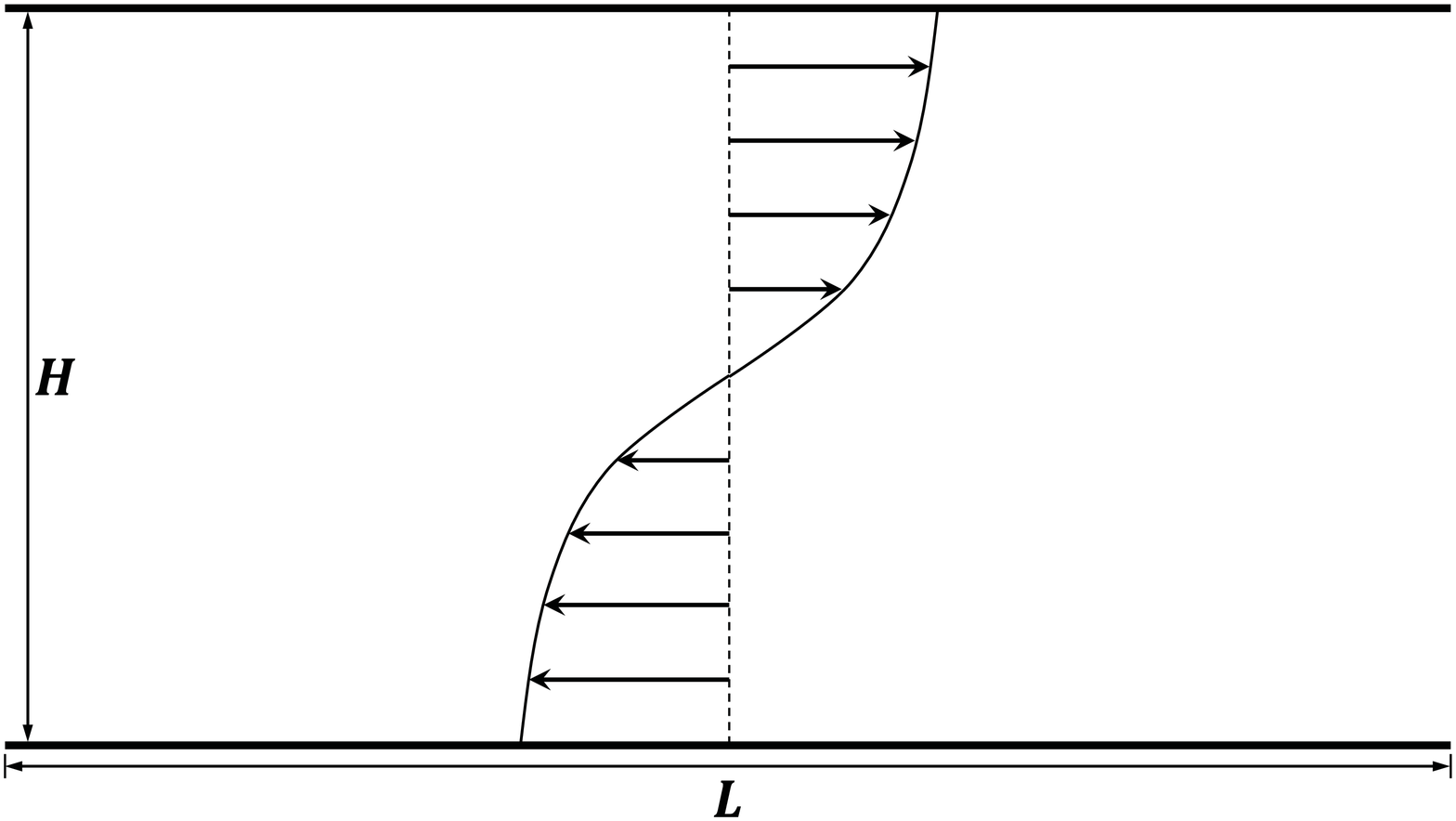}}
        \end{tabular}
 \end{center}
\caption{Schematic drawings of two-dimensional turbulent Rayleigh-B{\'e}nard convections in two totally different flow types: (a) typical vortical/roll-like flow and (b) zonal flow. The fluid layer between two parallel plates which are separated by a height $H$ obtains heat from the bottom boundary surface because of the constant temperature difference $\Delta T>0$, where $L$ is the horizontal length of computational domain and the downward direction of gravity acceleration $g$ is indicated.}
\label{schematic_drawings}
\end{figure}

As illustrated in Figure~\ref{schematic_drawings}, a thin layer of fluid is confined between two horizontal plates that are separated by a distance $H$, where $T_{0}$ and $T_{0}+\Delta T$ denote the temperatures of the top and bottom boundary surfaces, $L$ is the horizontal length of computational domain and $g$ is the gravity acceleration, respectively.
The typical vortical/roll-like motions of two-dimensional (2D) RBC, as illustrated in Figure~\ref{schematic_drawings}(a), and their corresponding turbulent states, have been studied extensively by means of the direct numerical simulation (DNS) \citep{saltzman1962finite, Fromm1965Numerical, Veronis1968Large, moore1973two, curry1984order, zienicke1998bifurcations, johnston2009comparison, Huang2013Counter, Zhang2017Statistics, Zhu2018Transition}. However, there is another type of flow in the 2D turbulent RBC in the case of the free-slip boundary conditions imposed on two horizontal parallel plates and the periodic boundary conditions on the left and right sides, namely zonal flow \citep{Goluskin2014Convectively, van2014Effect, Hardenberg2015Generation, wang2020From}, as shown in Figure~\ref{schematic_drawings}(b). It is worth noting that such kind of turbulent zonal flow has been widely found in nature and laboratory such as in the atmosphere of Jupiter \citep{Heimpel2005Simulation, Kaspi2018Jupiter} and some Jovian planets \citep{Sun1993Banded, Cho1996The, Yano2003Outer}, in the oceans \citep{Maximenko2005Observational, Richards2006Zonal}, in the Earth's outer core \citep{Miyagoshi2010Zonal}, in toroidal tokamak devices \citep{Diamond2005Zonal} and so on. Thus, here we choose the 2D turbulent RBC with the free-slip boundary conditions at the upper and lower plates and the periodic boundary conditions in the horizontal direction as our mathematical model for the sustained turbulence, governed by the NS equations.

Using the length scale $H$, velocity scale $\sqrt{g\alpha H\Delta T}$ and temperature scale $\Delta T$ as the characteristic scales, the corresponding dimensionless NS equations, combined with the Boussinesq approximation \citep{saltzman1962finite},  read
\begin{equation}
\frac{\partial}{\partial t}\nabla^{2}\psi+\frac{\partial(\psi,\nabla^{2}\psi)}{\partial(x,z)}-\frac{\partial\theta}{\partial x}-\sqrt{\frac{Pr}{Ra}}\nabla^{4}\psi=0,       \label{RB_psi}
\end{equation}
\begin{equation}
\frac{\partial\theta}{\partial t}+\frac{\partial(\psi,\theta)}{\partial(x,z)}-\frac{\partial\psi}{\partial x}-\frac{1}{\sqrt{PrRa}}\nabla^{2}\theta=0,       \label{RB_theta}
\end{equation}
where $\psi$ is a stream-function with the definition
\begin{equation}
u=-\frac{\partial\psi}{\partial z}, \hspace{1.0cm} w=\frac{\partial\psi}{\partial x},       \label{psi}
\end{equation}
in which $u$ and $w$ are the horizontal and vertical velocities, $\theta$ is the temperature departure from a linear variation background, i.e. the temperature is expressed as $T=\theta-z+1$ in the case of $T_0=0$, $t$ denotes the time,  $x\in[0,\Gamma]$ and $z\in[0,1]$ are the horizontal and vertical position coordinates, $\Gamma=L/H$ denotes the aspect ratio, $\nabla^{2}$ is the Laplace operator and thus $\nabla^{4}=\nabla^{2}\nabla^{2}$,
\begin{equation}
\frac{\partial(a,b)}{\partial(x,z)}=\frac{\partial a}{\partial x}\frac{\partial b}{\partial z}-\frac{\partial b}{\partial x}\frac{\partial a}{\partial z}       \label{Jacobi}
\end{equation}
is the Jacobi operator, the Rayleigh number $Ra$ and Prandtl number $Pr$ are defined by
\begin{equation}
Ra=\frac{g\alpha H^{3}\Delta T}{\nu\kappa}, \hspace{1.0cm} Pr=\frac{\nu}{\kappa},       \label{Ra_Pr}
\end{equation}
respectively, in which $\alpha$ is the thermal expansion coefficient and $\nu=\mu/\rho$ is the kinematic viscosity.

Note that the free-slip boundary conditions are adopted at the upper and lower plates where temperatures are assumed to be constant. Hence, the NS equations (\ref{RB_psi})-(\ref{RB_theta}) have the following boundary conditions:
\begin{equation}
\psi=\frac{\partial^{2}\psi}{\partial z^{2}}=\theta=0       \label{free-slip}
\end{equation}
at $z=0$ and $z=1$. On the other hand, since the fluid layer can extend to infinity in the horizontal direction, we adopt the periodic boundary conditions for $\psi$ and $\theta$ at the lateral boundaries in the horizontal direction, i.e. at $x=0$ and $x = \Gamma$.

Without loss of generality, in this paper let us consider the case with the aspect ratio $\Gamma=L/H=2\sqrt{2}$, which is large enough for the approximation of heat flux at an infinite aspect ratio \citep{saltzman1962finite,curry1984order,lin2017origin}, the Prandtl number $Pr=6.8$ (corresponding to the water at room temperature 20$^{\circ}$C), and the Rayleigh number $Ra=6.8\times10^{8}$ corresponding to a turbulent state, respectively. In addition, the initial temperature and velocity fields are randomly generated by the thermal fluctuations in Gaussian white noises \citep{lin2017origin}, with the temperature standard deviation $\sigma_T=10^{-10}$ and the velocity standard deviation $\sigma_u=10^{-9}$, respectively.

\section{Influence of numerical noises as artificial stochastic disturbances}    \label{ssec}

The deterministic NS equations (\ref{RB_psi})-(\ref{RB_theta}) are numerically solved here, separately, by means of a traditional algorithm (Runge-Kutta's method with double precision, marked by RKwD) whose numerical noises are mostly in the same order of magnitude as the ``true'' physical solution, and the CNS whose numerical noises are much less than the physical solution and thus negligible. By means of comparing the RKwD simulations with the CNS benchmark solution, it is found that the numerical noises indeed might lead to huge large-scale differences even in statistics and flow types of the 2D turbulent RBC, as described below.

First, we apply the clean numerical simulation (CNS) to greatly decrease the background numerical noises, i.e. the truncation errors and round-off errors, to such a tiny level that the numerical noises are much smaller than and thus negligible compared with the ``true'' physical solution of the 2D turbulent RBC in an interval of time that is long enough for statistics. In this way, a convergent (reproducible) solution of the 2D turbulent RBC can be obtained, which is used here as the ``clean''  benchmark solution. On the other hand, with the {\em same} initial/boundary conditions and the {\em same} physical parameters as described in the previous section, the NS equations (\ref{RB_psi})-(\ref{RB_theta}) are also numerically solved by a traditional algorithm, i.e. the fourth-order Runge-Kutta's method with double precision (marked by RKwD) using the time-step $\Delta t = 1\times10^{-4}$, whose numerical noises increase exponentially until to the same level of the ``true'' physical solution and thus are {\em not} negligible. By comparing these RKwD simulations with the CNS benchmark solution, we can investigate the influence of the numerical noises as tiny artificial stochastic disturbances on the 2D turbulent RBC in detail. In this section, we only briefly show some results of comparison. For details about the CNS algorithms, please refer to Appendix A.

Briefly speaking, to decrease the spatial truncation-error to a small enough level, we discretize the spatial domain of flow field by a uniform mesh $N_x\times N_z =1024\times1024$, and besides apply the Fourier spectral method with the $3/2$ rule for dealiasing \citep{pope2001turbulent}.
The corresponding spatial resolution is high enough for the considered turbulent RBC: the horizontal (maximum) grid spacing $\Delta_x=L/N_x=0.00276$ is less than the minimum Kolmogorov scale \citep{pope2001turbulent}, which will be shown later in details.
Besides, to decrease the temporal truncation-error to a small enough level for the CNS, we use the 45th-order (i.e. $M=45$) Taylor expansion with a time step $\Delta t=10^{-3}$. In addition, to decrease the round-off error to a small enough level, we use 70 significant digits (i.e. $N_{s}=70$) in multiple precision for all physical/numerical variables and parameters.
Similarly, we get another CNS result using the Fourier spectral method on the same uniform mesh $N_x\times N_z =1024\times1024$ with the even smaller background numerical noises by means of a higher order (i.e. $M=47$) Taylor expansion with the same time step ($\Delta t=10^{-3}$) and the higher multiple precision with more significant digits (i.e. $N_{s}=72$). Comparing these two CNS results, it is found that they have no distinct differences in an interval of time $0 \leq t\leq 500$, which is long enough for statistics. This verifies the convergence and reliability of our CNS result in $t\in[0,500]$ given by means of $M=45$, $\Delta t=10^{-3}$, and $N_{s}=70$, which is therefore used in the following parts as the ``clean'' benchmark solution.

\begin{figure}
    \begin{center}
        \begin{tabular}{cc}
            \includegraphics[width=2.5in]{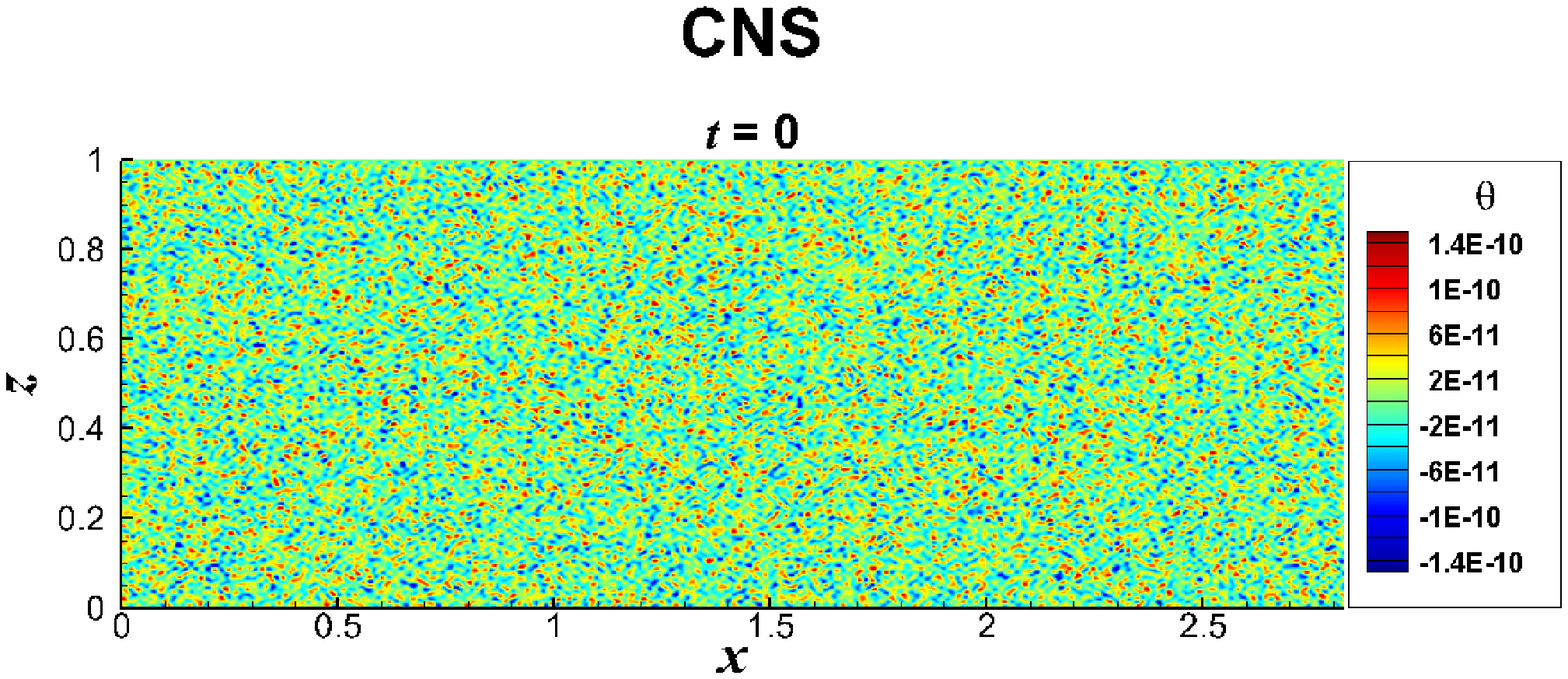}
            \includegraphics[width=2.5in]{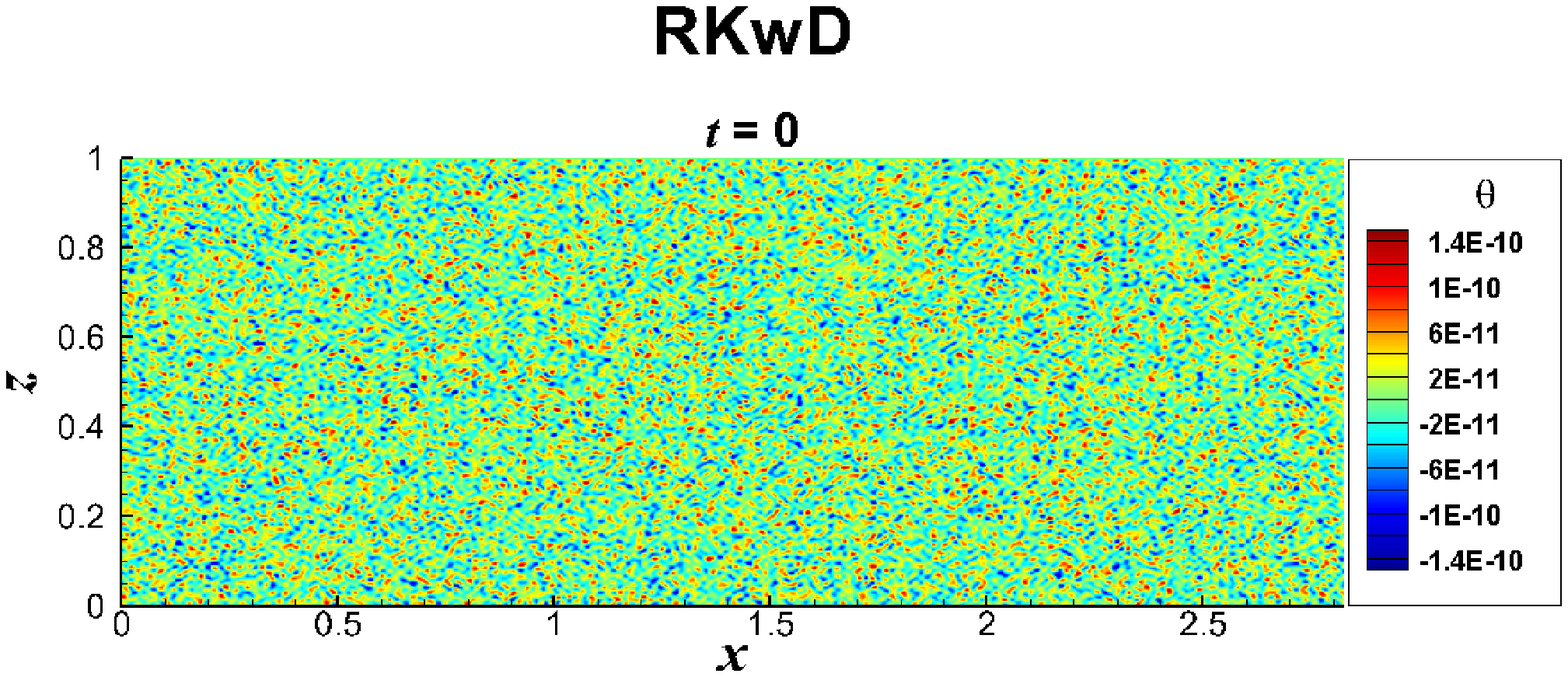} \\
            \includegraphics[width=2.5in]{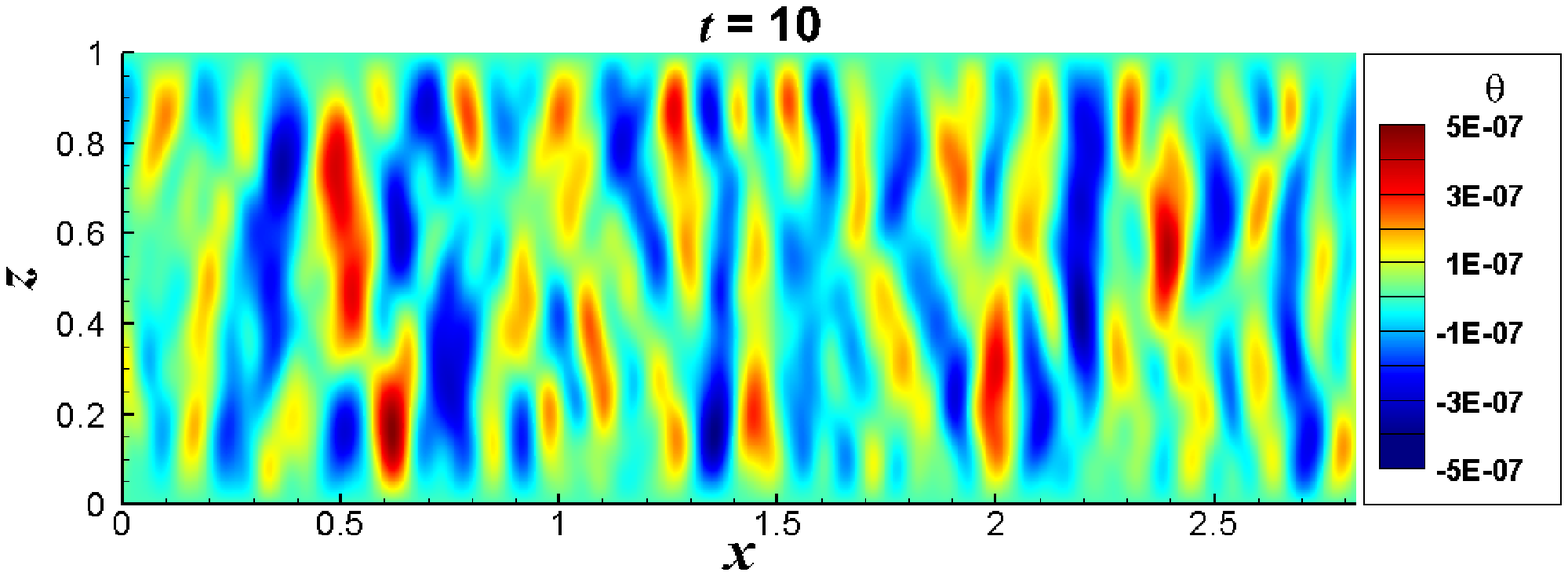}
            \includegraphics[width=2.5in]{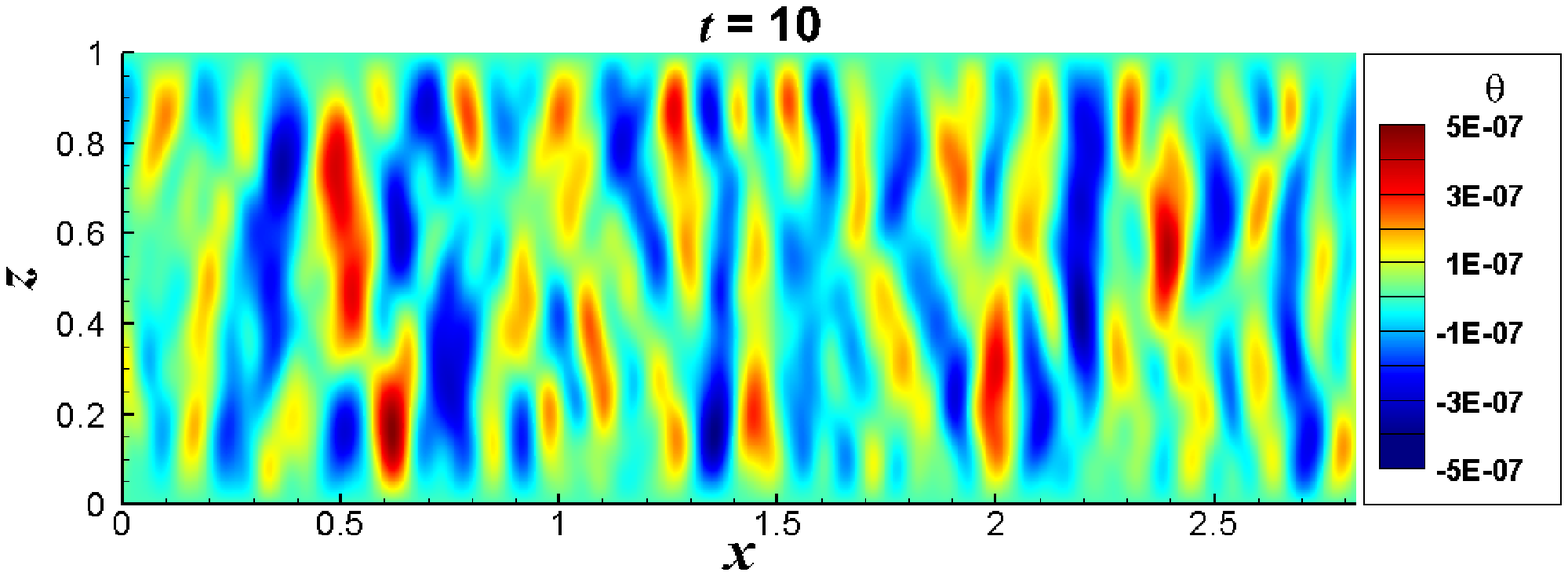} \\
            \includegraphics[width=2.5in]{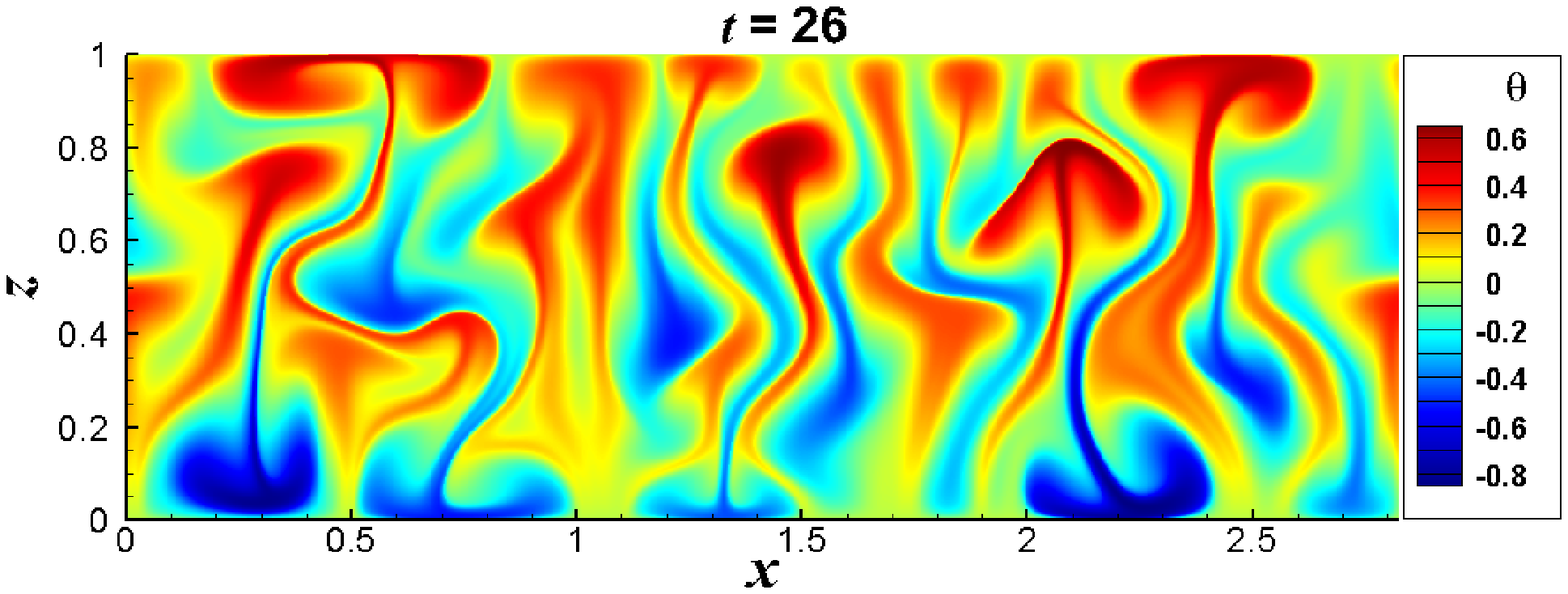}
            \includegraphics[width=2.5in]{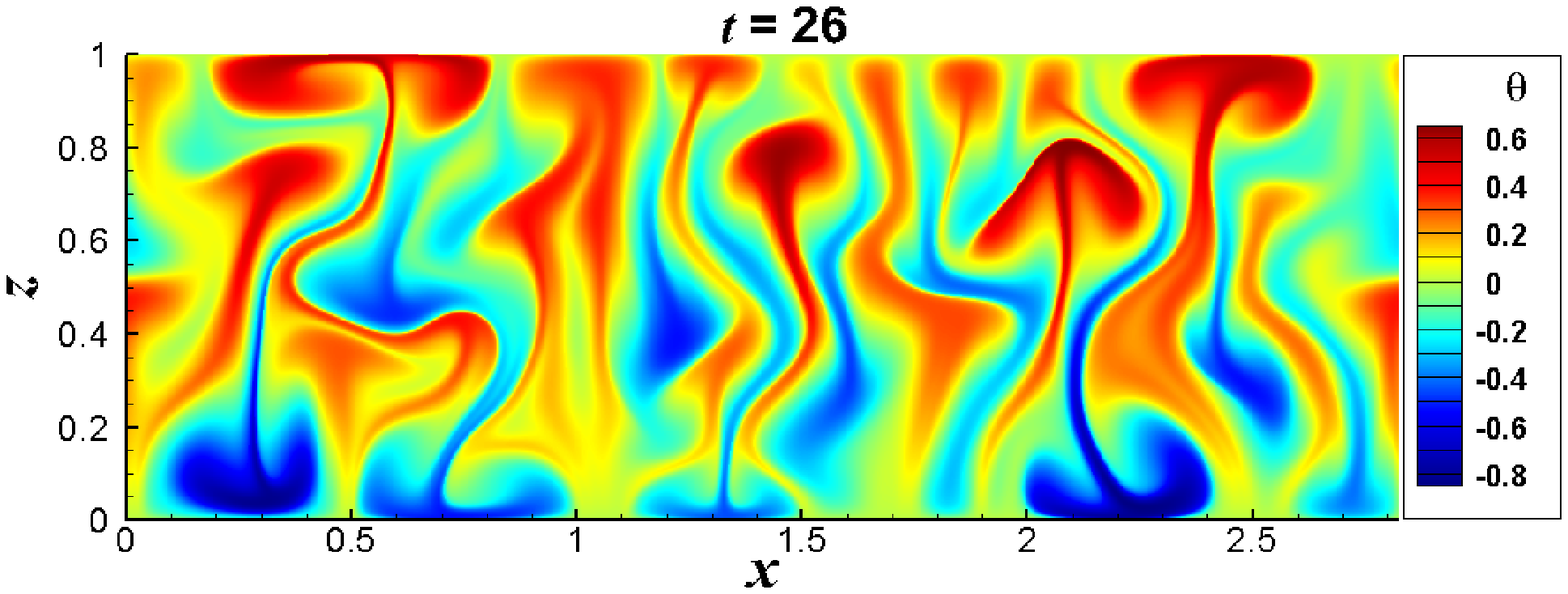} \\
            \includegraphics[width=2.5in]{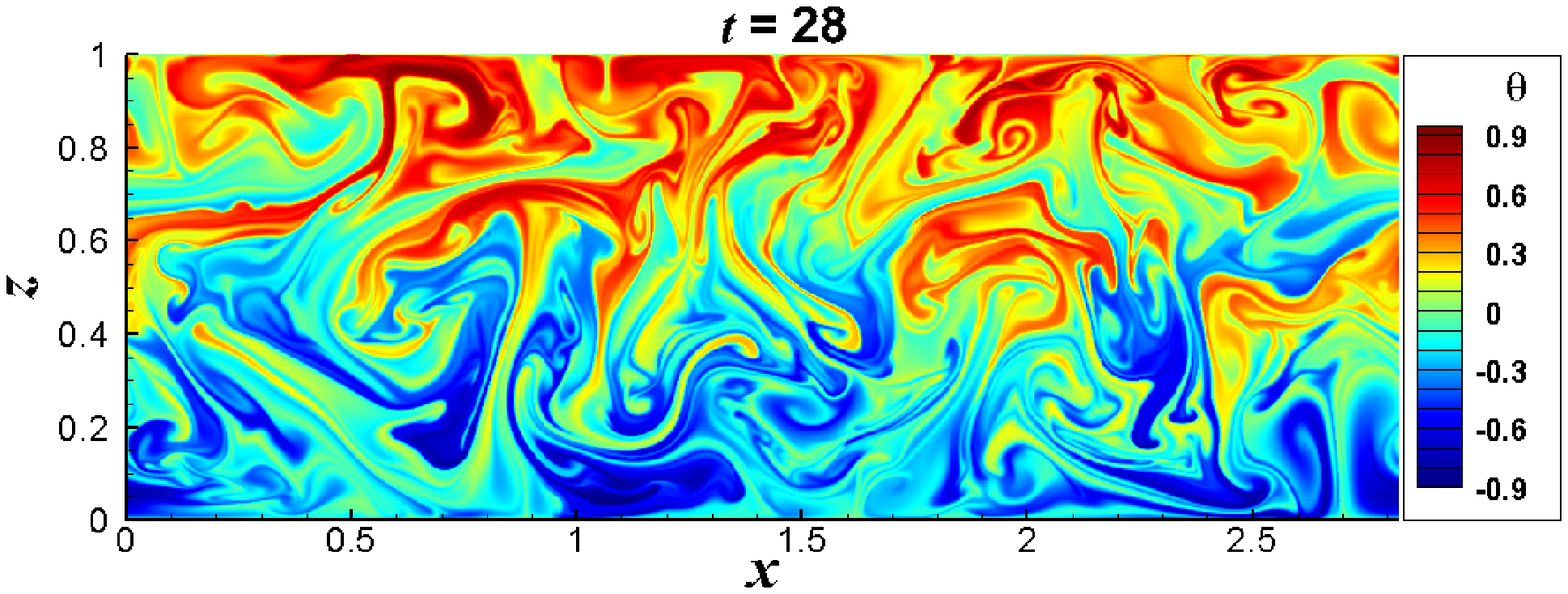}
            \includegraphics[width=2.5in]{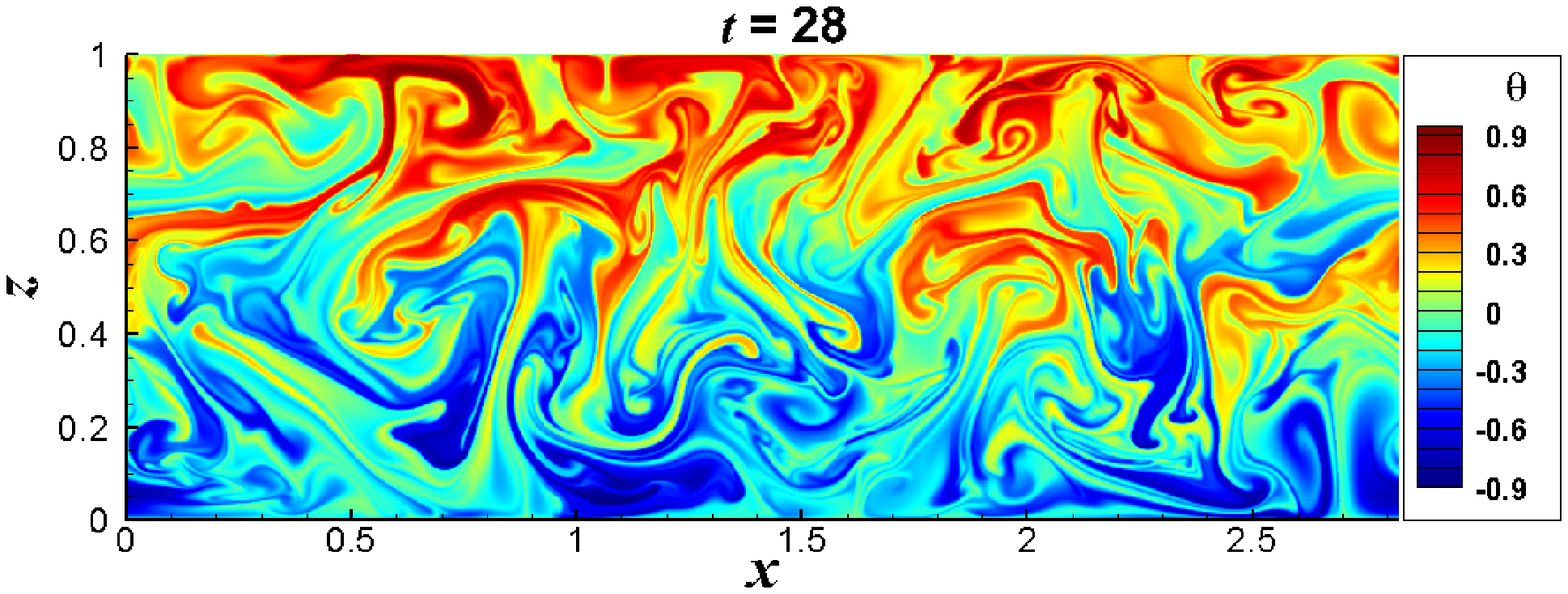} \\
            \includegraphics[width=2.5in]{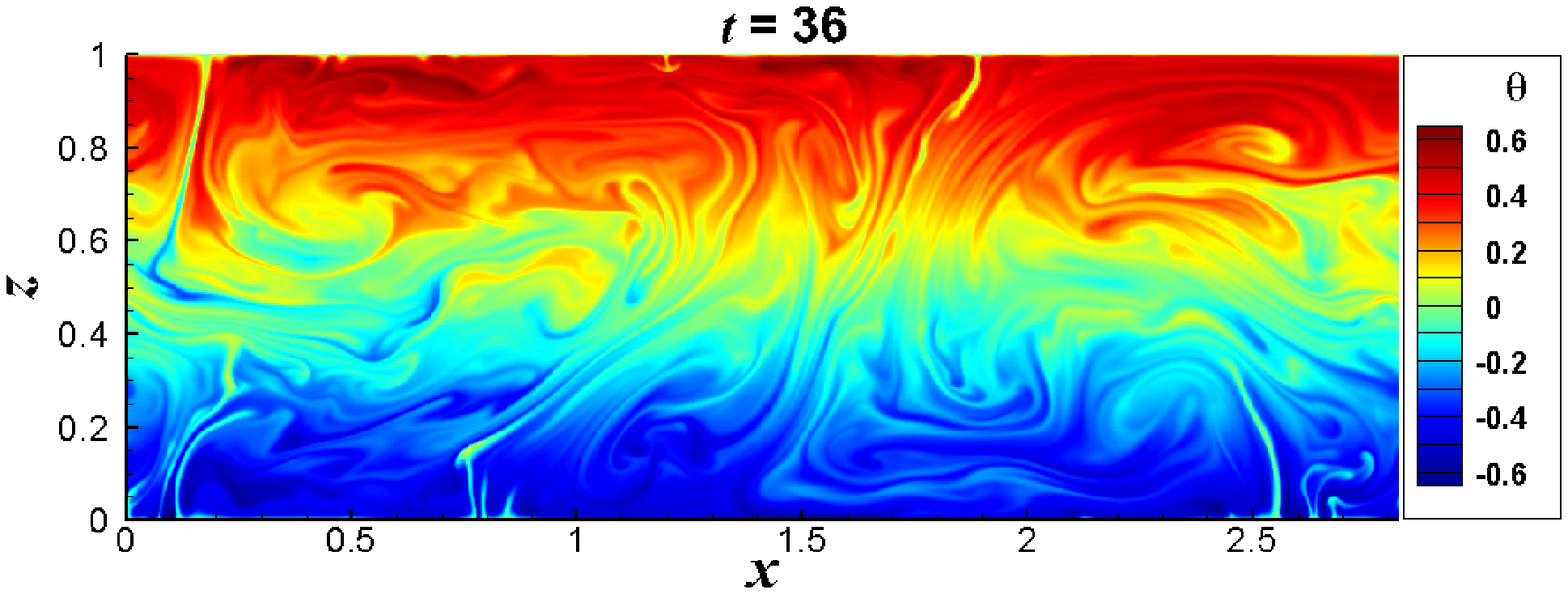}
            \includegraphics[width=2.5in]{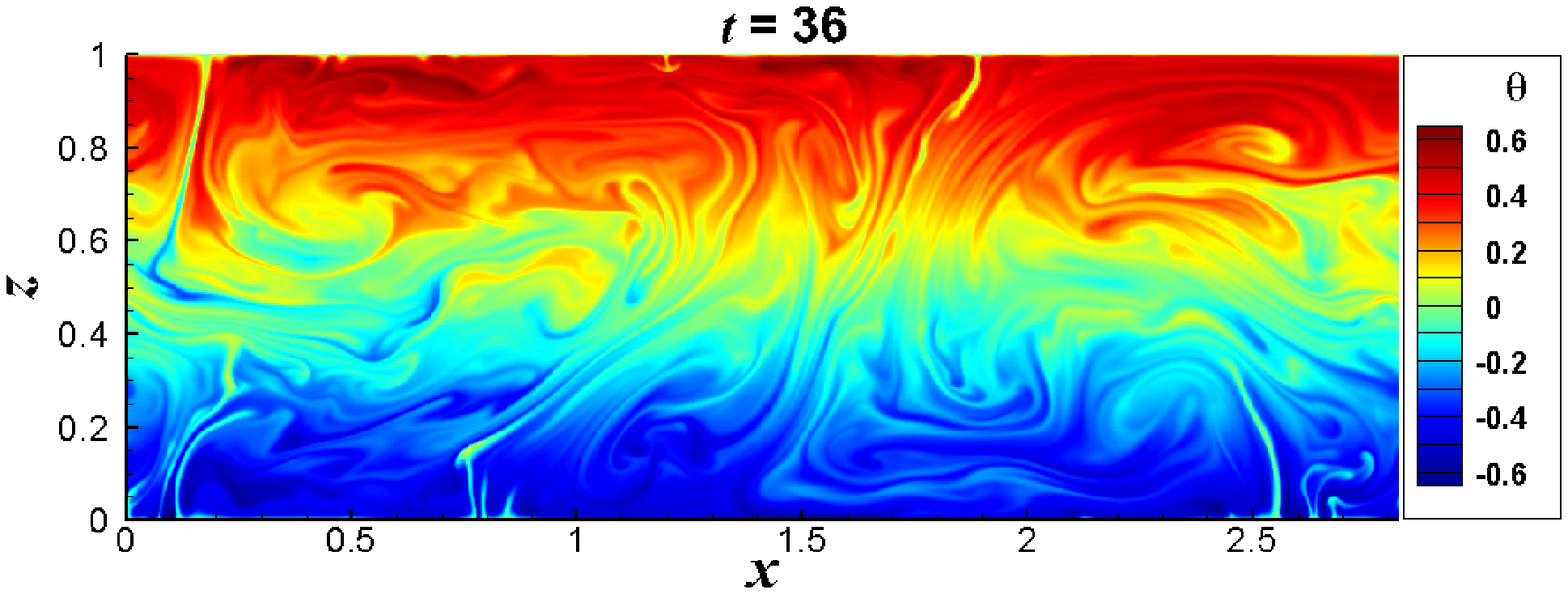} \\
            \includegraphics[width=2.5in]{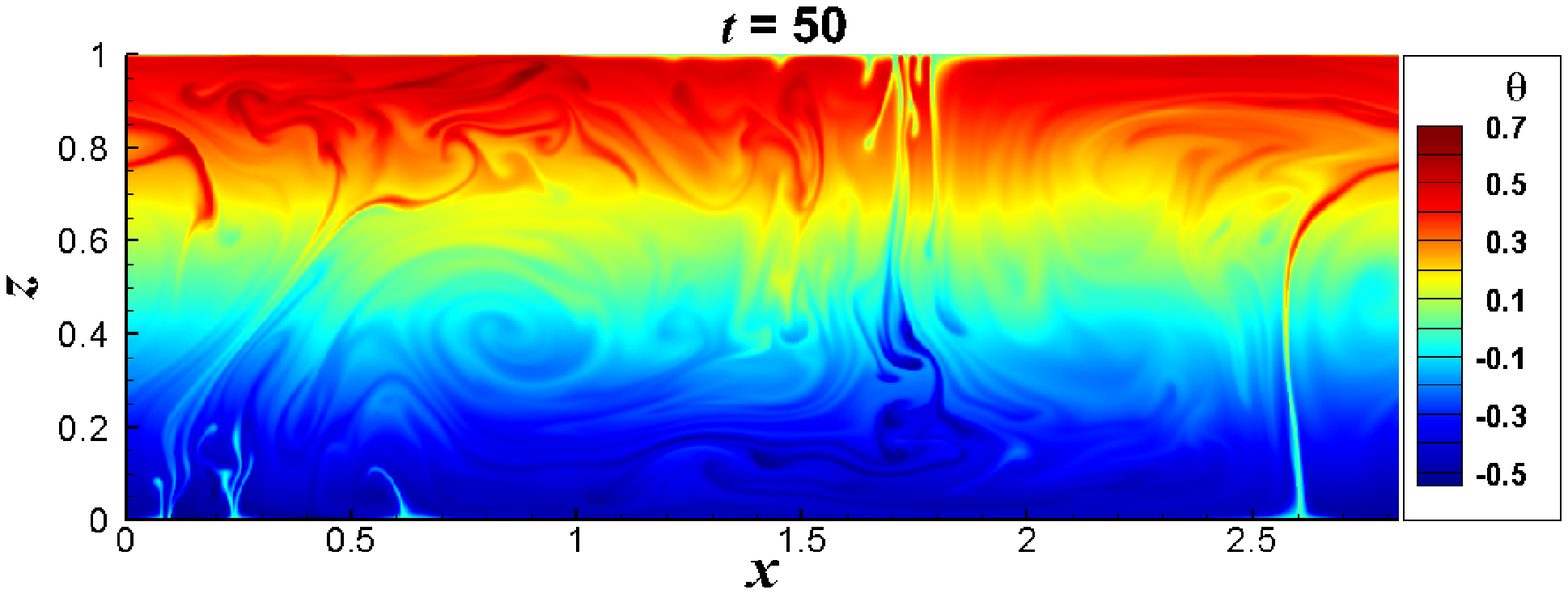}
            \includegraphics[width=2.5in]{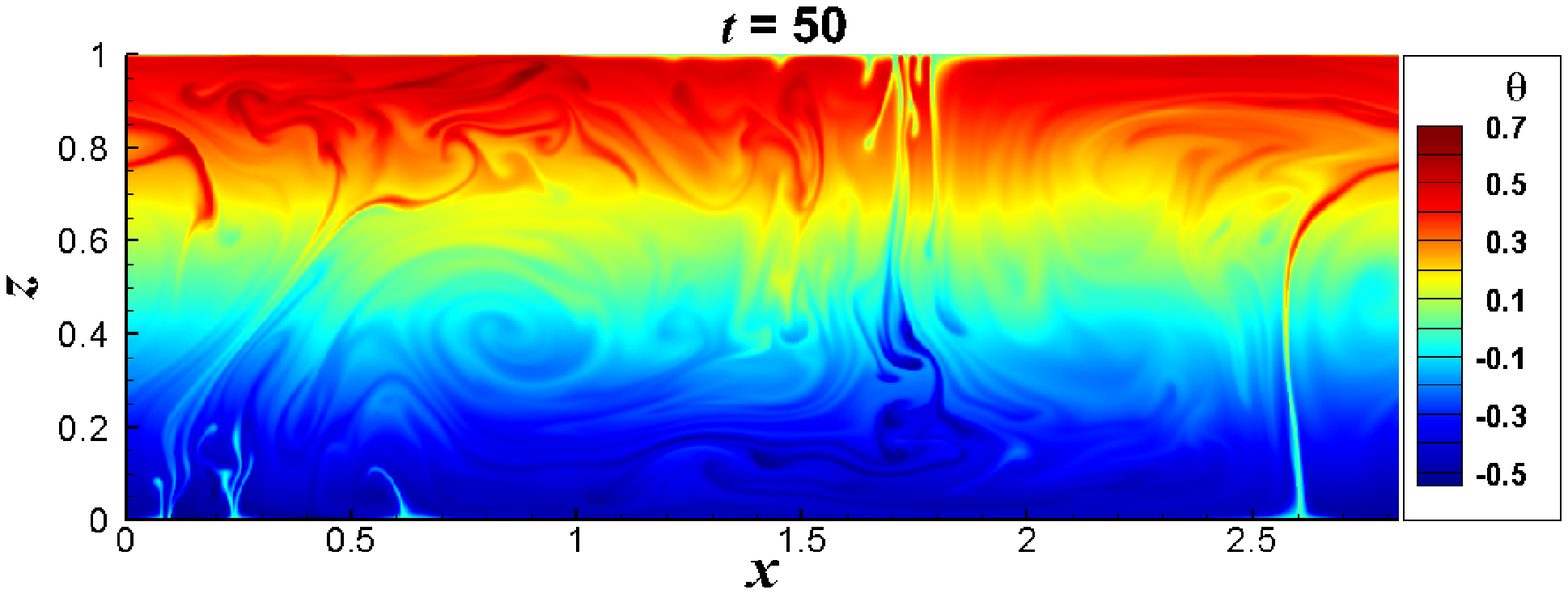} \\
            \includegraphics[width=2.5in]{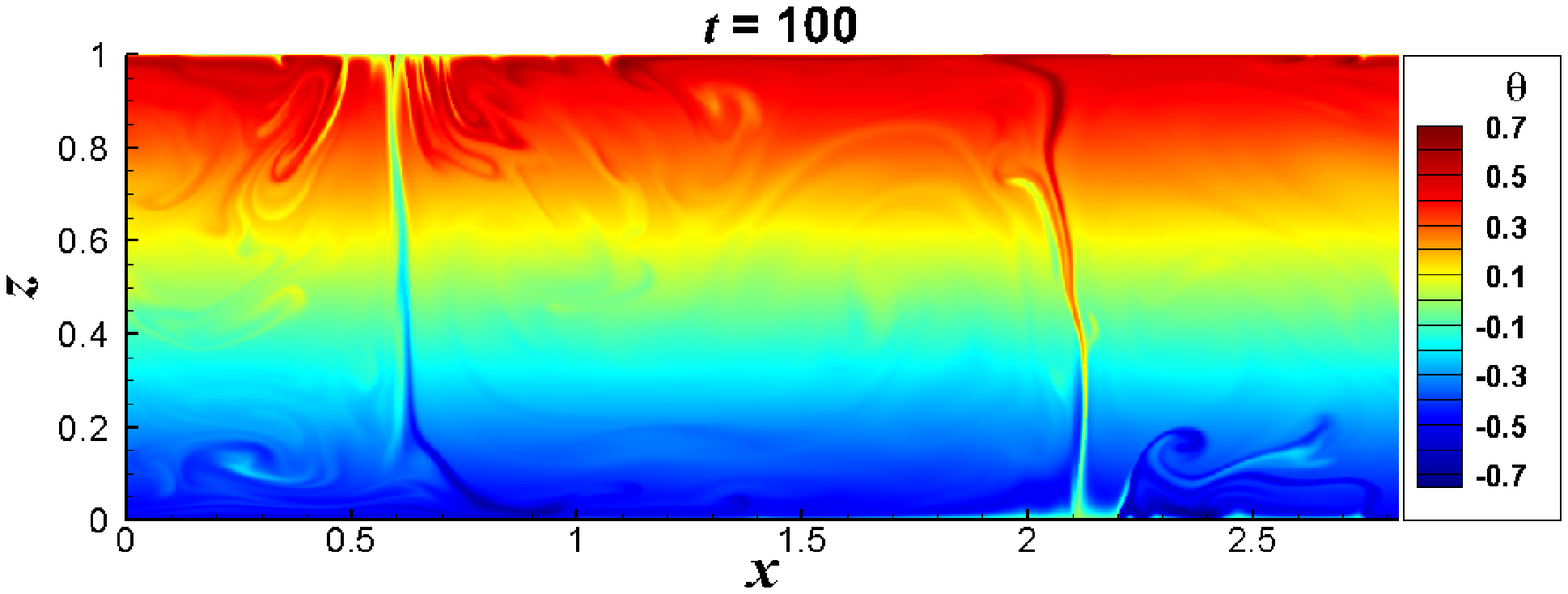}
            \includegraphics[width=2.5in]{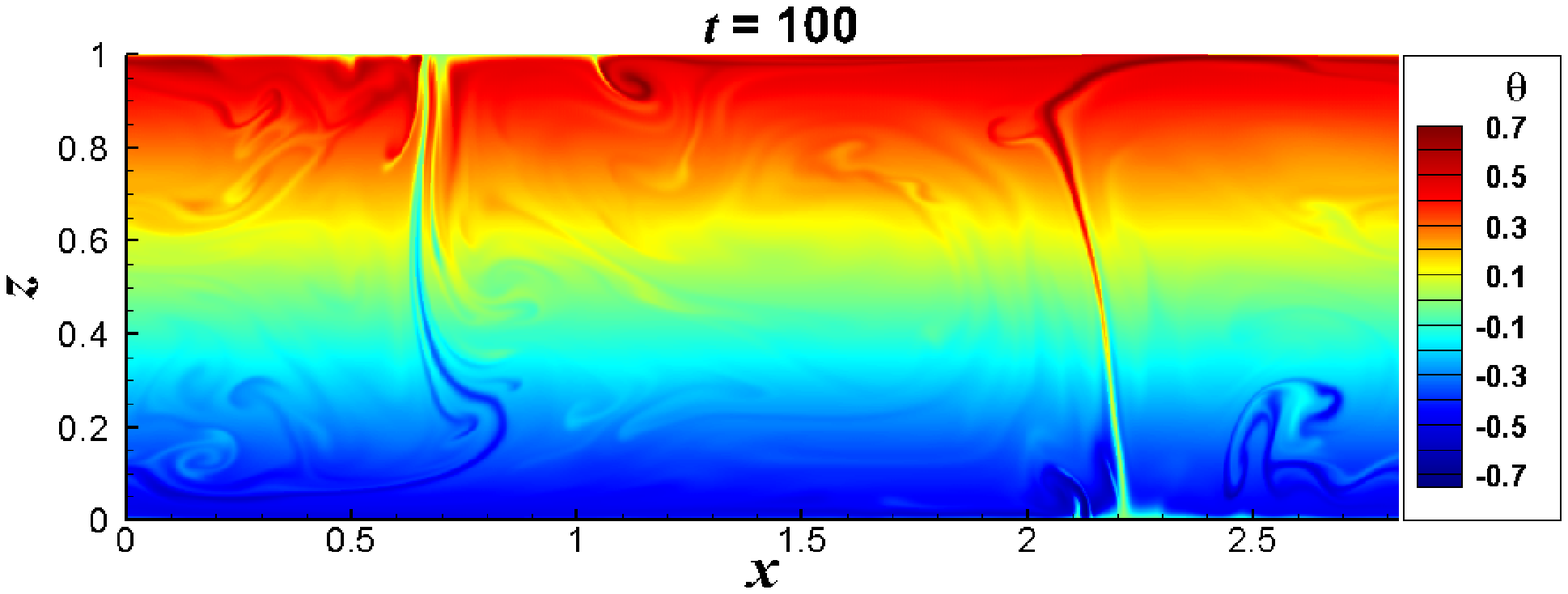} \\
            \includegraphics[width=2.5in]{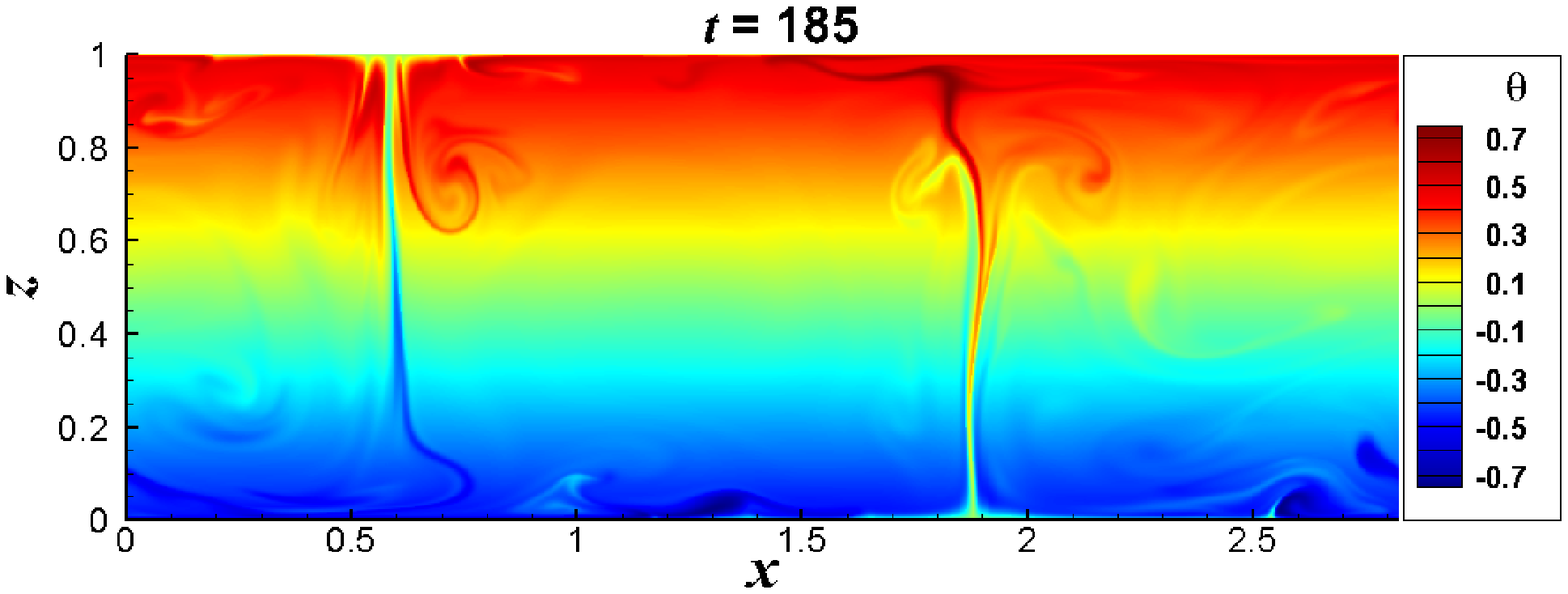}
            \includegraphics[width=2.5in]{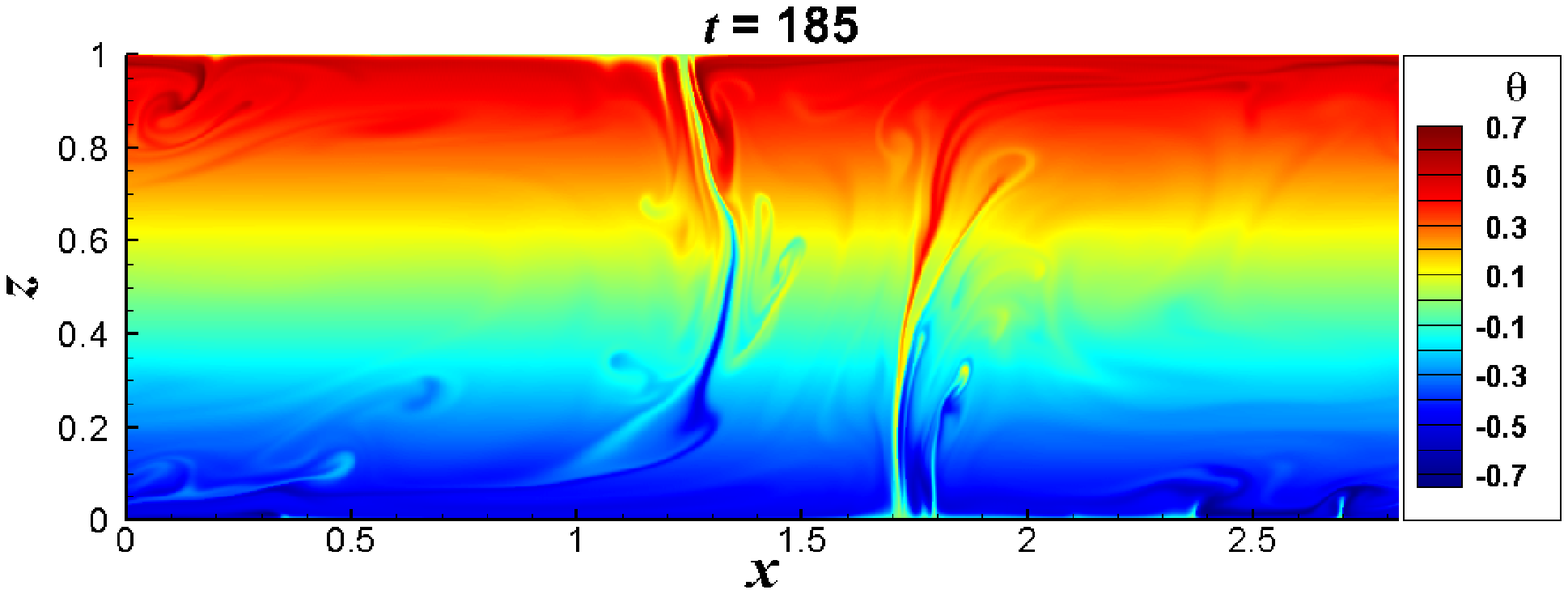}
        \end{tabular}
    \caption{Evolutions of the $\theta$ (dimensionless temperature departure from a linear variation background) field in the case of $Pr = 6.8$, $Ra = 6.8 \times 10^{8}$, and $L/H = 2\sqrt{2}$. Left: the CNS benchmark solution; Right: the RKwD simulation given by the Runge-Kutta's method with double precision using $\Delta t = 10^{-4}$. The corresponding movie is available at \href{https://github.com/sjtu-liao/RBC/blob/main/RBC-mv.mp4}{https://github.com/sjtu-liao/RBC/blob/main/RBC-mv.mp4}.}    \label{Contour}
    \end{center}
\end{figure}

\begin{figure}
    \begin{center}
        \begin{tabular}{cc}
            \includegraphics[width=2.5in]{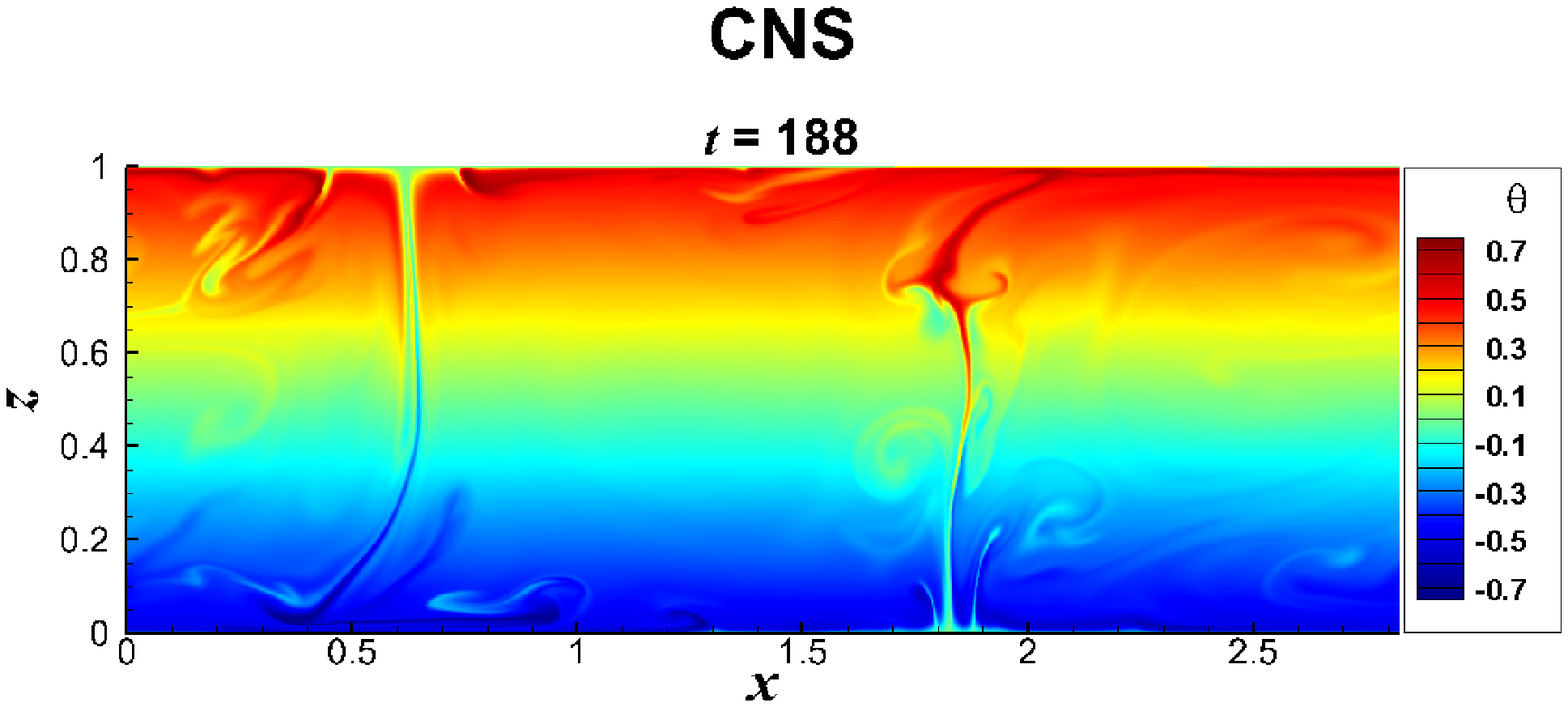}
            \includegraphics[width=2.5in]{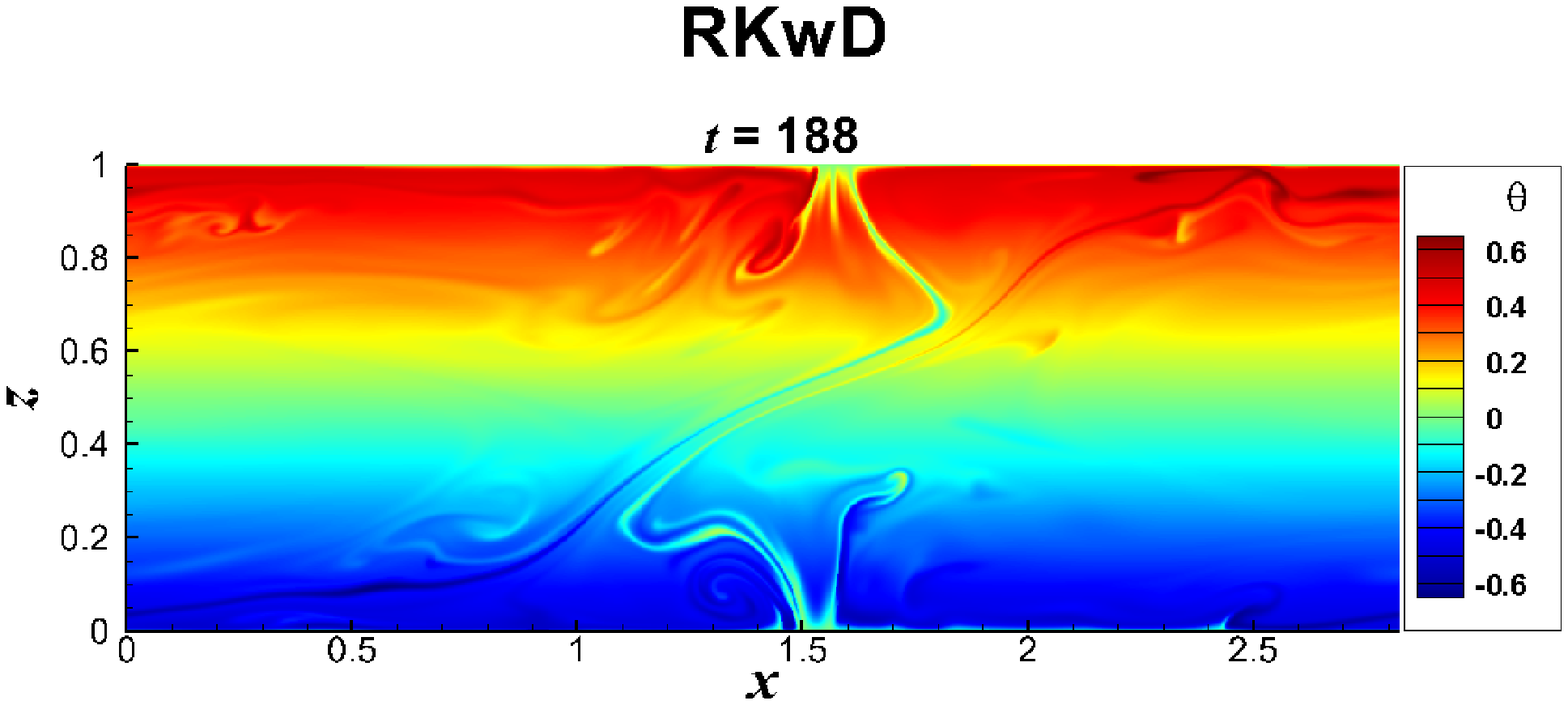} \\
            \includegraphics[width=2.5in]{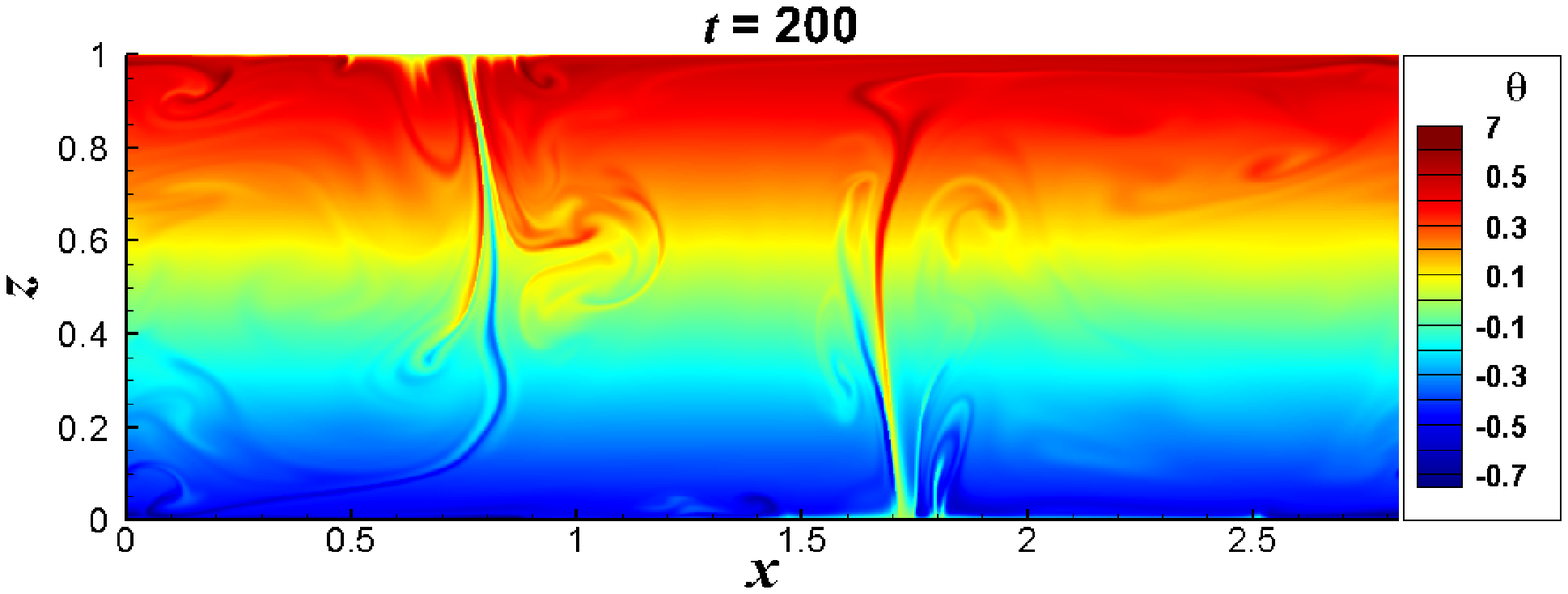}
            \includegraphics[width=2.5in]{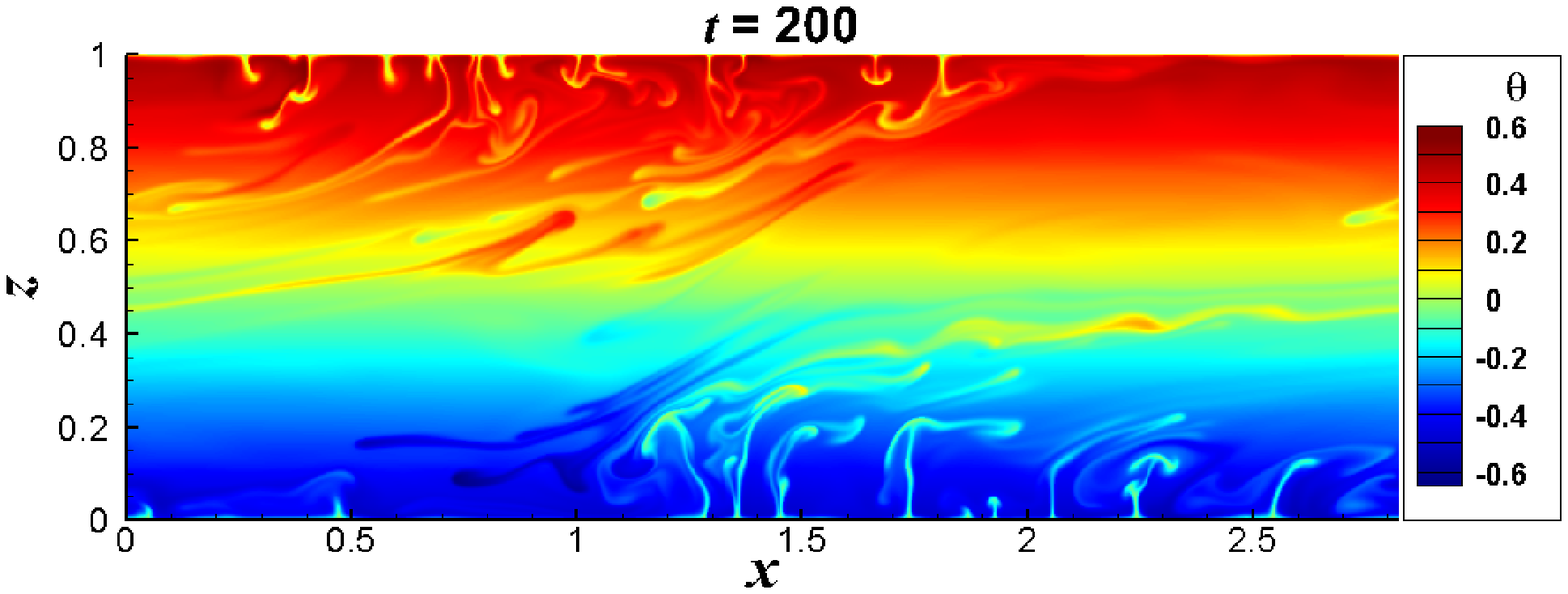} \\
            \includegraphics[width=2.5in]{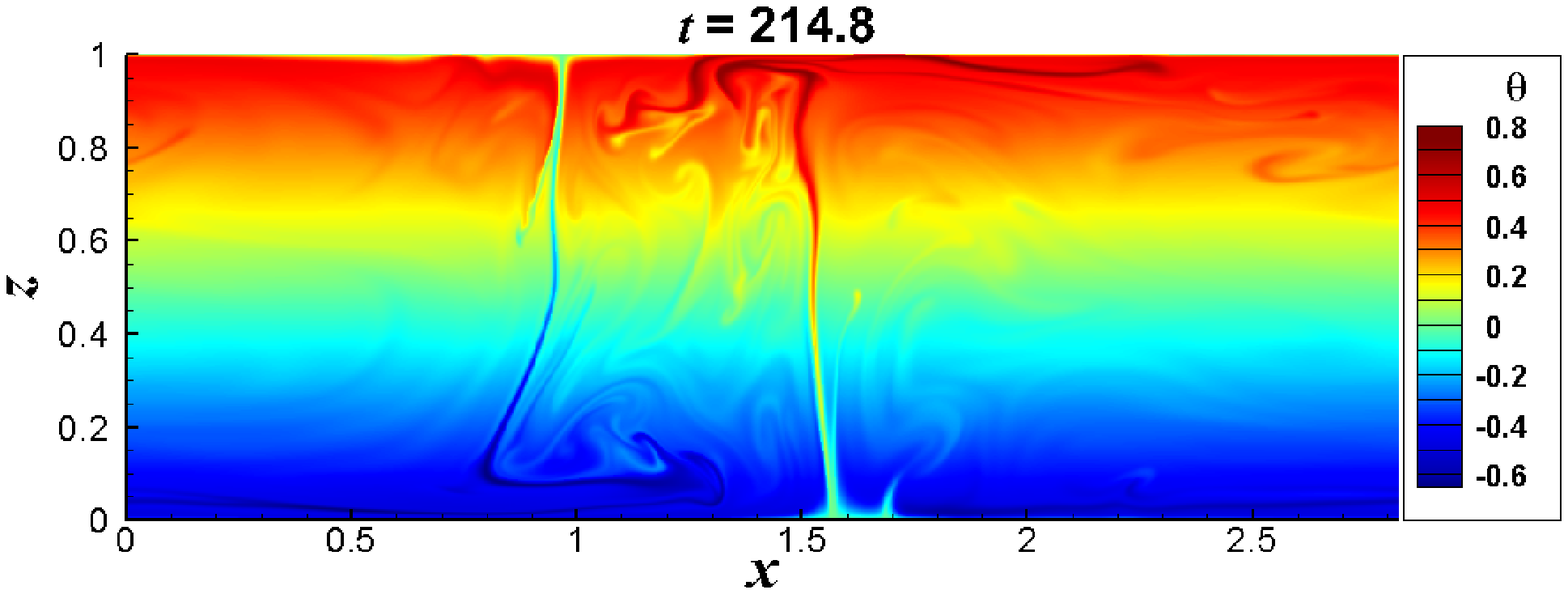}
            \includegraphics[width=2.5in]{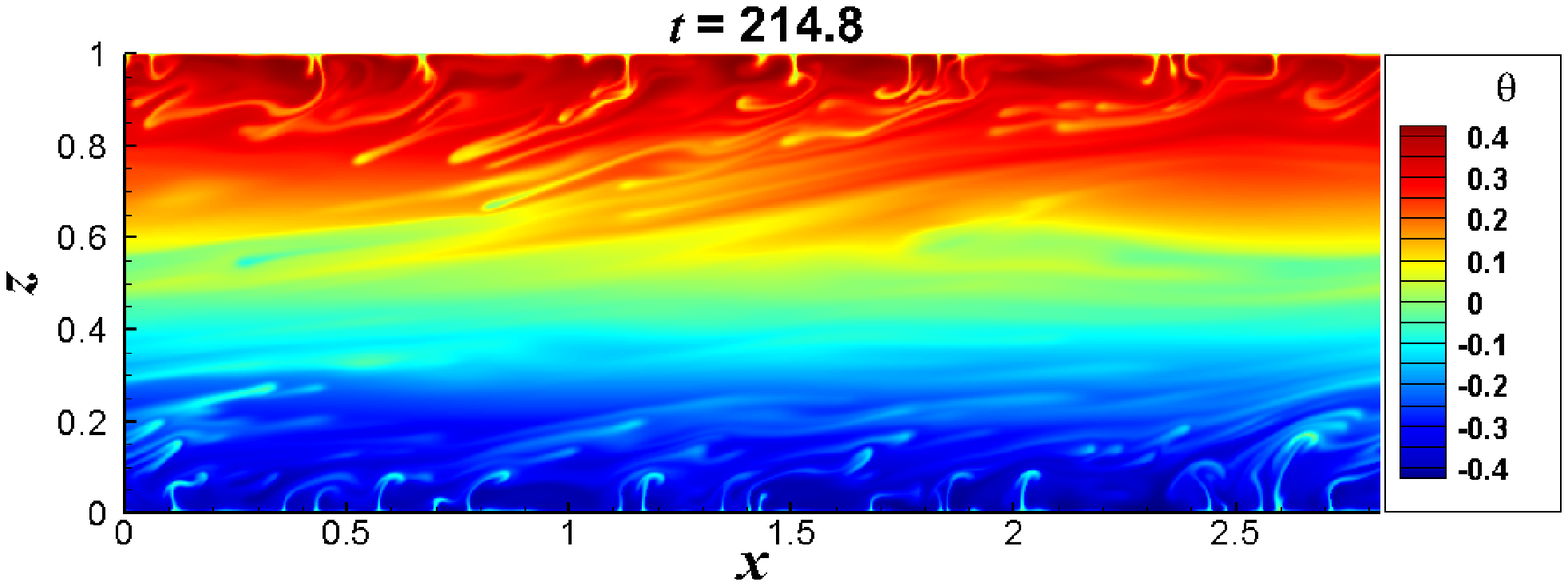} \\
            \includegraphics[width=2.5in]{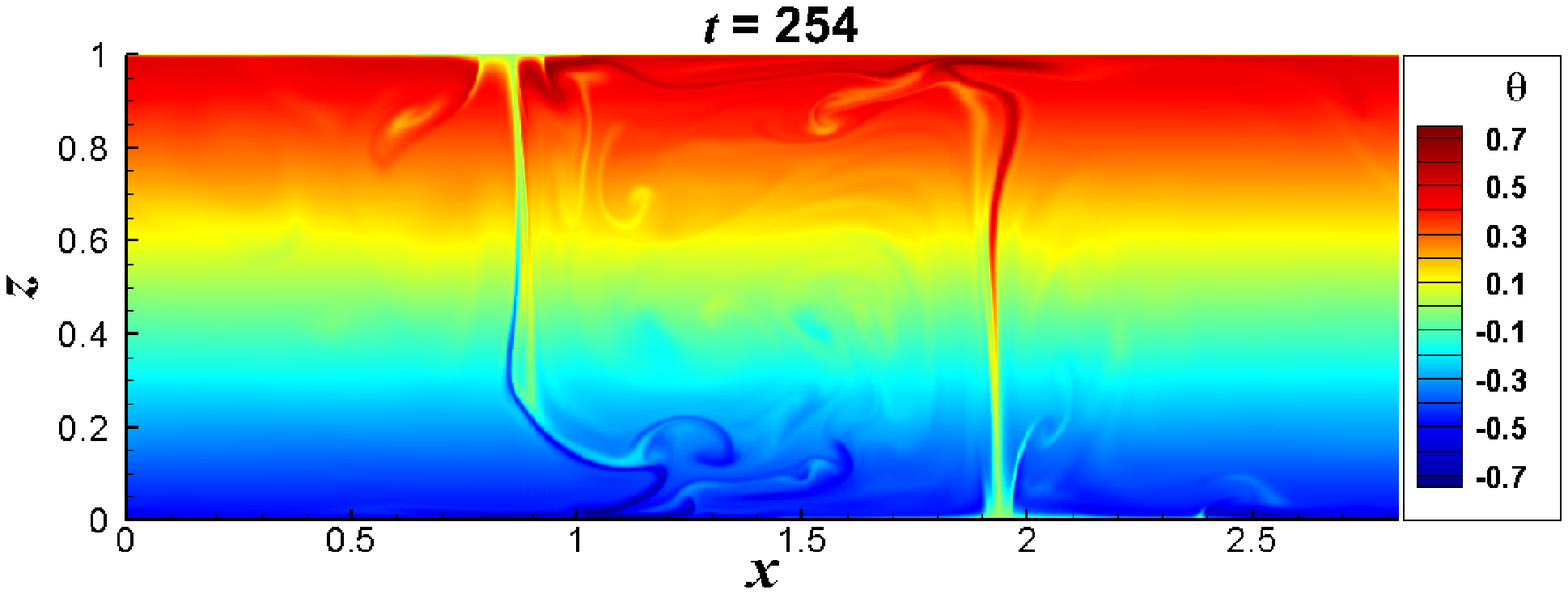}
            \includegraphics[width=2.5in]{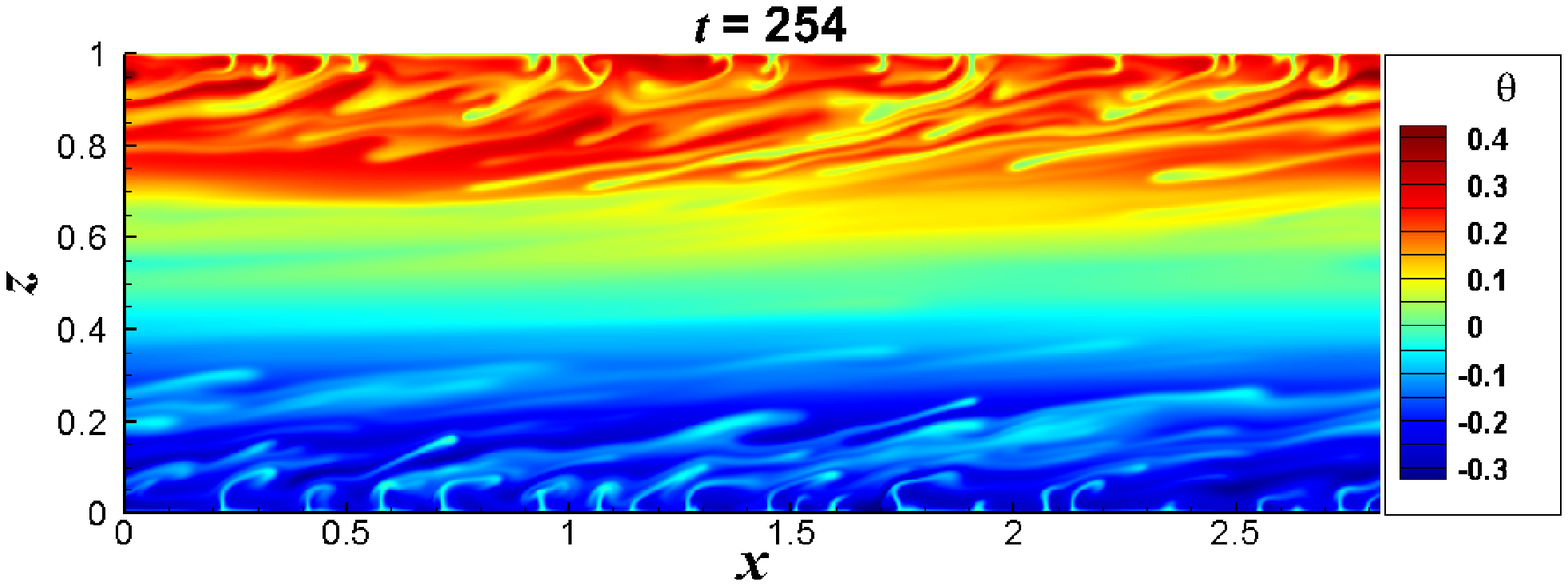} \\
            \includegraphics[width=2.5in]{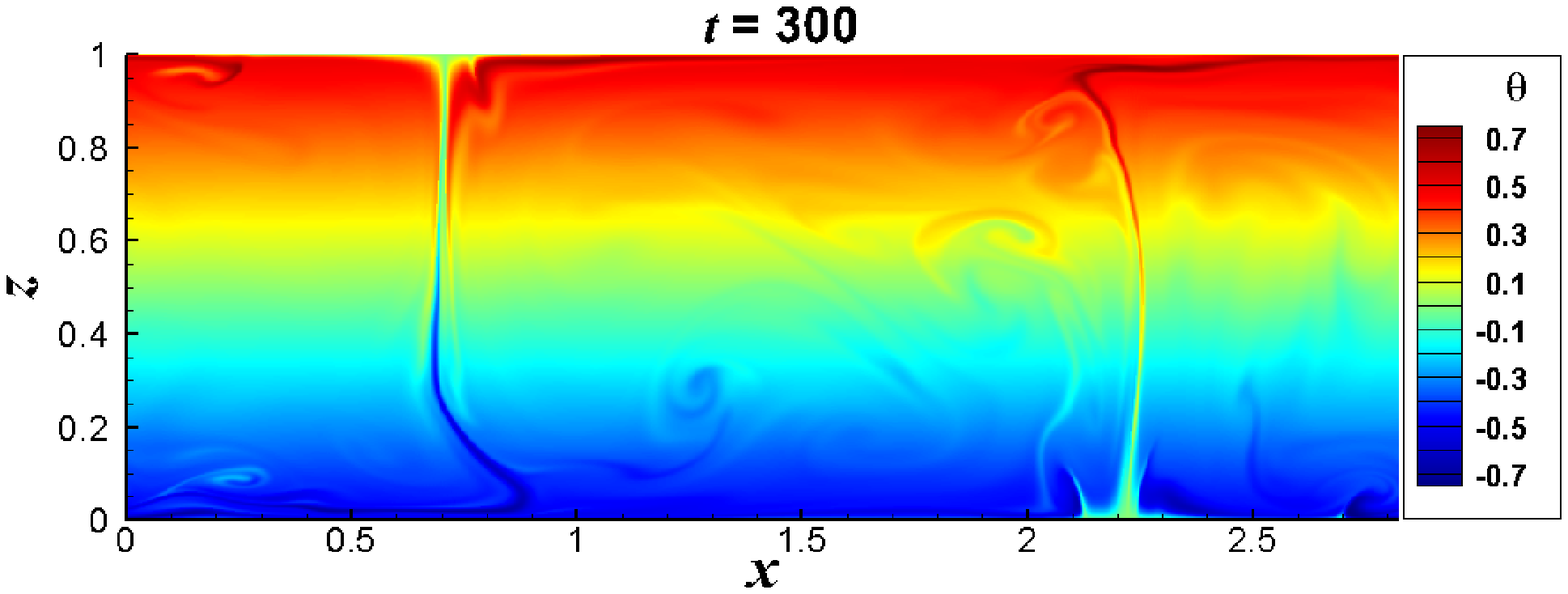}
            \includegraphics[width=2.5in]{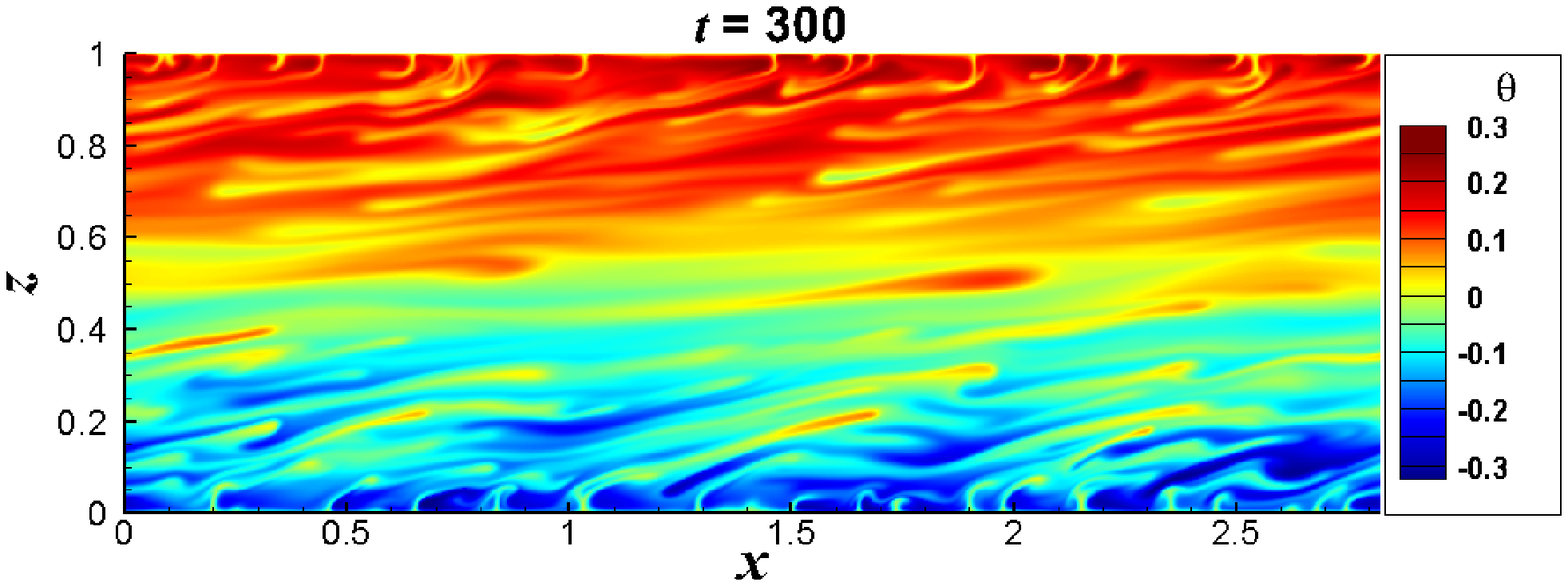} \\
            \includegraphics[width=2.5in]{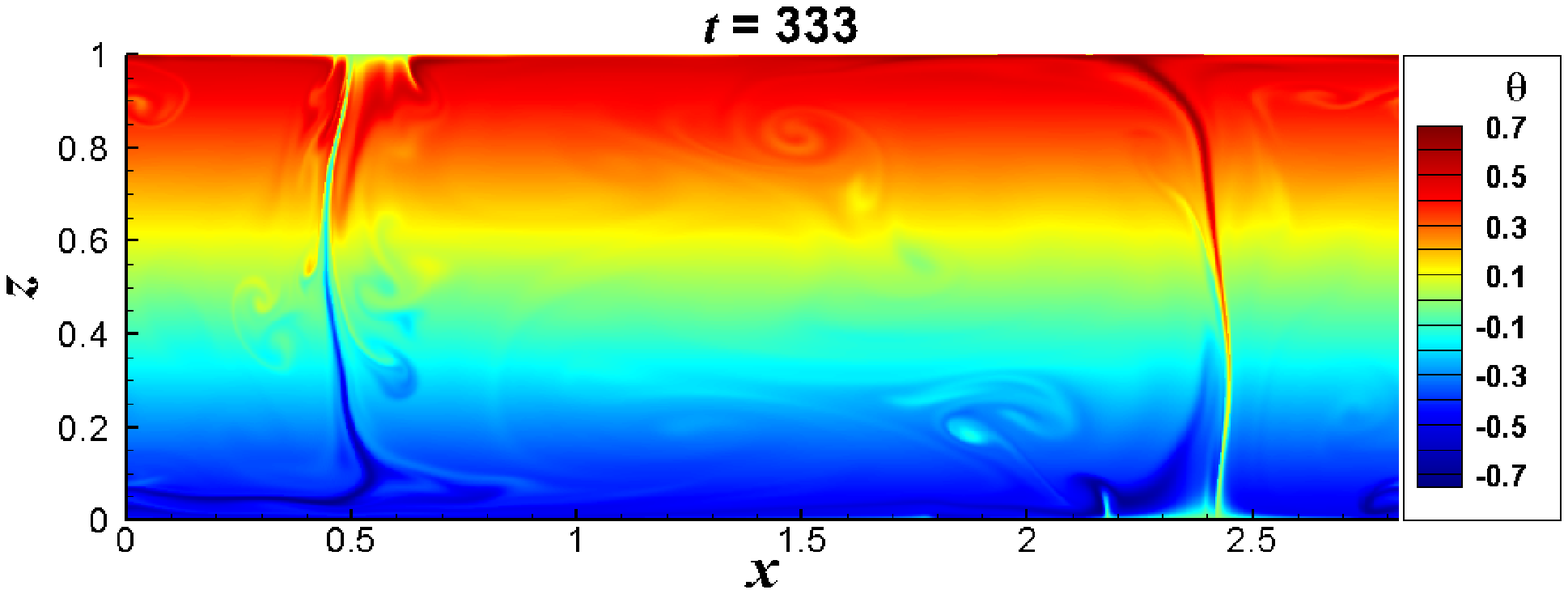}
            \includegraphics[width=2.5in]{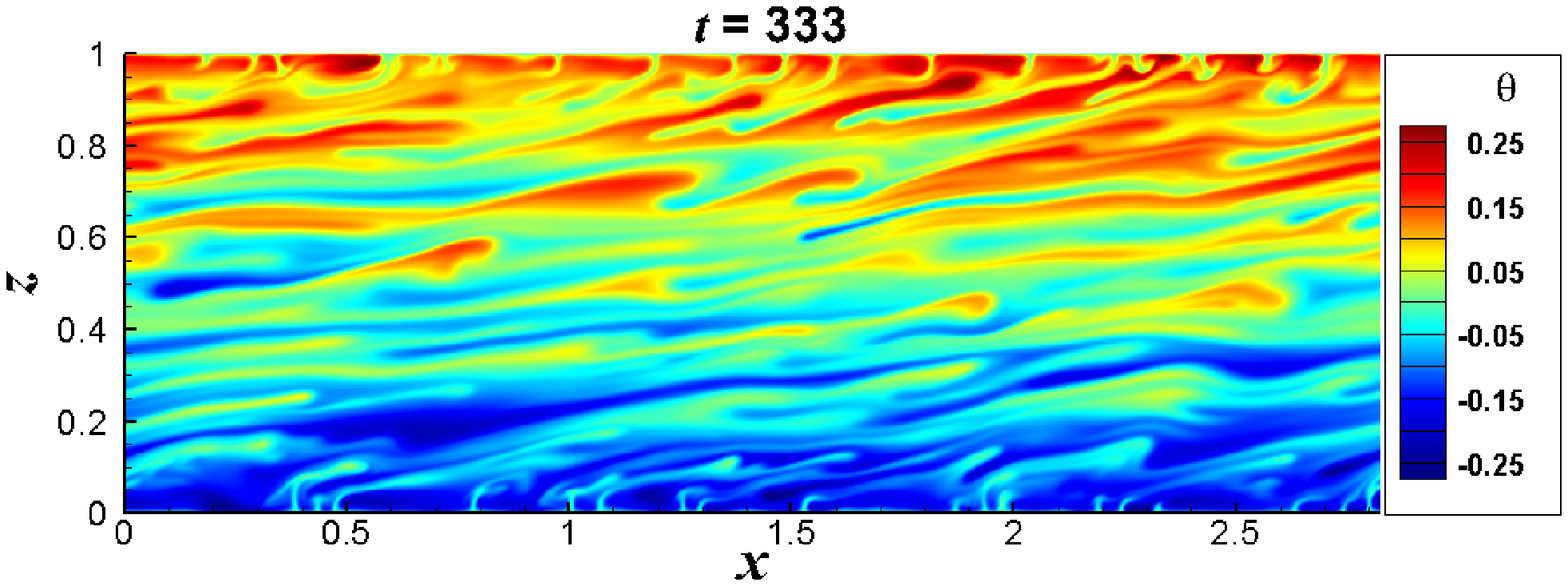} \\
            \includegraphics[width=2.5in]{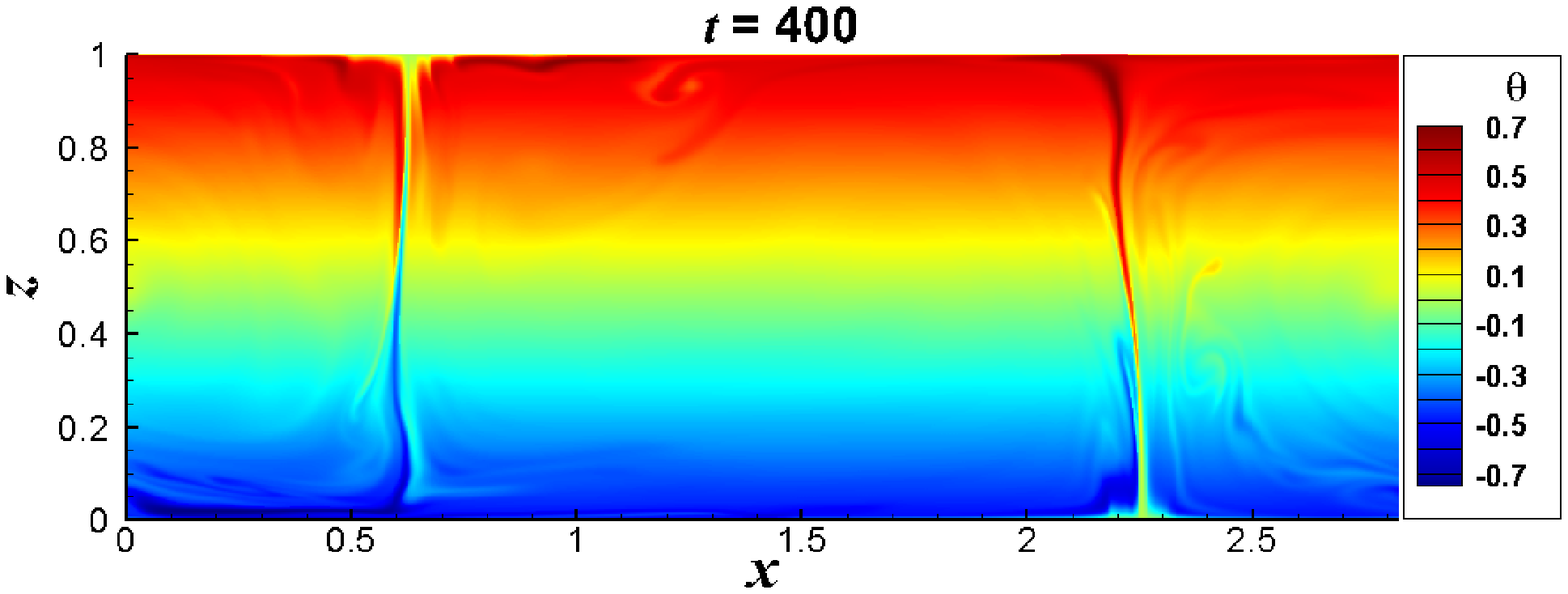}
            \includegraphics[width=2.5in]{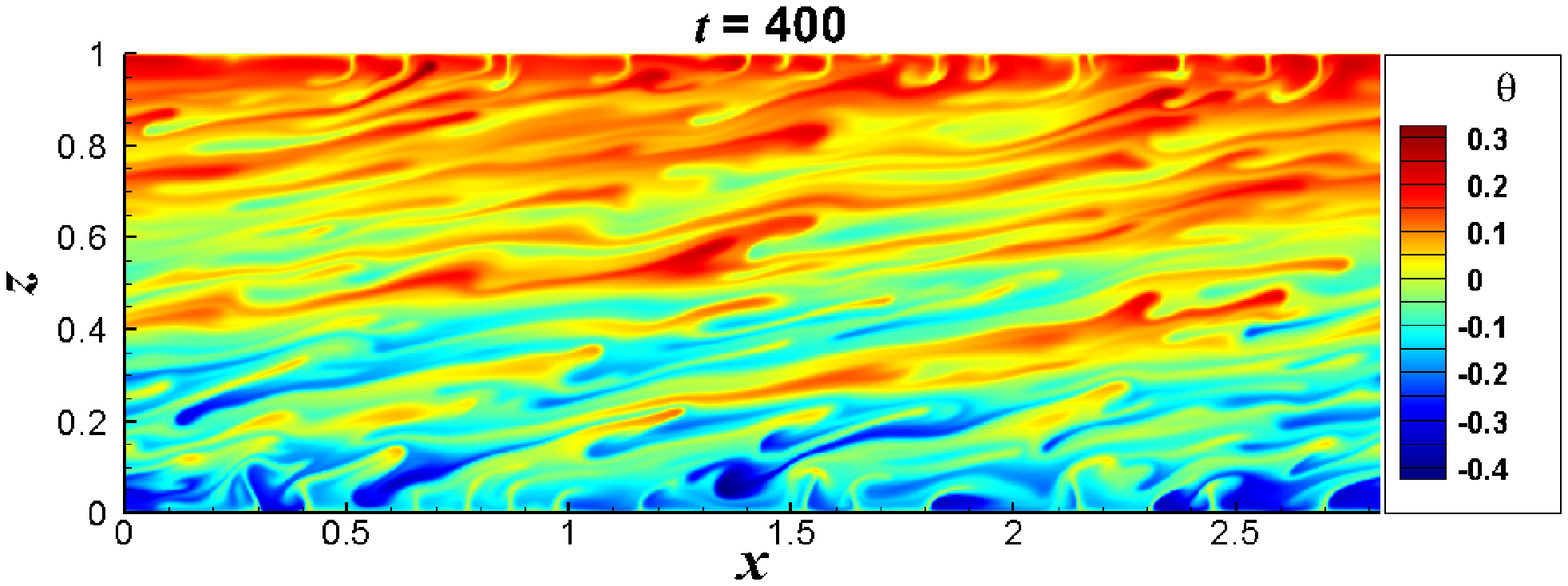} \\
            \includegraphics[width=2.5in]{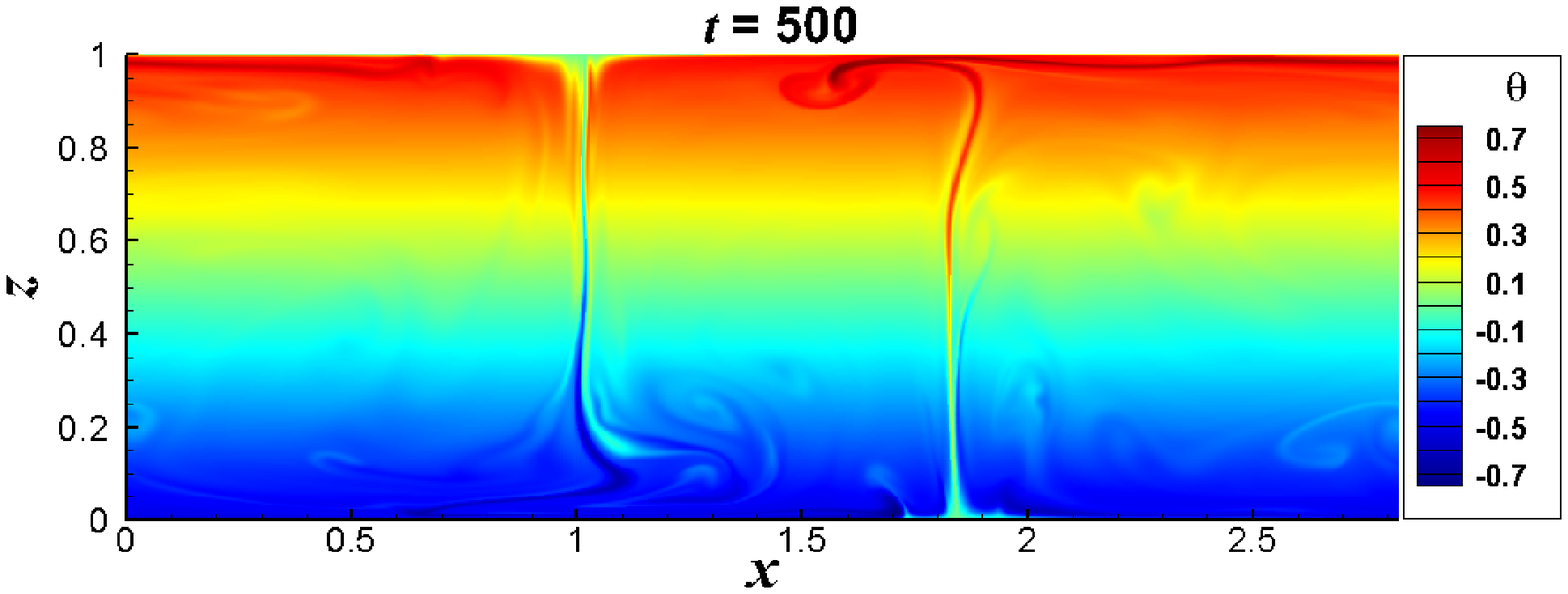}
            \includegraphics[width=2.5in]{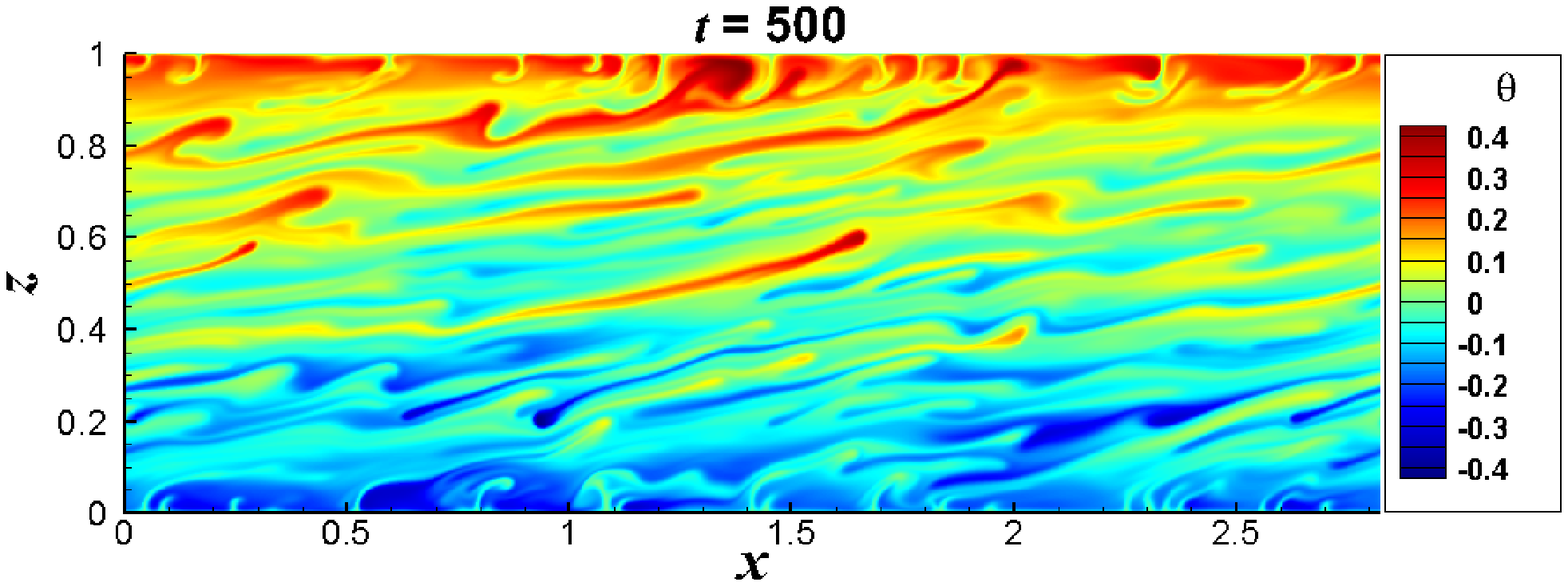}
        \end{tabular}
    \caption{Evolutions of the $\theta$ (dimensionless temperature departure from a linear variation background) field in the case of $Pr = 6.8$, $Ra = 6.8 \times 10^{8}$, and $L/H = 2\sqrt{2}$. Left: the CNS benchmark solution; Right: the RKwD simulation given by the Runge-Kutta's method with double precision using $\Delta t = 10^{-4}$. The corresponding movie is available at \href{https://github.com/sjtu-liao/RBC/blob/main/RBC-mv.mp4}{https://github.com/sjtu-liao/RBC/blob/main/RBC-mv.mp4}.}    \label{Contour-2}
    \end{center}
\end{figure}

As shown in Figures~\ref{Contour} and \ref{Contour-2}, the numerical simulation given by the RKwD is compared with the ``clean'' benchmark solution given by the CNS, and the corresponding movie is available at \href{https://github.com/sjtu-liao/RBC/blob/main/RBC-mv.mp4}{https://github.com/sjtu-liao/RBC/blob/main/RBC-mv.mp4}. Note that these two simulations have exactly the same initial conditions caused by the micro-level thermal fluctuations. For both of the CNS and RKwD simulations, the tiny initial disturbances of velocity and temperature evolve progressively from micro-level to macro-level until $t\approx 25$ when the transition from laminar flow to turbulence occurs, then the strong mixing occurs in $t\in[25,36]$ and the typical vortical/roll-like convection appears at $t\approx 50$, as shown in Figure~\ref{Contour}.
Thereafter, as the time increases, the RKwD simulation deviates from the CNS benchmark solution larger and larger, so that a distinct difference in large-scale between them can be observed, indicating that the numerical noises (as artificial stochastic disturbances) could indeed lead to some {\em large-scale} differences of flow fields of velocity and temperature in a macroscopic level, for example, as shown in Figure~\ref{Contour}  at $t=100$ and $t=185$. However, even so, the flow type of these two simulations keeps the same (i.e. vortical/roll-like turbulent convection) until $t\approx 188$ when the shearing convection occurs for the RKwD simulation and its corresponding flow field turns to a kind of zonal flow thereafter, as shown in Figures~\ref{Contour} and \ref{Contour-2}. On the contrary, the CNS benchmark solution thereafter always sustains the non-shearing vortical/roll-like convection during the whole process of simulation. Therefore, the RKwD simulation and the CNS benchmark solution have different {\em types} of turbulent convections after $t>188$. It should be emphasized that such kind of {\em qualitative} difference in large-scale is triggered only by the numerical noises (as artificial stochastic disturbances). All of these highly suggest that the numerical noises (as artificial stochastic disturbances) have quantitatively and qualitatively {\em large-scale} influences on the {\em sustained} turbulence, i.e. the two-dimensional turbulent RB convection considered in this paper.

\begin{figure}
    \begin{center}
        \begin{tabular}{cc}
            \subfigure[]{\includegraphics[width=2.55in]{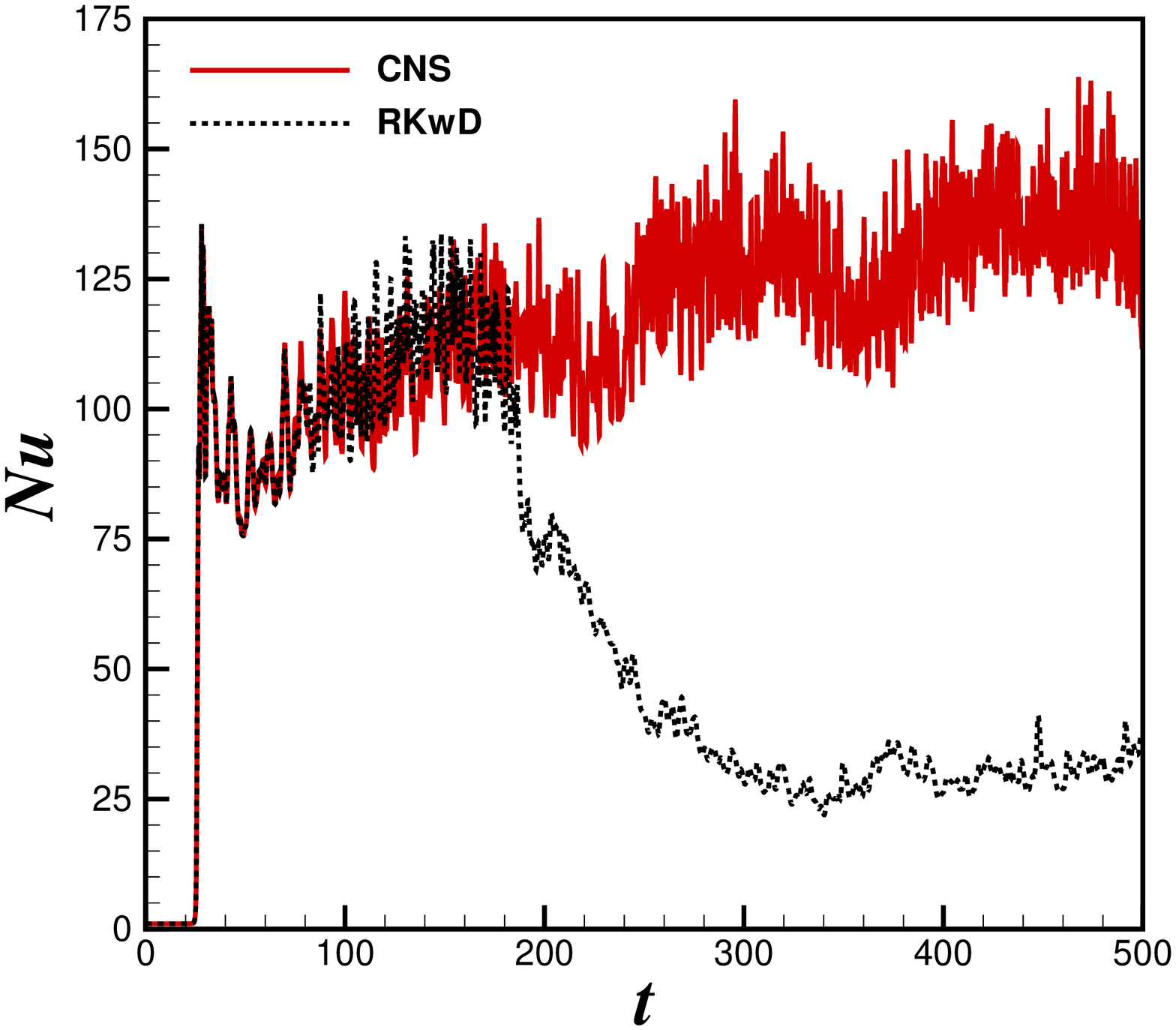}}
            \subfigure[]{\includegraphics[width=2.55in]{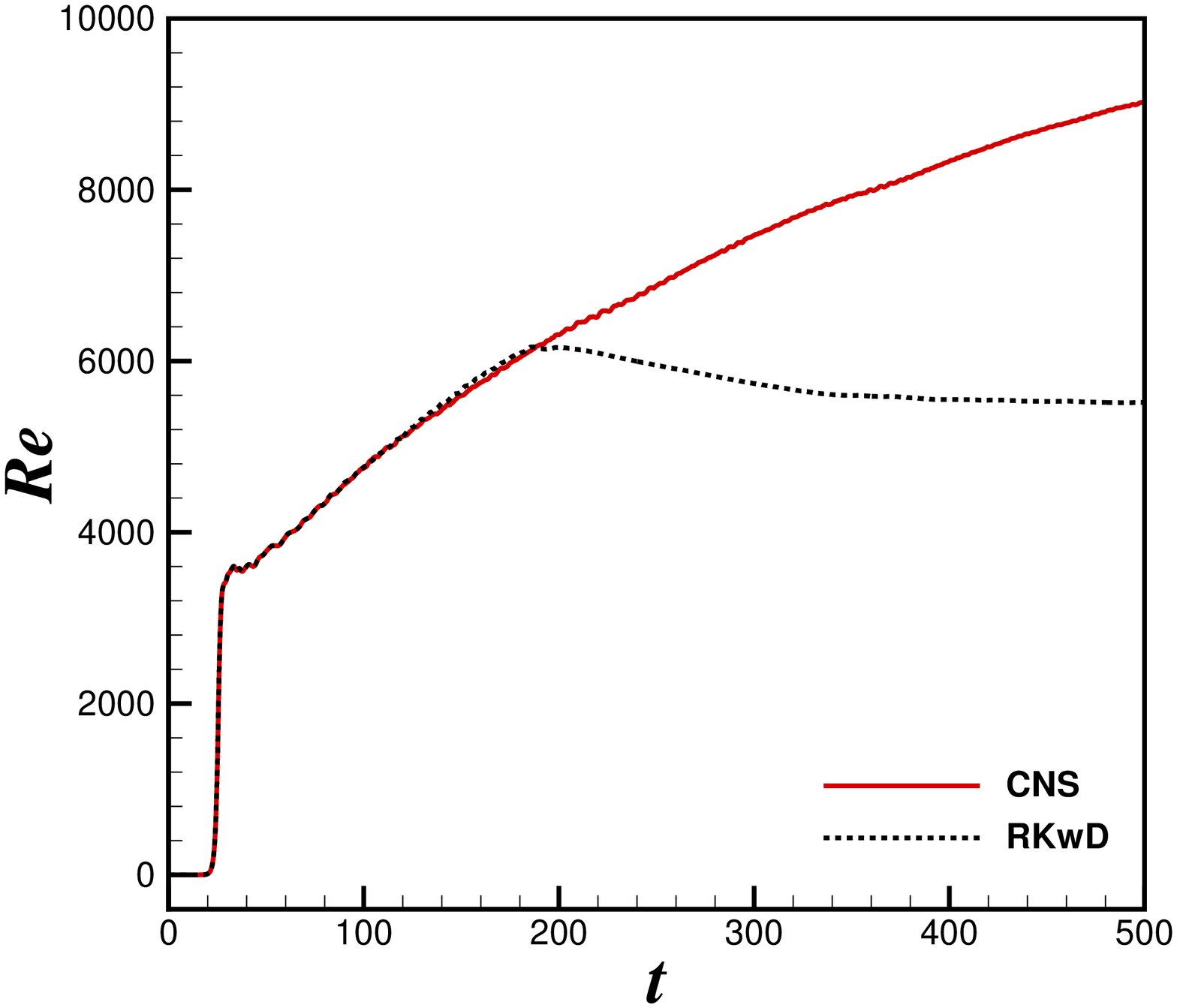}}
        \end{tabular}
    \caption{Comparisons of the instantaneous Nusselt number $N\hspace{-0.3mm}u$ and Reynolds number $Re$ in the case of $Pr = 6.8$, $Ra = 6.8 \times 10^{8}$, and $L/H = 2\sqrt{2}$. (a) The Nusselt number $N\hspace{-0.3mm}u$; (b) The Reynolds number $Re$. Solid line in red: the CNS benchmark solution; Dashed line in black: the RKwD simulation given by the Runge-Kutta's method with double precision using $\Delta t = 10^{-4}$.}    \label{Nu_t}
    \end{center}
\end{figure}

\begin{figure}
    \begin{center}
        \begin{tabular}{cc}
            \subfigure[]{\includegraphics[width=2.55in]{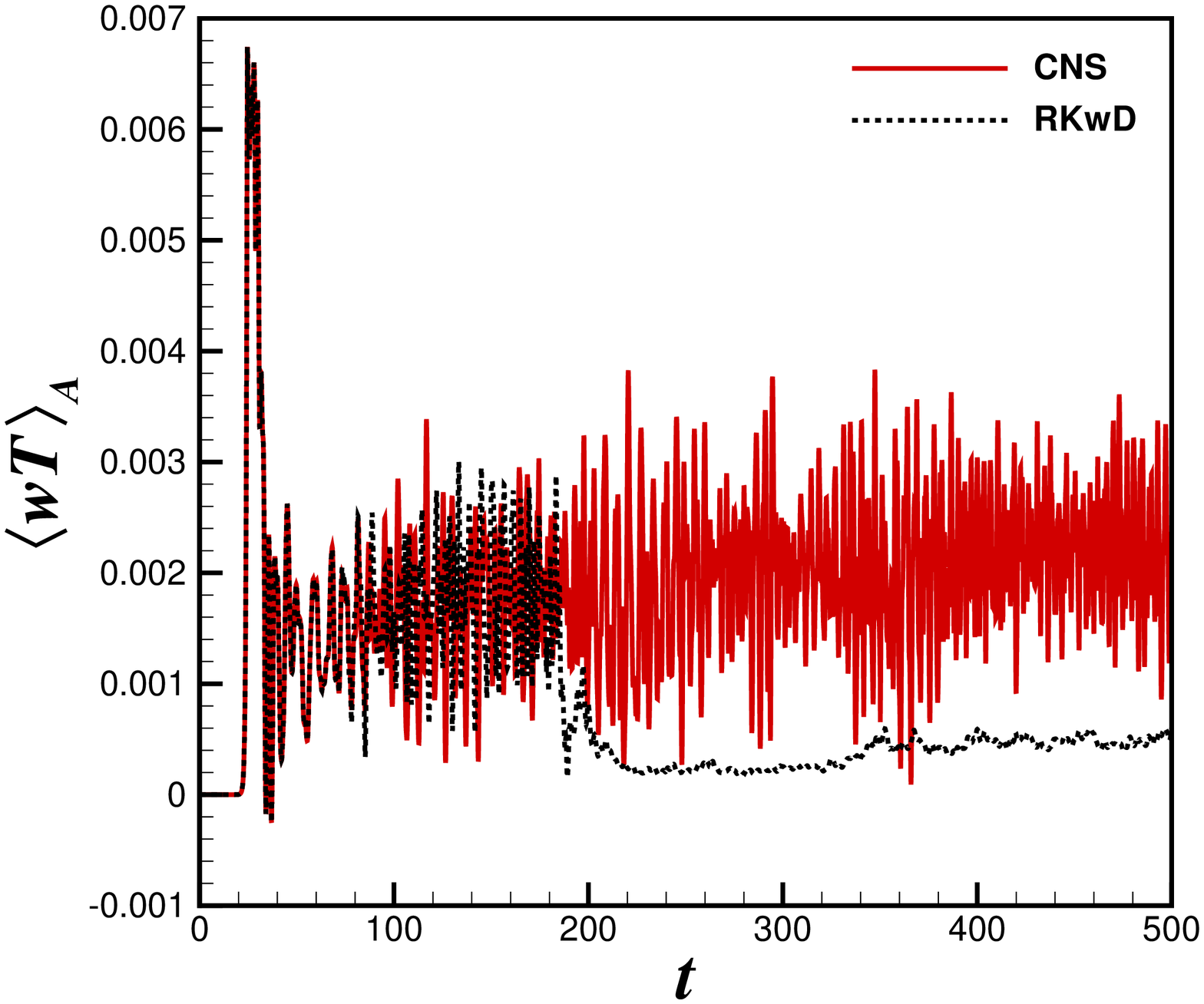}}
            \subfigure[]{\includegraphics[width=2.55in]{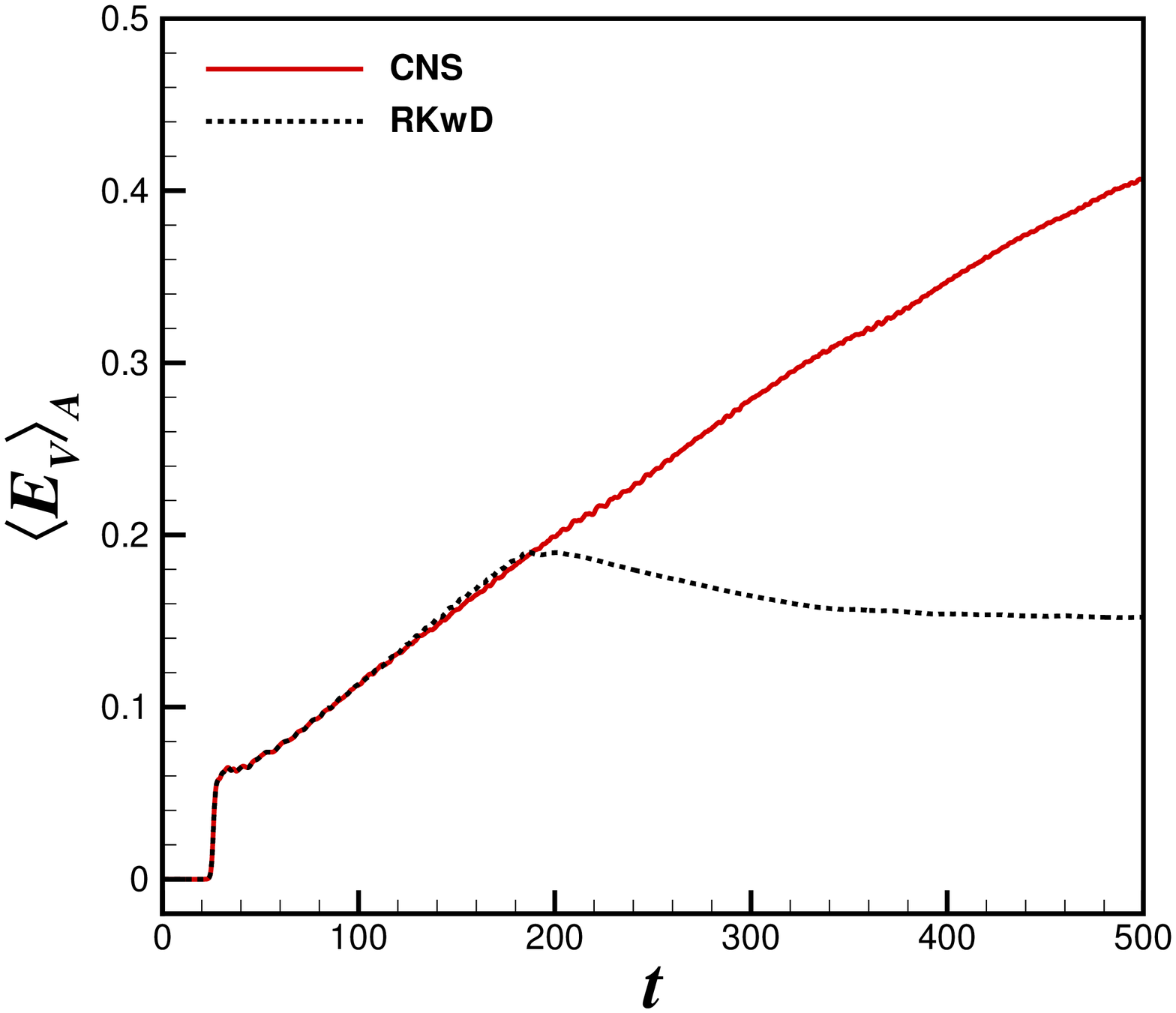}}
        \end{tabular}
    \caption{ Comparisons of the spatially averaged heat flux $\langle wT\rangle_{A}$ and kinetic energy $\langle E_V\rangle_{A}$ in the case of $Pr = 6.8$, $Ra = 6.8 \times 10^{8}$, and $L/H = 2\sqrt{2}$, where $\langle a\rangle_{A}=\int_{0}^{\Gamma}\int_{0}^{1}adxdz/\Gamma$ denotes the spatial average. (a) The heat flux $\langle wT\rangle_{A}$; (b) The kinetic energy $\langle E_V\rangle_{A}$. Solid line in red: the CNS benchmark solution; Dashed line in black: the RKwD simulation given by the Runge-Kutta's method with double precision using $\Delta t = 10^{-4}$.}    \label{wT_t}
    \end{center}
\end{figure}

How about the influence of numerical noises (as artificial stochastic disturbances) on statistical results? The heat transport of 2D turbulent RBC can be typically quantified by the Nusselt number defined by
\begin{equation}
N\hspace{-0.3mm}u(t)=1-\left. \frac{\partial\langle\theta(x,z,t)\rangle_{x}}{\partial z}\right|_{z=1},    \label{Nu}
\end{equation}
where $\langle a \rangle_{x}=\int_{0}^{\Gamma}a dx/\Gamma$ denotes the spatial average in horizontal direction. As shown in Figure~\ref{Nu_t}(a), the distinct deviation between the two time histories of $N\hspace{-0.3mm}u(t)$ given respectively by the CNS benchmark solution and the RKwD simulation happens at $t\approx80$
when the numerical noises of the RKwD simulation have been enlarged to a macroscopic level because of the butterfly-effect of chaos.
Note that the $N\hspace{-0.3mm}u(t)$ given by the RKwD simulation drops down greatly at $t\approx188$ until $N\hspace{-0.3mm}u(t) \approx$ 30 after $t>300$, one order of magnitude less than that of the CNS benchmark solution. Such a huge difference is not only quantitative but also qualitative, which is definitely due to the appearance of the zonal flow at $t \approx 188$ that is triggered by the numerical noises of the RKwD simulation (as artificial stochastic disturbances).
On the other hand, the Reynolds number is also calculated to measure the global convection strength, which is obtained via the root-mean-square (rms) velocity $U_{rms}$ \citep{sugiyama2009flow,Zhang2017Statistics}, i.e.
\begin{equation}
Re(t)=\sqrt{\frac{Ra}{Pr}}\hspace{0.03cm}U_{rms}    \label{Nu}
\end{equation}
with $U_{rms}=\sqrt{\langle u^2+w^2\rangle_{A}}$, where $\langle a\rangle_{A}=\int_{0}^{\Gamma}\int_{0}^{1}adxdz/\Gamma$ denotes the spatial average. As shown in Figure~\ref{Nu_t}(b), the CNS benchmark solution and RKwD simulation give almost the same values of $Re(t)$ when t < 188. However, the departure begins  at $t \approx 188$ when the shearing convection occurs, and thereafter the deviation between the RKwD simulation and the CNS benchmark solution becomes more and more obvious. Here, it should be emphasized that the Reynolds number $Re$ of the CNS benchmark solution at $t=500$ is about 3500 larger than that of the RKwD simulation! This is indeed a huge difference. Besides, the similar phenomena are observed in the comparisons of the spatially averaged heat flux $\langle wT\rangle_{A}$ and kinetic energy $\langle E_V\rangle_{A}$, given respectively by the CNS benchmark solution and RKwD simulation, as shown in Figure~\ref{wT_t}(a) and (b), where the maximum ratio of kinetic energy reaches about 2.5. Indeed, the numerical noises (as artificial stochastic disturbances) might lead to qualitatively huge deviations of the 2D turbulent RBC.

\begin{figure}
    \begin{center}
        \begin{tabular}{cc}
             \subfigure[]{\includegraphics[width=2.55in]{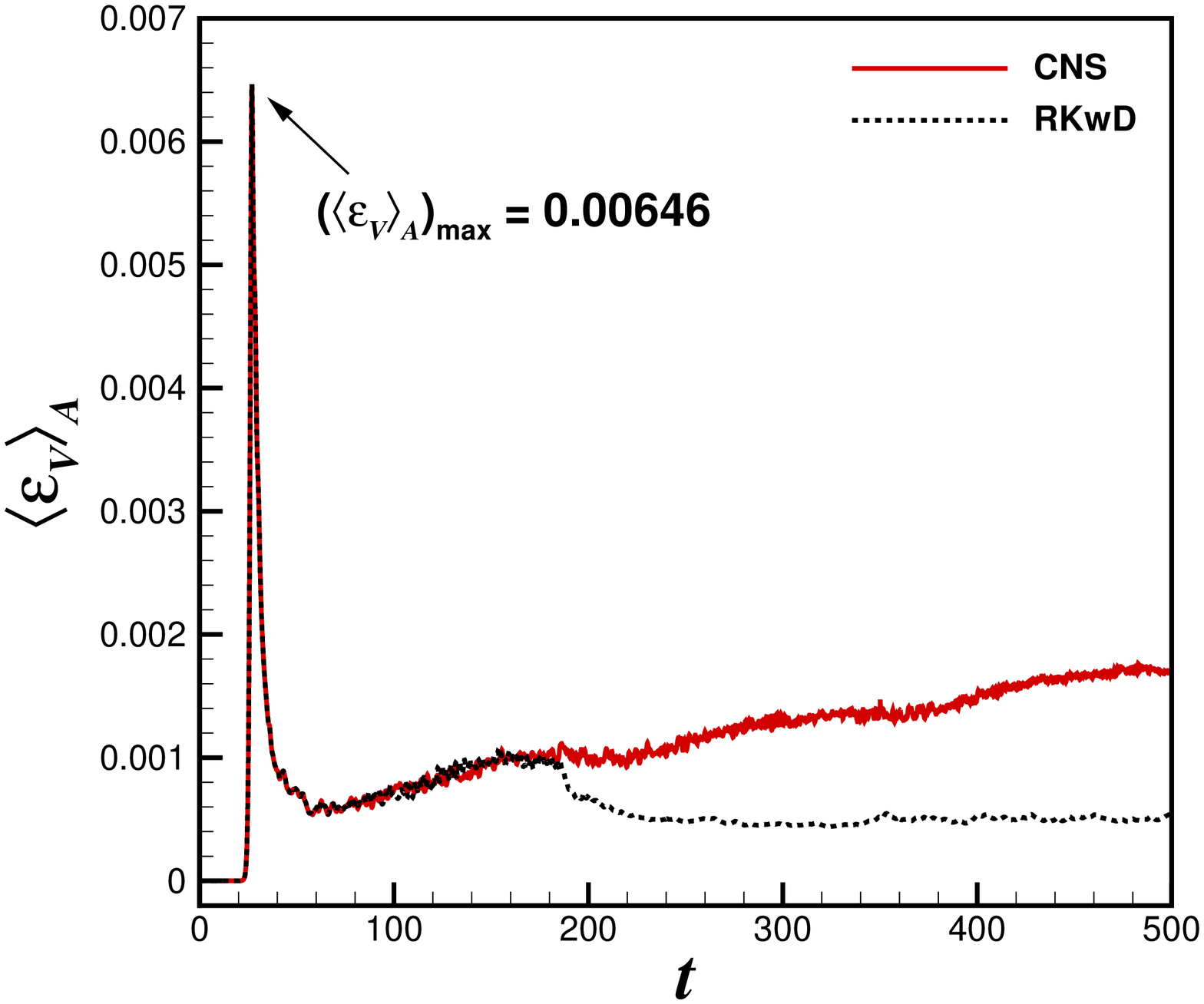}}
             \subfigure[]{\includegraphics[width=2.55in]{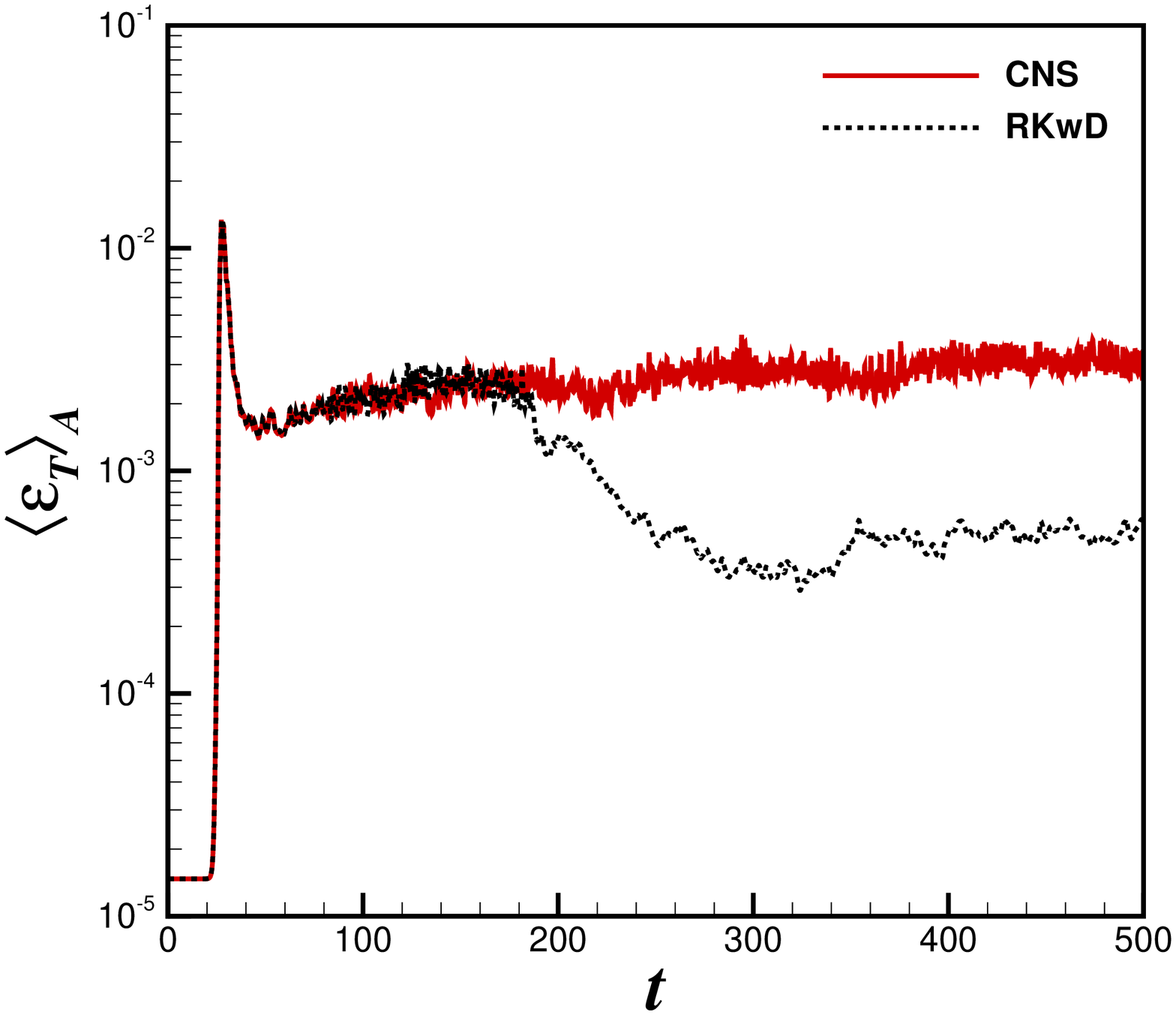}}
        \end{tabular}
    \caption{Comparisons of the spatially averaged kinetic energy dissipation rate $\langle\varepsilon_{V}\rangle_{A}$ and thermal dissipation rate $\langle\varepsilon_{T}\rangle_{A}$ in the case of $Pr = 6.8$, $Ra = 6.8 \times 10^{8}$, and $L/H = 2\sqrt{2}$, where $\langle a\rangle_{A}=\int_{0}^{\Gamma}\int_{0}^{1}adxdz/\Gamma$ denotes the spatial average. (a) The kinetic energy dissipation rate $\langle\varepsilon_{V}\rangle_{A}$; (b) The thermal dissipation rate $\langle\varepsilon_{T}\rangle_{A}$. Solid line in red: the CNS benchmark solution; Dashed line in black: the RKwD simulation given by the Runge-Kutta's method with double precision using $\Delta t = 10^{-4}$.}    \label{e_eT_t}
    \end{center}
\end{figure}

Besides the above-mentioned large-scale quantities, the comparisons of some small-scale properties of fluid flows such as the kinetic energy dissipation rate $\langle\varepsilon_{V}\rangle_{A}$ and the thermal dissipation rate $\langle\varepsilon_{T}\rangle_{A}$ of this 2D turbulent RBC given by the CNS solution and the RKwD simulation are as shown in Figure~\ref{e_eT_t}, where
\begin{equation}
\varepsilon_{V}(x,z,t)=\frac{1}{2}\sqrt{\frac{Pr}{Ra}}\sum_{ij}\left[ \partial_iu_j(x,z,t)+\partial_ju_i(x,z,t) \right]^2    \label{e}
\end{equation}
and
\begin{equation}
\varepsilon_{T}(x,z,t)=\frac{1}{\sqrt{Pr\,Ra}} \hspace{0.03cm} \Big| \nabla \left[ \theta(x,z,t)-z \right] \Big|^2    \label{eT}
\end{equation}
with $i,\,j=1,\,2$, $u_1(x,z,t)=u(x,z,t)$, $u_2(x,z,t)=w(x,z,t)$, $\partial_1=\partial /\partial x$, $\partial_2=\partial /\partial z$, and that $\nabla$ is the Hamilton operator and $\langle \;\; \rangle_{A}$ denotes the spatial average, respectively.  Note that, for the RKwD simulation,  both of the kinetic energy dissipation rate $\langle\varepsilon_{V}\rangle_{A}$ and thermal dissipation rate $\langle\varepsilon_{T}\rangle_{A}$  greatly drop down at $t\approx 188$ when the shearing convection (i.e. the zonal flow) occurs, which is triggered only by the numerical noises (as artificial stochastic disturbances).

By the way, as shown in Figure~\ref{e_eT_t}(a), we have the maximum kinetic energy dissipation rate $(\langle\varepsilon_{V}\rangle_{A})_{max}=0.00646$ at $t=26.9$ when the transition from the laminar flow to turbulence occurs, corresponding to the minimum Kolmogorov scale
\[\left(\langle\eta\rangle_{A}\right)_{min}\approx(Pr/Ra)^{3/8} \big[ (\langle\varepsilon_{V}\rangle_{A})_{max} \big] ^{-1/4}=0.00353.\]
Thus, the criteria condition that the maximum horizontal grid spacing $\Delta_x=\Gamma/N_x=0.00276 < 0.8(\langle\eta\rangle_{A})_{min}=0.00282$ is indeed satisfied, say, the spatial resolution used in this paper is fine enough for the 2D turbulent RBC under consideration.

\begin{figure}
    \begin{center}
        \begin{tabular}{cc}
            \subfigure[]{\includegraphics[width=2.55in]{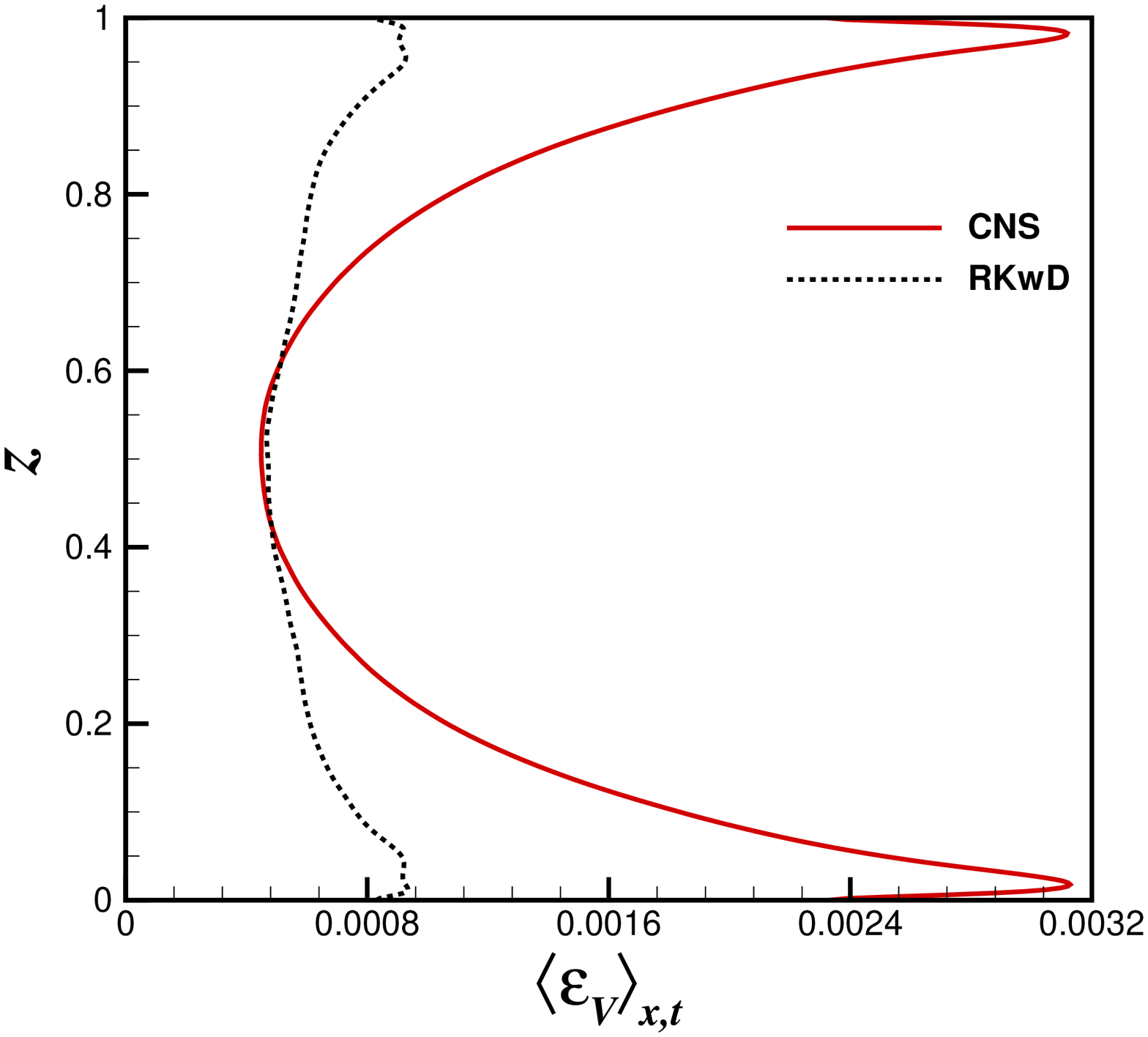}}
            \subfigure[]{\includegraphics[width=2.55in]{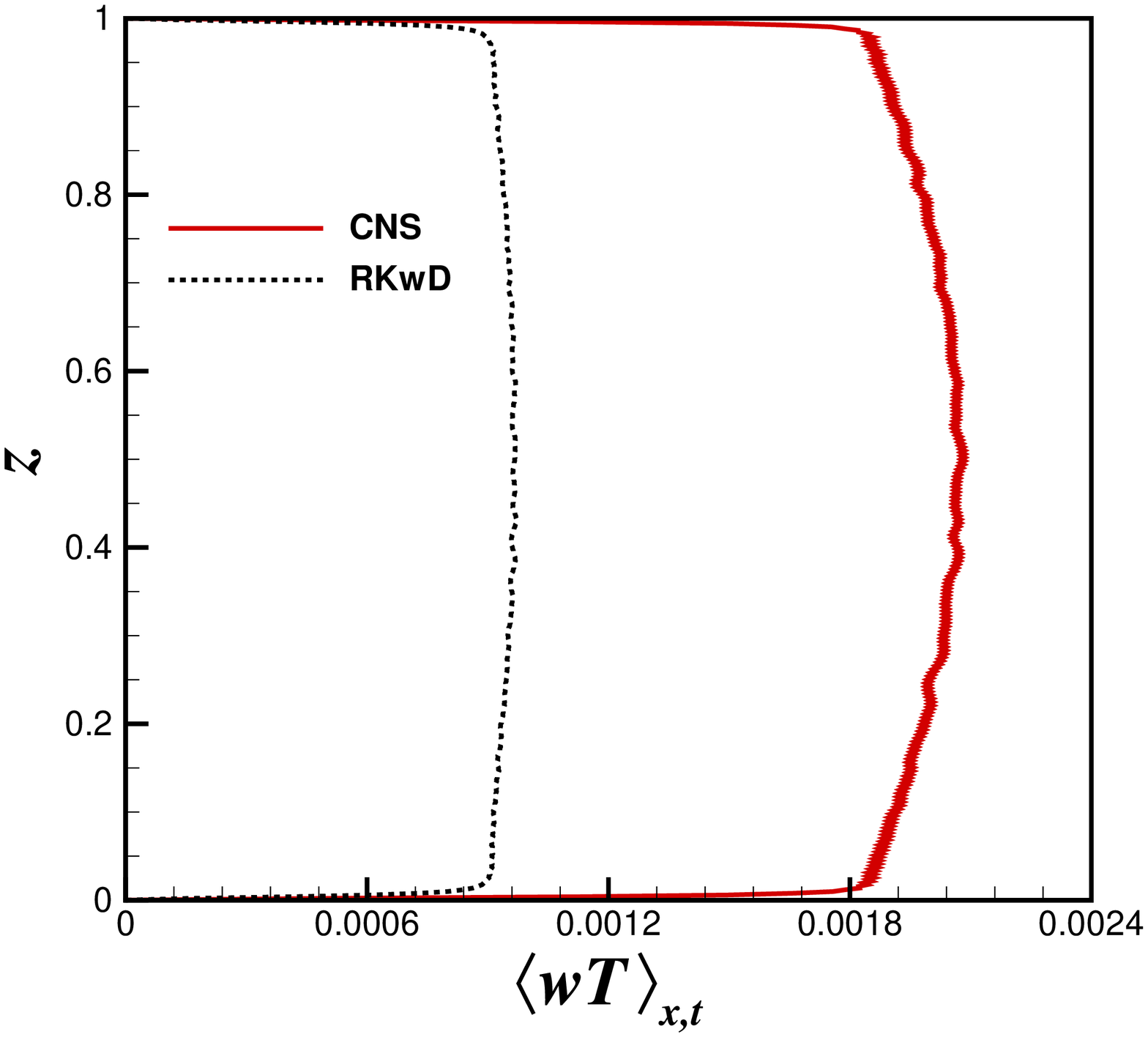}}
        \end{tabular}
    \end{center}
\caption{Comparisons of the horizontally and temporally averaged kinetic energy dissipation rate $\langle\varepsilon_{V}\rangle_{x,t}(z)$ and the heat flux $\langle wT\rangle_{x,t}(z)$ in the case of $Pr = 6.8$, $Ra = 6.8 \times 10^{8}$, and $L/H = 2\sqrt{2}$, where $\langle a\rangle_{x,t}=\int_{0}^{\Gamma}\int_{0}^{500}adxdt/\Gamma/500$ denotes the horizontal and temporal average. (a) The kinetic energy dissipation rate $\langle\varepsilon_{V}\rangle_{x,t}(z)$; (b) The heat flux $\langle wT\rangle_{x,t}(z)$. Solid line in red: the CNS benchmark solution; Dashed line in black: the RKwD simulation given by the Runge-Kutta's method with double precision using $\Delta t = 10^{-4}$.}    \label{e_wT_z}
\end{figure}

Figure~\ref{e_wT_z} shows the comparisons of the kinetic energy dissipation rate $\langle\varepsilon_{V}\rangle_{x,t}(z)$ and the heat flux $\langle wT\rangle_{x,t}(z)$, where $\langle a\rangle_{x,t}=\int_{0}^{\Gamma}\int_{0}^{500}a \; dxdt/\Gamma/500$ denotes the horizontally spatial and temporal average. Obviously, both of the kinetic energy dissipation rate $\langle\varepsilon_{V}\rangle_{x,t}$ and the heat flux $\langle wT\rangle_{x,t}$ of the CNS benchmark solution are significantly larger than those given by the RKwD simulation. Especially, near the lower and upper plates, the $\langle\varepsilon_{V}\rangle_{x,t}$ given by the CNS benchmark solution has a much sharper peak than that given by the RKwD simulation, as shown in Figure~\ref{e_wT_z}(a). These indicate that the numerical noises (as artificial stochastic disturbances) can lead to large-scale deviations in statistics of the 2D turbulent RBC under consideration.

\begin{figure}
    \begin{center}
        \begin{tabular}{cc}
             \subfigure[]{\includegraphics[width=2.55in]{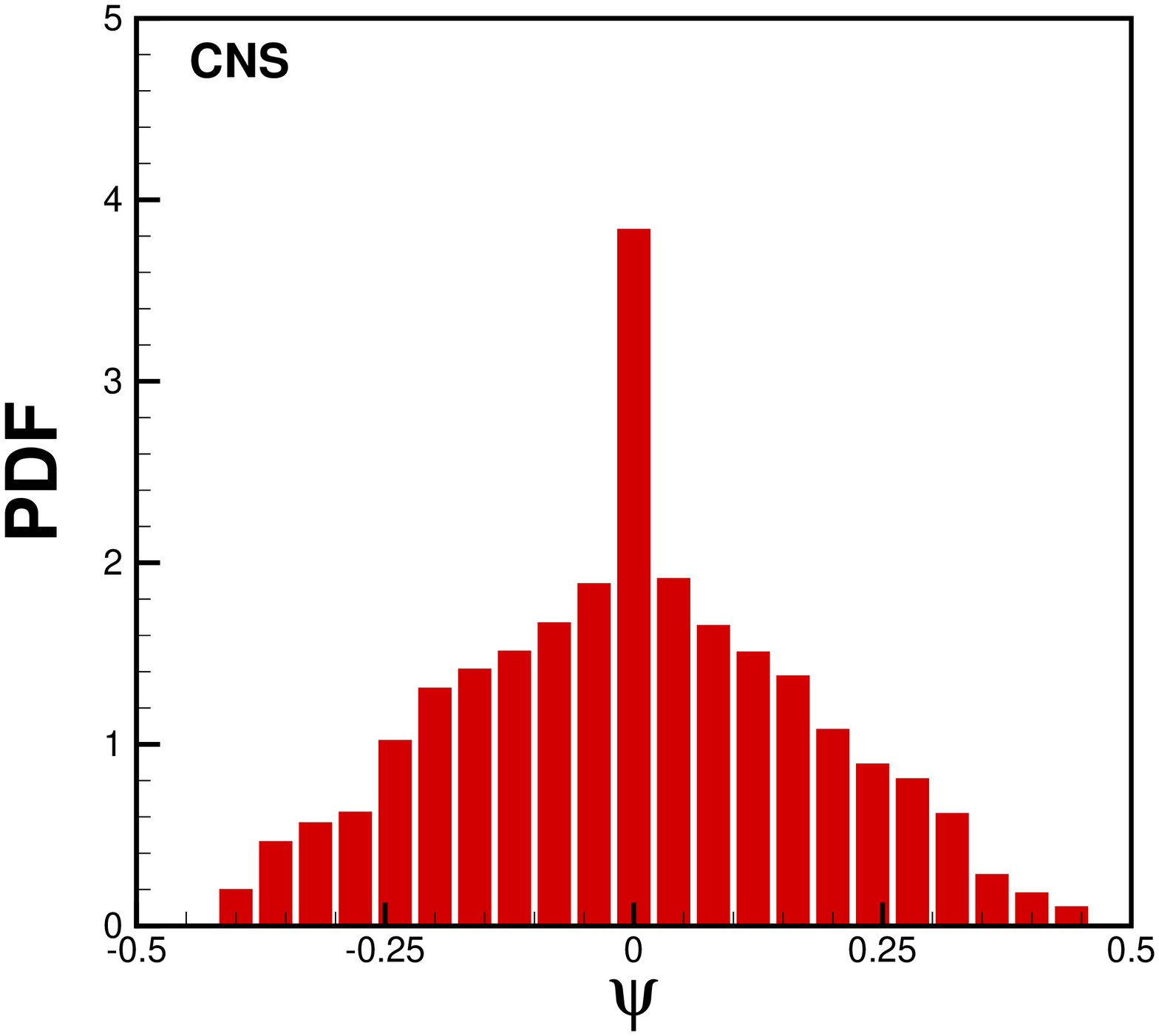}}
             \subfigure[]{\includegraphics[width=2.55in]{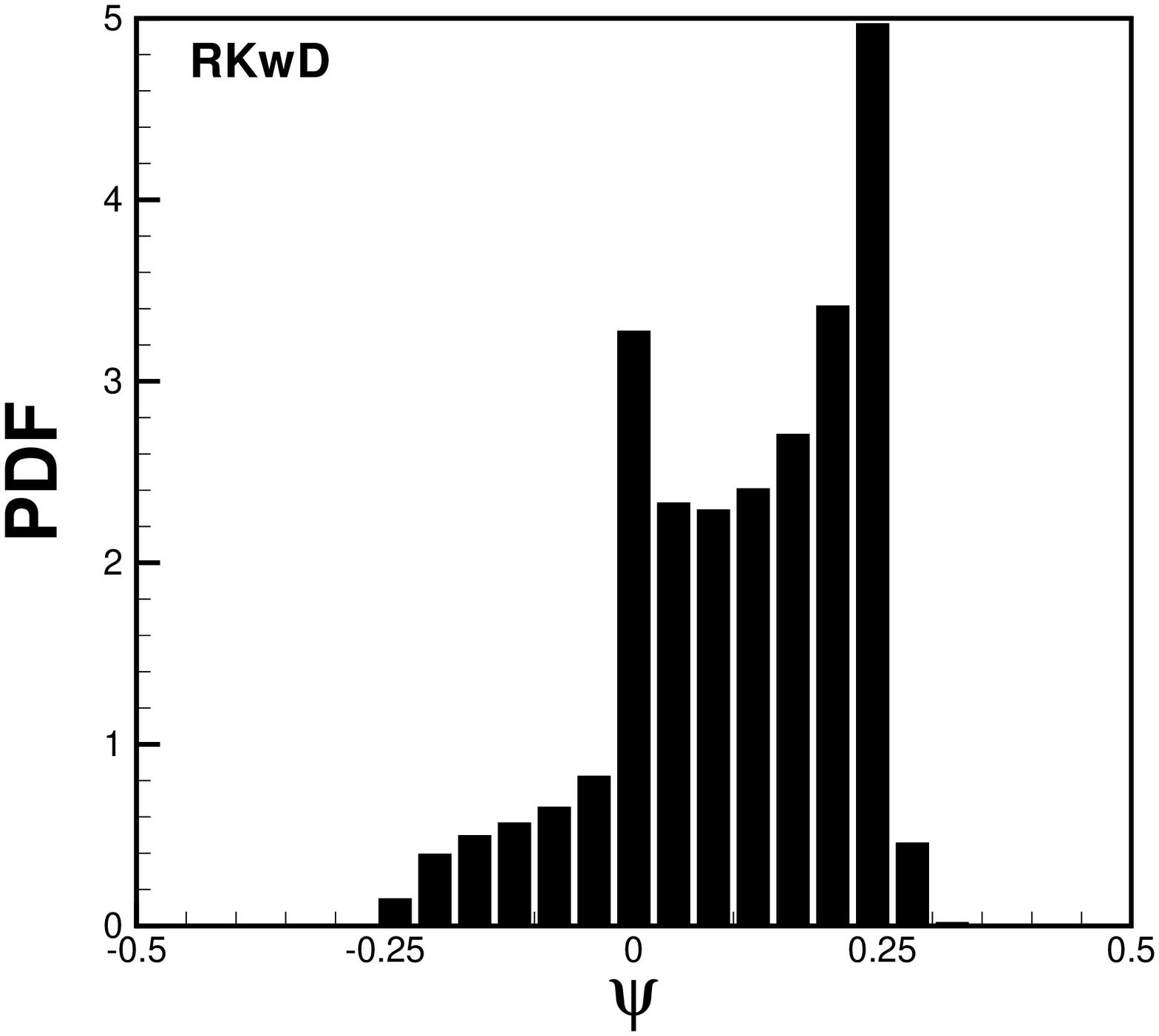}}
        \end{tabular}
    \caption{Probability density functions (PDFs) of $\psi(x,z,t)$ in $0 \leq x < \Gamma$, $0 \leq z \leq 1$ and $0 \leq t \leq 500$ in the case of $Pr = 6.8$, $Ra = 6.8 \times 10^{8}$, and $L/H = 2\sqrt{2}$. (a) The PDF given by the CNS benchmark solution; (b) The PDF given by the RKwD simulation obtained via the Runge-Kutta's method with double precision using $\Delta t = 10^{-4}$.}     \label{PDF_psi}
    \end{center}
%\end{figure}

%\begin{figure}
    \begin{center}
        \begin{tabular}{cc}
             \subfigure[]{\includegraphics[width=2.55in]{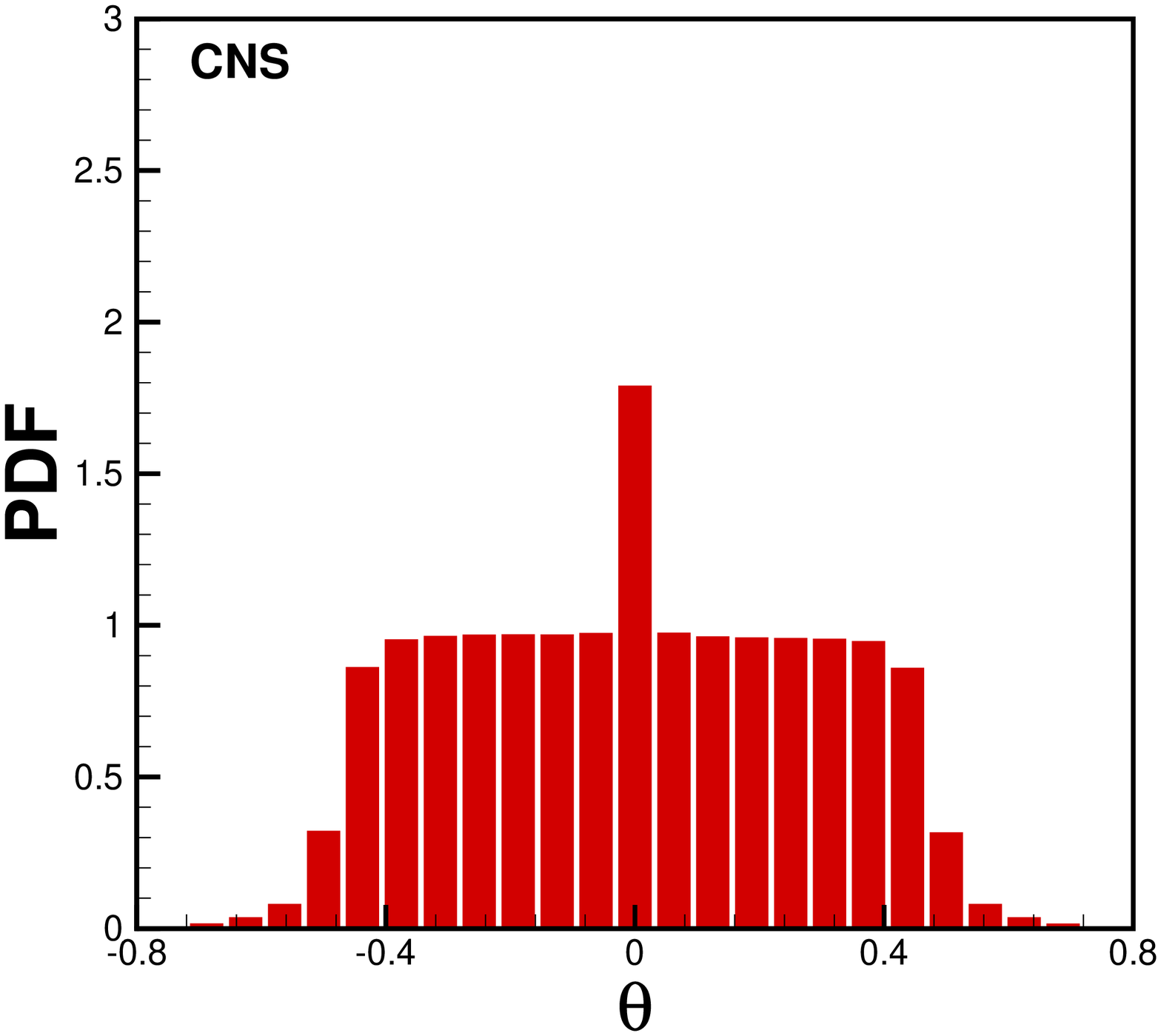}}
             \subfigure[]{\includegraphics[width=2.55in]{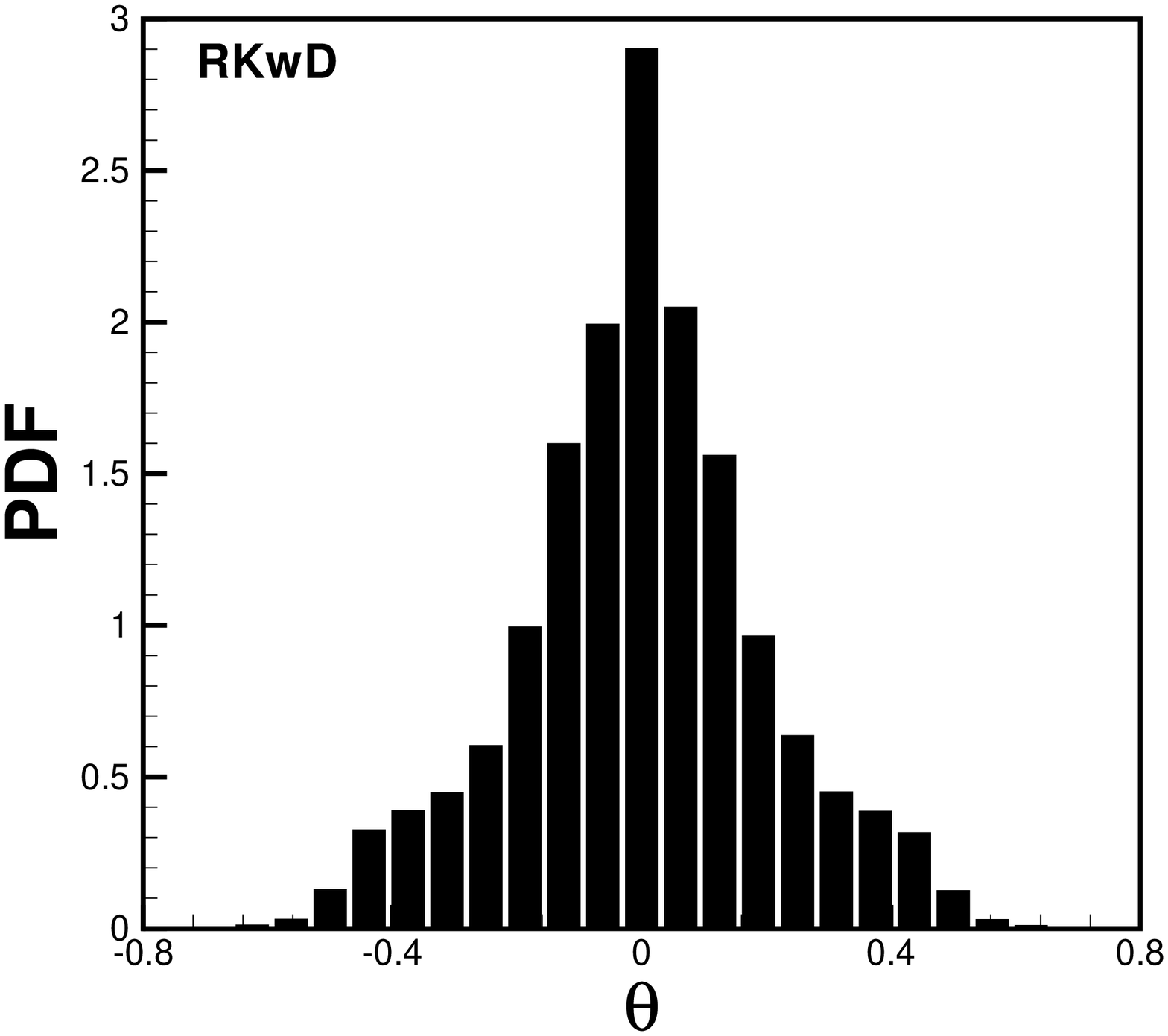}}
        \end{tabular}
    \caption{Probability density functions (PDFs) of $\theta(x,z,t)$ in $0 \leq x < \Gamma$, $0 \leq z \leq 1$ and $0 \leq t \leq 500$ in the case of $Pr = 6.8$, $Ra = 6.8 \times 10^{8}$, and $L/H = 2\sqrt{2}$. (a) The PDF given by the CNS benchmark solution; (b) The PDF given by the RKwD simulation obtained via the Runge-Kutta's method with double precision using $\Delta t = 10^{-4}$.}     \label{PDF_theta}
    \end{center}
\end{figure}

The comparison of the probability density functions (PDFs) of the stream function $\psi(x,z,t)$ in $0 \leq x < \Gamma$, $0 \leq z \leq 1$ and $0 \leq t \leq 500$ given by the CNS benchmark solution and the RKwD simulation is as shown in Figure~\ref{PDF_psi}. Unlike the PDF of the CNS benchmark solution that has a kind of asymmetry about $\psi = 0$, the PDF of the RKwD simulation has no such kind of asymmetry but two peaks at $\psi\approx0$ and $\psi \approx 0.25$. Furthermore, the comparison of the PDFs of $\theta(x,z,t)$ given by the CNS benchmark solution and RKwD simulation is as shown in Figure~\ref{PDF_theta}. As shown in Figure~\ref{PDF_theta}(a), except at $\theta\approx0$, the PDF of $\theta(x,z,t)$ given by the CNS benchmark solution remains almost the same value. In contrast, the PDF of $\theta$ given by the RKwD simulation is relatively more typical, as shown in Figure~\ref{PDF_theta}(b). Thus, the numerical noises (as artificial stochastic disturbances) indeed lead to large-scale deviations even in the PDFs of the 2D turbulent RBC under consideration.

% \textcolor{blue}{In addition, another numerical solution given by the CNS, without any numerical noise but added with tiny artificial stochastic disturbances generated by the Gaussian random numbers, is also obtained by means of numerically solving the above-mentioned NS equations with the same initial/boundary conditions using the same values of physical/numerical parameters. Similar to the above-mentioned RKwD result (that contains the tiny numerical noises), this kind of solution shows the phenomenon that the shearing convection occurs, say, the typical vortical/roll-like flow turns to the zonal flow, and thus has the large-scale differences compared with the solution given by CNS without any stochastic disturbance. It indicates that the numerical noises indeed belong to a kind of artificial stochastic disturbances.}

All of the above-mentioned comparisons indicate that the micro-level background numerical noises as a kind of artificial stochastic disturbances can lead to large-scale differences not only in spatio-temporal trajectories but also even in {\em flow types} of the 2D turbulent RBC, which further affects the statistics of the Nusselt number, the Reynolds number, the kinetic energy, the kinetic energy dissipation rate, the thermal dissipation rate and so on.
Note that it is currently reported by \cite{mcmullen2022navier} that ``the Navier-Stokes equations do not describe turbulent gas flows in the dissipation range because they neglect thermal fluctuations'', say, tiny stochastic disturbances resulting from thermal fluctuations might influence the {\em small-scale} properties of the freely {\em decaying} turbulent flows under their consideration. In this paper, the detailed comparisons between the CNS benchmark solution and the RKwD simulation provide us the  rigorous  evidence that numerical noises as a kind of small-scale artificial stochastic disturbances might influence the {\em large-scale} properties of a {\em sustained} turbulence, i.e. the 2D turbulent RB convection considered in this paper.

\section{Concluding remarks and discussions}

All numerical algorithms have the background numerical noises, i.e. the truncation errors and round-off errors, which are tiny and random. It was reported that, for a chaotic dynamic system, the random numerical noises increase exponentially due to the butterfly-effect of chaos, until up to the {\em same} order of magnitude as its ``true'' physical solution \citep{hu2020risks, qin2020influence}. Therefore, numerical simulations of a {\em deterministic} chaotic system are mostly a mixture of the ``true'' physical solution, which is {\em deterministic} in physics, and the  ``false'' numerical noises, which is however {\em stochastic}. This is the reason why numerical simulations of a {\em deterministic} chaotic system given by traditional algorithms in single/double precision often look like {\em stochastic}, and why it has been wrongly believed that a {\em deterministic} chaotic system can lead to {\em randomness}. This is also the reason why \citet{Teixeira2007Time} made a rather pessimistic conclusion that ``for chaotic systems, numerical convergence {\em cannot} be guaranteed {\em forever}''. Obviously, even given an accurate initial condition, this kind of randomness of chaotic system comes from the randomness of artificial background numerical noises. Thus, the background numerical noises can be naturally regarded as a kind of {\em artificial stochastic disturbances} to a chaotic system when it is solved numerically.

In this paper, the so-called clean numerical simulation (CNS) is adopted to {\em accurately} investigate the influence of numerical noises as a kind of tiny artificial stochastic disturbances on the two-dimensional turbulent RB convection under consideration. This is mainly because the CNS can reduce the background numerical noises (i.e. round-off errors and truncation errors) to any a required level, which can be so small that the ``false'' numerical noises are negligible compared with its ``true'' physical solution and thus the numerical simulations of turbulence are convergent/reproducible in an interval of time long enough for statistics, as illustrated in this paper. Strictly speaking, the CNS solution is also a mixture of the ``true'' physical solution and the ``false'' numerical noises. However, unlike the simulation given by a traditional algorithm in single/double precision, the ``false'' numerical noises of a CNS solution are often several orders of magnitude smaller than its ``true'' physical solution in a region $t\in[0,T_{c}]$ long enough for statistics, so that its numerical noises are negligible and the CNS result is ``convergent'' and ``reproducible''. Such kind of CNS solution in  $t\in[0,T_{c}]$ can be used as a ``clean'' benchmark solution for comparison with those given by traditional algorithms in single/double precision so as to investigate the influence of numerical noises as tiny artificial stochastic disturbances on the two-dimensional turbulent RB convection under consideration.  So, unlike the Taylor series method,  the key point of the CNS is the so-called  ``critical predictable time''    $T_c$  that determines a temporal interval  $[0,T_c]$ in which the numerical simulations are ``reliable''  and  ``clean'',  since their ``false'' numerical noises are much smaller than the ``true'' physical solution and thus are negligible.   

It was currently reported by \citet{mcmullen2022navier} that tiny stochastic disturbances resulting from thermal fluctuations might influence the {\em small-scale} properties of the freely {\em decaying} turbulence under their consideration, which is in agreement with the conclusions given by \citet{Gallis2021-PRF}, \citet{bandak2022dissipation}, \citet{bell2022thermal}, \citet{Eyink2022} and so on.
In this paper, we investigate the large-scale influence of numerical noises as a kind of tiny artificial stochastic disturbances on a sustained turbulence. Using the two-dimensional (2D) turbulent Rayleigh-B{\'e}nard convection (RBC) as an example, we illustrate that the numerical noises as a kind of micro-level artificial stochastic disturbances could indeed lead to {\em large-scale} deviations not only in spatio-temporal trajectories but also even in statistics of the {\em sustained} turbulence considered in this paper. Especially, such kind of tiny artificial stochastic disturbances even leads to different types of flows: the shearing convection occurs for the RKwD simulations, and its corresponding flow field turns to a kind of zonal flow thereafter, however the CNS benchmark solution always sustains the non-shearing vortical/roll-like convection during the whole process of simulation, as shown in Figures~\ref{Contour} and \ref{Contour-2}. Thus, we provide a {\em rigorous}  evidence that numerical noises as a kind of tiny artificial stochastic disturbances have not only quantitatively but also qualitatively {\em large-scale} influences on a {\em sustained} turbulence, i.e. the 2D turbulent RB convection considered in this paper.  Of course, for various types of turbulent flows governed by the NS equations, more investigations are needed in the future.

\begin{figure}
    \begin{center}
        \begin{tabular}{cc}
            \subfigure[]{\includegraphics[width=2.55in]{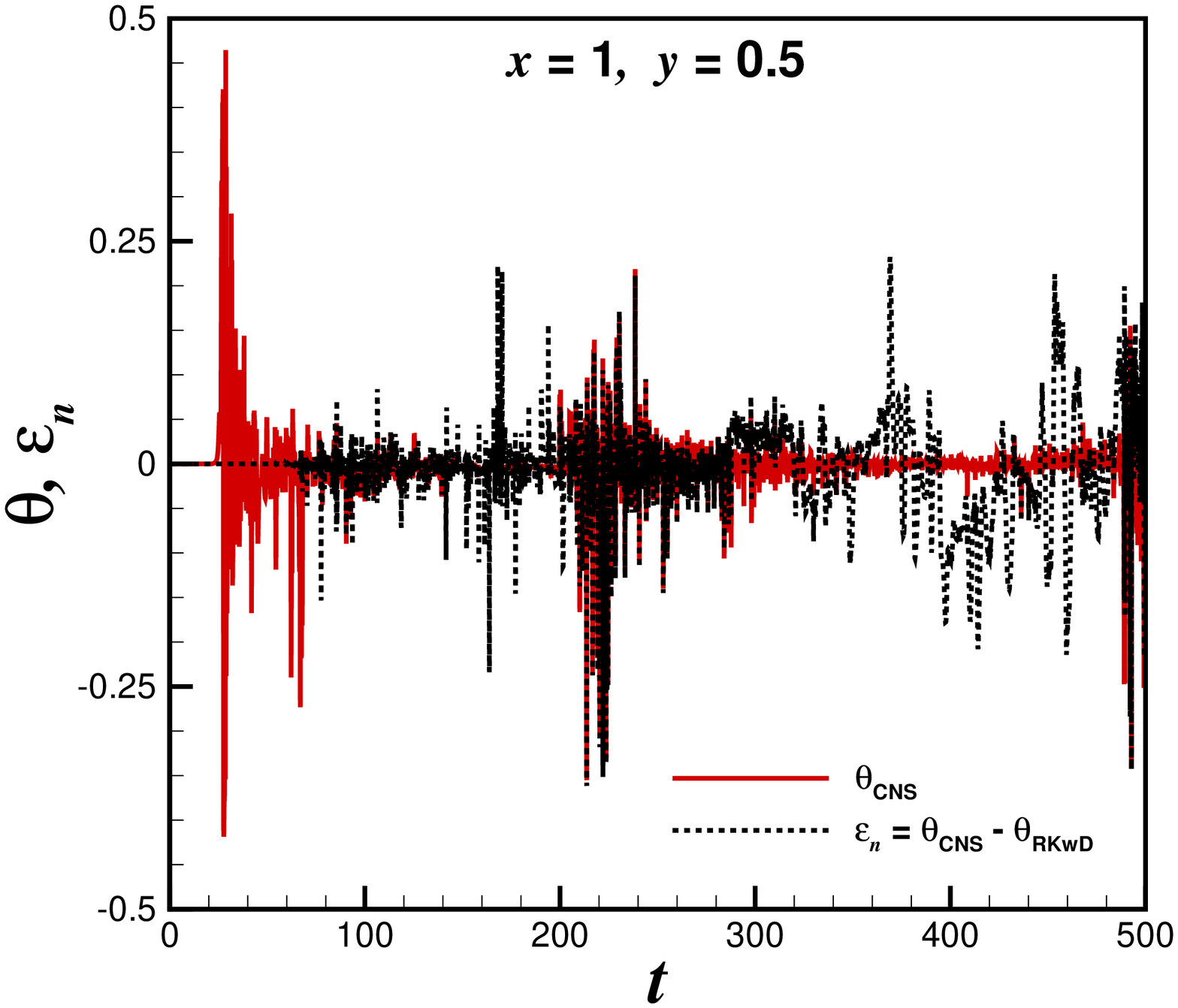}}
            \subfigure[]{\includegraphics[width=2.55in]{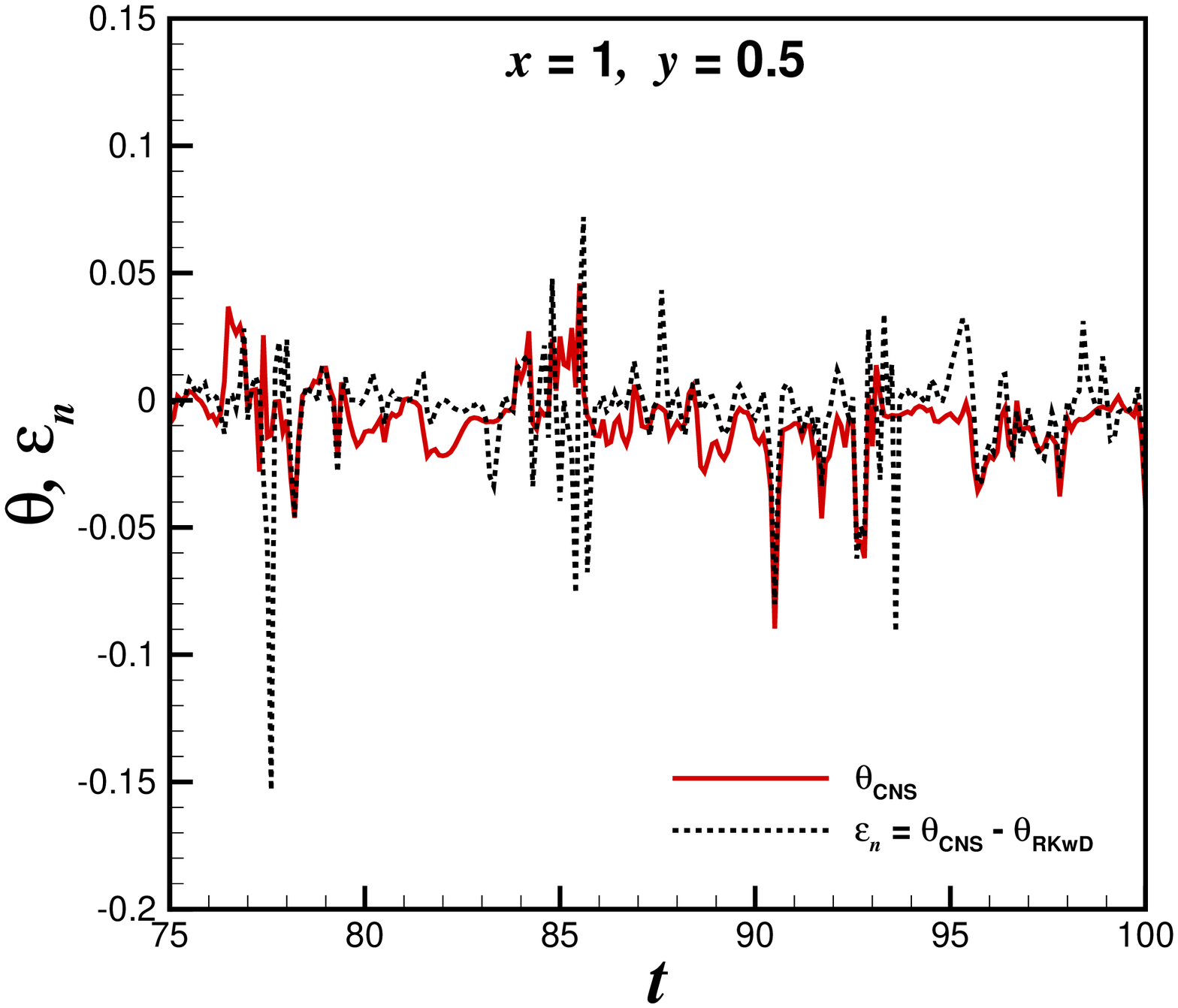}}
        \end{tabular}
    \caption{Comparison between the CNS benchmark solution $\theta_{CNS}$ and the numerical noises $\varepsilon_{n} = \theta_{CNS} - \theta_{RKwD}$ at the point $x=1$ and $y=1/2$, where $\theta_{RKwD}$ is the RKwD simulation.  Solid line: the CNS benchmark solution; Dashed line: the  numerical noises.  (a)  $1\leq t\leq 500$ ; (b) $75\leq t\leq 100$.}    \label{comparison-error}
    \end{center}
\end{figure}

\begin{figure}
    \begin{center}
        \begin{tabular}{cc}
            \subfigure[]{\includegraphics[width=2.55in]{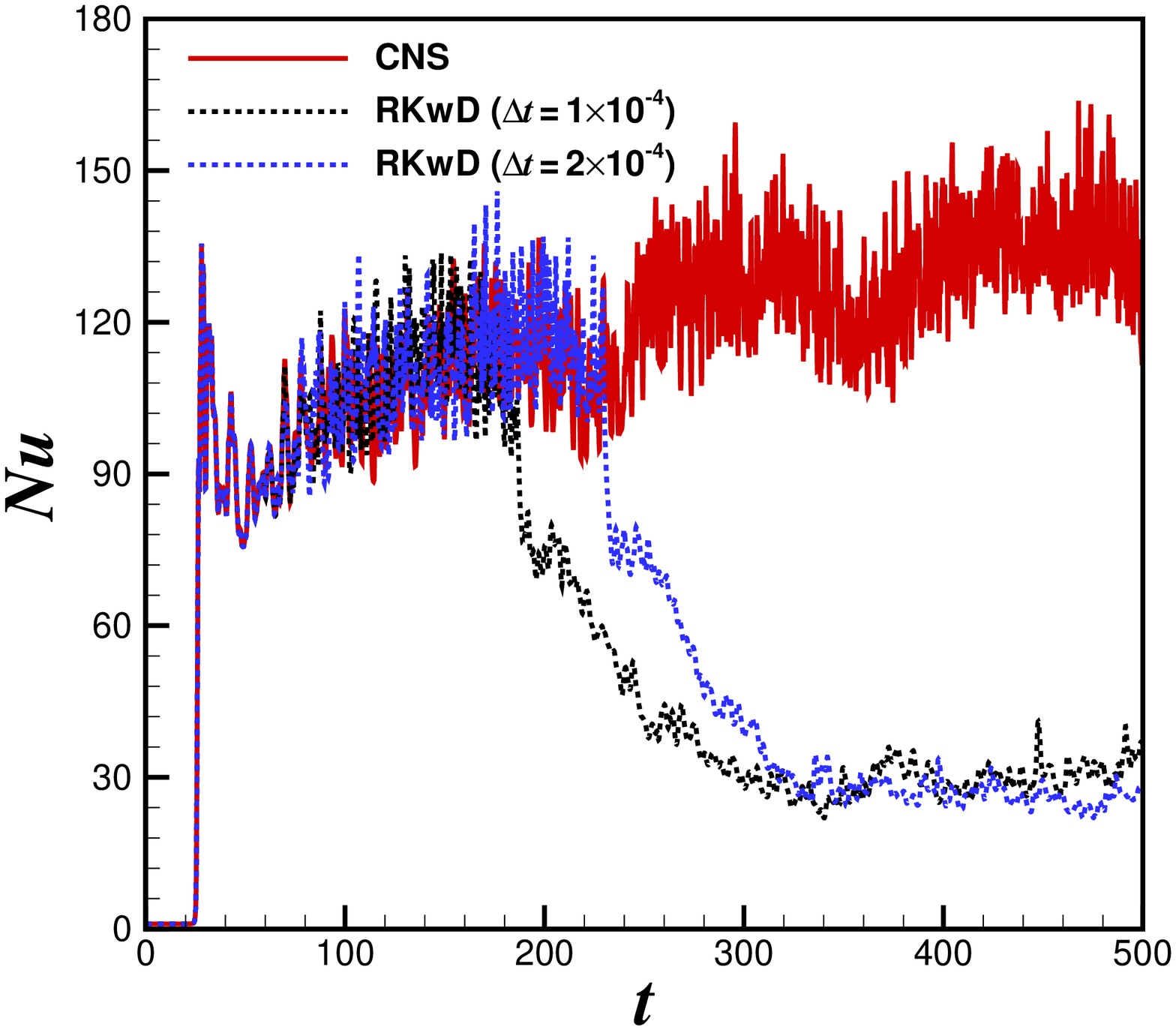}}
            \subfigure[]{\includegraphics[width=2.55in]{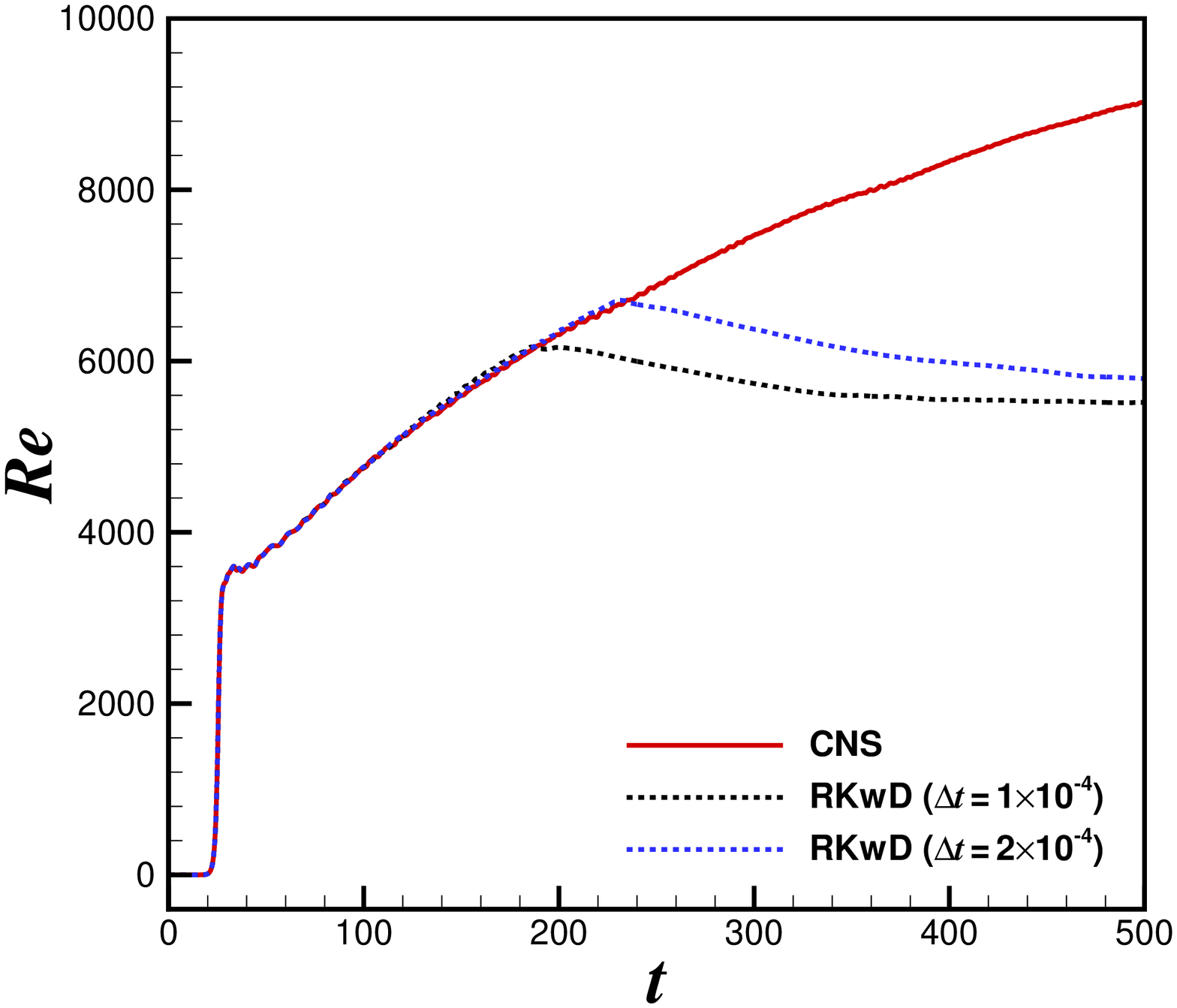}}
        \end{tabular}
    \caption{Temporal evolution of the Nusselt number $N\hspace{-0.3mm}u$ and the Reynolds number $Re$ in the case of $Pr = 6.8$, $Ra = 6.8 \times 10^{8}$, and $L/H = 2\sqrt{2}$. (a) Nusselt number $N\hspace{-0.3mm}u$; (b) Reynolds number $Re$; Solid line in red: the CNS benchmark solution; Dashed line in black: the RKwD simulation (with $\Delta t =  10^{-4}$); Dashed line in blue:  the RKwD$^{\prime}$ simulation (with $\Delta t = 2 \times 10^{-4}$).}    \label{Nu_t-2}
    \end{center}
\end{figure}

Why? We try to give an explanation. This 2D turbulent RBC system might have two possible final states: the vortical/roll-like flow and the zonal flow, which can be seen as two minima of a double-well potential. Once the CNS benchmark solution falls in one of these two minima (e.g. the roll-like flow), it remains there forever, since the corresponding ``false'' numerical noises are much less than the ``true'' physical solution and thus can not trigger off the transition to another minimum. On the contrary, the RKwD simulation can perform the transition from one state to the other, because its ``false'' numerical noises as artificial stochastic disturbances might be at the same order of magnitude as the ``true'' physical solution, as shown in Figure~\ref{comparison-error}, so that the RKwD simulation might depart very far away from its ``true'' physical solution and thus fall in another minimum.
If so, the transition of the RKwD simulation from the vortical/roll-like flow to the zonal flow of the 2D turbulent RBC should occur at random times for different numerical noises as a kind of artificial stochastic disturbances. In order to check this hypothesis, we perform a new RKwD simulation of the 2D turbulent RBC with the identical initial/boundary conditions and physical parameters of the previous one, but using a different time-step $\Delta t = 2\times10^{-4}$, marked by RKwD$^{\prime}$. The corresponding numerical noises as a different kind of artificial stochastic disturbances are verified to be at the same tiny level as the previous ones. As shown in Figure~\ref{Nu_t-2} (that describes the temporal evolutions of the Nusselt number $N{u}$ and the Reynolds number $Re$), the transition from the vortical/roll-like flow to the zonal flow of the RKwD$^{\prime}$ simulation with $\Delta t = 2 \times 10^{-4}$ occurs indeed at a different time $t\approx 230$, compared to $t\approx 188$ of the previous RKwD simulation with $\Delta t = 10^{-4}$. Thus, our above-mentioned explanation in terms of a stochastic dynamical system in the double-well potential should be reasonable. This test provides useful information to better understand the origin of the phenomenon reported in this paper.

Thus, generally speaking, if a chaotic system has $N$-well potential, where $N\geq 2$, its numerical simulations given by traditional algorithms in single/double precision should fall {\em randomly} in one of the $N$ minimums, and besides the transition between different minimums should occur frequently and randomly. Similarly, if a turbulent flow has multiple states in a large scale, small disturbances might lead to the transition between different states. Such kind of small disturbances can be either natural (such as the environmental perturbations) or artificial (such as the background numerical noises). In this case, it is {\em impossible} to make a correct prediction of flow states by means of traditional algorithms in single/double precision. By contrast, if a turbulent flow has an {\em unique} state, the background numerical noises should not have a large-scale influence on the flow. This should be a piece of good news for researchers in the field of CFD.

In theory, even if a turbulent flow has multiple states in large-scale, we can still give a {\em deterministic} prediction of its large-scale states by means of the CNS, since the {\em artificial} background numerical noises are negligible in a long enough interval of time for CNS solution. However, a {\em natural} disturbance, say, thermal fluctuation, always exists in fluid flows, which is random, small-scale, but {\em unavoidable} in practice. It is an open question whether or not, like tiny artificial numerical noises, the micro-scale thermal fluctuation might have a large-scale influence on some turbulent flows with multiple states. Note that the influence of thermal fluctuation is {\em not} considered in the NS equations. So, it is highly suggested to investigate the large-scale influence of thermal fluctuation on turbulent flows with multiple states by means of the Landau-Lifshitz-Navier-Stokes (LLNS) equations \citep{LLNS1959}. It is also worthwhile seriously discussing which one is better for turbulent flows (especially those with large-scale multiple states), either the {\em deterministic} NS equations or the {\em stochastic} LLNS equations. Obviously, more investigations are necessary in the future.

It should be emphasized that the external noises considered in most of related articles are mostly about ten orders of magnitude larger than the numerical noises considered in this paper, which might greatly change the characteristics of the chaotic/turbulent systems under their consideration. Besides, even when their external noises are zero, their numerical simulations are in fact a mixture of the ``true'' physical solution and the ``false'' numerical noises: both of them might be at the same level, as shown in this paper. Thus, strictly speaking, the conclusions based on this kind of ``mixtures'' should be {\em not} rigorous in theory. Thus, the CNS provides us, for the first time, a {\em rigorous} way to investigate the influence of external disturbances and very tiny artificial numerical noises on chaotic dynamical systems and turbulent flows.       

In summary, numerical noises as weak, small-scale stochastic perturbations increase exponentially to a macro level of numerical simulations, and besides might have a large influence on the macroscopic statistics of turbulent flows. Therefore, we should pay more attentions to the influences of small-scale stochastic perturbations on turbulence. Finally, this work also illustrates the validity, novelty, and great potential of the CNS as a reliable and accurate tool in theoretical studies of turbulence. Hopefully, the CNS might provide a brand-new, extremely accurate numerical tool to study turbulent flows.

\backsection[Acknowledgements]{Thanks to the anonymous reviewers for their valuable suggestions and constructive comments.}

\backsection[Funding]{This work is partly supported by the National Natural Science Foundation of China (No. 91752104) and Shanghai Pilot Program for Basic Research - Shanghai Jiao Tong University (No. 21TQ1400202).}

\backsection[Declaration of Interests]{The authors report no conflict of interest.}

\backsection[Data availability statement]{The data that support the findings of this study are available on request from the corresponding author.   The movie related to Figures~\ref{Contour} and \ref{Contour-2} is available at \href{https://github.com/sjtu-liao/RBC/blob/main/RBC-mv.mp4}{https://github.com/sjtu-liao/RBC/blob/main/RBC-mv.mp4} }

\backsection[Author ORCID]{Shijie Qin, https://orcid.org/0000-0002-0809-1766; Shijun Liao, https://orcid.org/0000-0002-2372-9502}

\backsection[Author contributions]{Liao conceived and designed the analysis. Qin performed the analysis. Both wrote the manuscript.}

\appendix

\section{The CNS algorithm for 2D turbulent RBC}    \label{app}

\cite{lin2017origin} combined the clean numerical simulation (CNS) with the traditional Fourier-Galerkin spectral method (in spectral space) to solve a two-dimensional (2D) turbulent Rayleigh-B{\'e}nard convection (RBC) with free-slip boundary conditions. However, their approach is rather time-consuming. Currently, \cite{hu2020risks} and \cite{qin2020influence} proposed an efficient CNS algorithm in physical space for spatio-temporal chaos to overcome the shortcoming of the CNS algorithm in spectral space. Here, the basic idea of this kind of efficient CNS algorithm in physical space is briefly described by using the 2D turbulent RBC as an example.

\subsection{The CNS algorithm in physical space}    \label{app1}

Applying the coordinate transformations $\tilde{x}=\lambda x$ and $\tilde{z}=\mu z$ to the NS equations (\ref{RB_psi}) and (\ref{RB_theta}), where $\lambda=2\pi/\Gamma$ and $\mu=\pi$, we obtain the following governing equations:
\begin{eqnarray}
\frac{\partial}{\partial t}(\lambda^{2}\psi_{xx}+\mu^{2}\psi_{zz})=\lambda\mu\hspace{0.3mm}\psi_{z}\left(\lambda^{2}\psi_{xxx}+\mu^{2}\psi_{xzz}\right)
-\lambda\mu\hspace{0.3mm}\psi_{x}\left(\lambda^{2}\psi_{xxz}+\mu^{2}\psi_{zzz}\right) \nonumber \\
+\lambda\hspace{0.3mm}\theta_{x}
+\sqrt{\frac{Pr}{Ra}}\left(\lambda^{4}\psi_{xxxx}+2\lambda^{2}\mu^{2}\psi_{xxzz}+\mu^{4}\psi_{zzzz}\right),       \label{RB_psi_0}
\end{eqnarray}
\begin{equation}
\frac{\partial\theta}{\partial t}=\lambda\mu\hspace{0.3mm}(\psi_{z}\theta_{x}-\psi_{x}\theta_{z})+\lambda\hspace{0.3mm}\psi_{x}
+\frac{1}{\sqrt{PrRa}}(\lambda^{2}\theta_{xx}+\mu^{2}\theta_{zz}),       \label{RB_theta_0}
\end{equation}
with $t\geq0$, $x\in[0,2\pi]$ and $z\in[0,\pi]$, where $x$ and $z$ as subscripts denote the spatial derivatives and the overhead tildes are omitted in the remainder of this article.

We prolongate the computational domain from  $z\in[0,\pi]$ to  $z\in[0,2\pi]$ so as to easily  satisfy the free-slip boundary conditions at lower ($z=0$) and upper ($z=\pi$) plates by means of the Fourier series. We use $N_x\times N_z$ equidistant points, i.e.
\begin{equation}
x_{j}=\frac{2\pi}{N_x}j,\hspace{1.0cm} z_{k}=\frac{2\pi}{N_z}k,    \label{points}
\end{equation}
where $j=0, \,1, \,2, \,..., \,N_x-1$ and $k=0, \,1, \,2, \,..., \,N_z-1$, to discretize $\psi$ and $\theta$, respectively.

To reduce truncation errors in the temporal dimension, the high-order Taylor expansions are adopted, i.e.
\begin{equation}
\psi(x_{j},z_{k},t+\Delta t)\approx\sum^{M}_{m=0}\psi^{[m]}(x_{j},z_{k},t)(\Delta t)^{m},  \label{Taylor_psi}
\end{equation}
\begin{equation}
\theta(x_{j},z_{k},t+\Delta t)\approx\sum^{M}_{m=0}\theta^{[m]}(x_{j},z_{k},t)(\Delta t)^{m},  \label{Taylor_theta}
\end{equation}
where $\Delta t$ is the time step, $M$ is the order of Taylor expansion,  with the definitions
\begin{equation}
\psi^{[m]}(x_{j},z_{k},t)=\frac{1}{m!}\frac{\partial^{m}\psi(x_{j},z_{k},t)}{\partial t^{m}},\hspace{1.0cm}
\theta^{[m]}(x_{j},z_{k},t)=\frac{1}{m!}\frac{\partial^{m}\theta(x_{j},z_{k},t)}{\partial t^{m}}.      \label{Taylor_details}
\end{equation}
Here, the order $M$ should be large enough so as to reduce the truncation errors (in the temporal dimension) to a required tiny level.

Differentiating $(m-1)$ times both sides of  (\ref{RB_psi_0}) and (\ref{RB_theta_0}) with respect to $t$ and then dividing them by $m!$, we obtain the governing equations of $\psi^{[m]}$ and $\theta^{[m]}$:
\begin{eqnarray}
\lambda^{2}\psi^{[m]}_{xx}(x_{j},z_{k},t)+\mu^{2}\psi^{[m]}_{zz}(x_{j},z_{k},t)=\frac{1}{m} \Big\{
\sqrt{\frac{Pr}{Ra}} \big[ 2\lambda^{2}\mu^{2}\psi^{[m-1]}_{xxzz}(x_{j},z_{k},t) \nonumber  \\
+\lambda^{4}\psi^{[m-1]}_{xxxx}(x_{j},z_{k},t)+\mu^{4}\psi^{[m-1]}_{zzzz}(x_{j},z_{k},t) \big] \nonumber \\
+\sum^{m-1}_{r=0}\lambda\mu\hspace{0.3mm}\psi^{[r]}_{z}(x_{j},z_{k},t) \big[ \lambda^{2}\psi^{[m-1-r]}_{xxx}(x_{j},z_{k},t)
+\mu^{2}\psi^{[m-1-r]}_{xzz}(x_{j},z_{k},t) \big ] \nonumber \\
-\sum^{m-1}_{r=0}\lambda\mu\hspace{0.3mm}\psi^{[r]}_{x}(x_{j},z_{k},t) \big[ \lambda^{2}\psi^{[m-1-r]}_{xxz}(x_{j},z_{k},t)
+\mu^{2}\psi^{[m-1-r]}_{zzz}(x_{j},z_{k},t) \big] \nonumber \\
+\lambda\hspace{0.3mm}\theta^{[m-1]}_{x}(x_{j},z_{k},t)
\Big\} ,       \label{RB_psi_m}
\end{eqnarray}
\begin{eqnarray}
\theta^{[m]}(x_{j},z_{k},t)=\frac{1}{m} \Big \{
\frac{1}{\sqrt{PrRa}} \big [ \lambda^{2}\theta^{[m-1]}_{xx}(x_{j},z_{k},t)
+\mu^{2}\theta^{[m-1]}_{zz}(x_{j},z_{k},t) \big ] \nonumber \\
+\lambda\mu \sum^{m-1}_{r=0}\psi^{[r]}_{z}(x_{j},z_{k},t)\theta^{[m-1-r]}_{x}(x_{j},z_{k},t) \nonumber \\
-\lambda\mu \sum^{m-1}_{r=0}\psi^{[r]}_{x}(x_{j},z_{k},t)\theta^{[m-1-r]}_{z}(x_{j},z_{k},t) \nonumber \\
+\lambda\hspace{0.3mm}\psi^{[m-1]}_{x}(x_{j},z_{k},t)
\Big \} ,      \label{RB_theta_m}
\end{eqnarray}
where $m\geq1$.

Note that there exist some spatial partial derivatives (denoted by subscripts) in (\ref{RB_psi_m}) and (\ref{RB_theta_m}), such as $\partial^{s_1+s_2}\psi^{[r]}/(\partial x^{s_1}\partial z^{s_2})$ and $\partial^{s_1+s_2}\theta^{[r]}/(\partial x^{s_1}\partial z^{s_2})$ with $r,\,s_1,\,s_2\geq0$. In order to approximate these spatial partial derivative terms with high computational efficiency and precision from the known discrete variables $\psi^{[r]}(x_{j},z_{k},t)$ and $\theta^{[r]}(x_{j},z_{k},t)$, we adopt the spatial Fourier series
\begin{equation}
\psi^{[r]}(x,z,t)\approx\sum^{\frac{N_{\hspace{-0.3mm} x}}{2}-1}_{n_{\hspace{-0.15mm} x}=-\frac{N_{\hspace{-0.3mm} x}}{2}+1}\,
\sum^{\frac{N_{\hspace{-0.2mm} z}}{2}-1}_{n_{\hspace{-0.1mm} z}=-\frac{N_{\hspace{-0.2mm} z}}{2}+1}
\Psi^{[r]}(n_x,n_z,t)
e^{i \hspace{0.03cm} n_{\hspace{-0.15mm} x}x}e^{i \hspace{0.03cm} n_{\hspace{-0.1mm} z}z},   \label{Appro_psi}
\end{equation}
\begin{equation}
\theta^{[r]}(x,z,t)\approx\sum^{\frac{N_{\hspace{-0.3mm} x}}{2}-1}_{n_{\hspace{-0.15mm} x}=-\frac{N_{\hspace{-0.3mm} x}}{2}+1}\,
\sum^{\frac{N_{\hspace{-0.2mm} z}}{2}-1}_{n_{\hspace{-0.1mm} z}=-\frac{N_{\hspace{-0.2mm} z}}{2}+1}
\Theta^{[r]}(n_x,n_z,t)
e^{i \hspace{0.03cm} n_{\hspace{-0.15mm} x}x}e^{i \hspace{0.03cm} n_{\hspace{-0.1mm} z}z},   \label{Appro_theta}
\end{equation}
where $i=\sqrt{-1}$ is the imaginary unit and
\begin{equation}
\Psi^{[r]}(n_x,n_z,t)=\frac{1}{N_x\hspace{0.03cm}N_z}\sum^{N_{\hspace{-0.3mm} x}-1}_{j=0}\sum^{N_{\hspace{-0.2mm} z}-1}_{k=0}\psi^{[r]}(x_{j},z_{k},t)
e^{-i \hspace{0.03cm} n_{\hspace{-0.15mm} x} x_{\hspace{-0.15mm} j}}e^{-i \hspace{0.03cm} n_{\hspace{-0.1mm} z} z_{\hspace{-0.05mm} k}},   \label{Coe_psi}
\end{equation}
\begin{equation}
\Theta^{[r]}(n_x,n_z,t)=\frac{1}{N_x\hspace{0.03cm}N_z}\sum^{N_{\hspace{-0.3mm} x}-1}_{j=0}\sum^{N_{\hspace{-0.2mm} z}-1}_{k=0}\theta^{[r]}(x_{j},z_{k},t)
e^{-i \hspace{0.03cm} n_{\hspace{-0.15mm} x} x_{\hspace{-0.15mm} j}}e^{-i \hspace{0.03cm} n_{\hspace{-0.1mm} z} z_{\hspace{-0.05mm} k}},   \label{Coe_theta}
\end{equation}
are determined by the known $\psi^{[r]}(x_{j},z_{k},t)$ and $\theta^{[r]}(x_{j},z_{k},t)$, respectively, at discrete points ($x_j,z_k$) with $j=0, \,..., \,N_x-1$ and $k=0, \,..., \,N_z-1$. Then, we can obtain the rather accurate approximations of the spatial partial derivative terms in  (\ref{RB_psi_m}) and (\ref{RB_theta_m}), say,
\begin{eqnarray}
&& \frac{\partial^{s_1+s_2}\psi^{[r]}}{\partial x^{s_1}\partial z^{s_2}}(x_{j},z_{k},t)\nonumber\\
&\approx &  i^{s_1+s_2} \sum^{\frac{N_{\hspace{-0.3mm} x}}{2}-1}_{n_{\hspace{-0.15mm} x}=-\frac{N_{\hspace{-0.3mm} x}}{2}+1}\,
\sum^{\frac{N_{\hspace{-0.2mm} z}}{2}-1}_{n_{\hspace{-0.1mm} z}=-\frac{N_{\hspace{-0.2mm} z}}{2}+1}(n_x)^{s_1}(n_z)^{s_2}\hspace{0.03cm}
\Psi^{[r]}(n_x,n_z,t)
e^{i \hspace{0.03cm} n_{\hspace{-0.15mm} x}x_{\hspace{-0.15mm} j}}e^{i \hspace{0.03cm} n_{\hspace{-0.1mm} z}z_{\hspace{-0.05mm} k}},        \label{Appro_psi_xz}
\end{eqnarray}
\begin{eqnarray}
&& \frac{\partial^{s_1+s_2}\theta^{[r]}}{\partial x^{s_1}\partial z^{s_2}}(x_{j},z_{k},t)\nonumber\\
&\approx & i^{s_1+s_2} \sum^{\frac{N_{\hspace{-0.3mm} x}}{2}-1}_{n_{\hspace{-0.15mm} x}=-\frac{N_{\hspace{-0.3mm} x}}{2}+1}\,
\sum^{\frac{N_{\hspace{-0.2mm} z}}{2}-1}_{n_{\hspace{-0.1mm} z}=-\frac{N_{\hspace{-0.2mm} z}}{2}+1}(n_x)^{s_1}(n_z)^{s_2}\hspace{0.03cm}
\Theta^{[r]}(n_x,n_z,t)
e^{i \hspace{0.03cm} n_{\hspace{-0.15mm} x}x_{\hspace{-0.15mm} j}}e^{i \hspace{0.03cm} n_{\hspace{-0.1mm} z}z_{\hspace{-0.05mm} k}}.        \label{Appro_theta_xz}
\end{eqnarray}
Here, the fast Fourier transform (FFT) algorithm is used to increase computational efficiency.  In this way, the spatial truncation error can be decreased to a required tiny level, as long as the mode numbers $N_x$ and $N_z$ are large enough.

Note that, if the order $M$ of the Taylor expansions (\ref{Taylor_psi}) and (\ref{Taylor_theta}) is large enough, the temporal truncation errors can be controlled below a required tiny level. Besides, if spatial discretizations are fine enough, say, the mode numbers $N_x$ and $N_z$ are large enough, the spatial truncation errors in Fourier expressions (\ref{Appro_psi}) and (\ref{Appro_theta}) and besides the corresponding spatial derivative terms in (\ref{RB_psi_m}) and (\ref{RB_theta_m}) can be reduced to a required tiny level. However, this is {\em not} enough, since there always exist some round-off errors, which might be larger than the truncation errors. So, in addition, {\em all} physical/numerical variables and parameters are expressed in multiple precision (MP) with a large enough number $N_s$ of significant digits so that the round-off errors can also be controlled below a required tiny level. In this way, the background numerical noises, i.e. {\em both} of the spatial/temporal truncation errors and the round-off error as a whole, can be controlled below a required tiny level. This is different from other numerical algorithms including the DNS. In theory, the CNS results are more accurate than those given by the DNS, since the CNS adopts the multiple precision, however the DNS mostly uses the double precision. In addition, note that the CNS results are useful only in a {\em limited} interval of time $t\in[0,T_{c}]$, in which numerical noises can be neglected.

\subsection{Realization of free-slip boundary conditions}    \label{app2}

Considering the Fourier expansions (\ref{Appro_psi}) and (\ref{Appro_theta}), the complex coefficients $\Psi^{[r]}(n_x,n_z,t)$ and $\Theta^{[r]}(n_x,n_z,t)$ can be expressed in terms of their real and imaginary parts, respectively, as follows:
\begin{equation}
\Psi^{[r]}(n_x,n_z,t)=\Psi_1^{[r]}(n_x,n_z,t)+i \hspace{0.03cm} \Psi_2^{[r]}(n_x,n_z,t),
\end{equation}
\begin{equation}
\Theta^{[r]}(n_x,n_z,t)=\Theta_1^{[r]}(n_x,n_z,t)+i \hspace{0.03cm} \Theta_2^{[r]}(n_x,n_z,t),
\end{equation}
where $r\geq0$, $-N_x/2+1 \leq n_x \leq N_x/2-1$ and $-N_z/2+1\leq n_z \leq  N_z/2-1$. Using the conjugate symmetry of $\Psi^{[r]}(n_x,n_z,t)$ and $\Theta^{[r]}(n_x,n_z,t)$, we enforce
\begin{equation}
\Psi_{1}^{[r]}(n_x,n_z,t)=\Psi_{1}^{[r]}(-n_x,-n_z,t)=-\Psi_{1}^{[r]}(n_x,-n_z,t)=-\Psi_{1}^{[r]}(-n_x,n_z,t),  \label{BC1}
\end{equation}
\begin{equation}
\Psi_{2}^{[r]}(n_x,n_z,t)=-\Psi_{2}^{[r]}(-n_x,-n_z,t)=-\Psi_{2}^{[r]}(n_x,-n_z,t)=\Psi_{2}^{[r]}(-n_x,n_z,t),  \label{BC2}
\end{equation}
\begin{equation}
\Theta_{1}^{[r]}(n_x,n_z,t)=\Theta_{1}^{[r]}(-n_x,-n_z,t)=-\Theta_{1}^{[r]}(n_x,-n_z,t)=-\Theta_{1}^{[r]}(-n_x,n_z,t),  \label{BC3}
\end{equation}
\begin{equation}
\Theta_{2}^{[r]}(n_x,n_z,t)=-\Theta_{2}^{[r]}(-n_x,-n_z,t)=-\Theta_{2}^{[r]}(n_x,-n_z,t)=\Theta_{2}^{[r]}(-n_x,n_z,t),  \label{BC4}
\end{equation}
so as to automatically satisfy the free-slip boundary conditions at $z=0$ and $z =\pi$, corresponding to the upper and lower boundaries, respectively.
Besides, there should exist
\begin{equation}
\Psi_{1}^{[r]}(0,n_z,t)=\Psi_{1}^{[r]}(n_x,0,t)=\Psi_{2}^{[r]}(n_x,0,t)=0,    \label{BCS1}
\end{equation}
\begin{equation}
\Theta_{1}^{[r]}(0,n_z,t)=\Theta_{1}^{[r]}(n_x,0,t)=\Theta_{2}^{[r]}(n_x,0,t)=0.    \label{BCS2}
\end{equation}
For more details, please refer to \cite{saltzman1962finite}. Actually, the equations (\ref{BC1})-(\ref{BCS2}) imply that $\psi^{[r]}(x,z,t)$ and $\theta^{[r]}(x,z,t)$ in  (\ref{Appro_psi}) and (\ref{Appro_theta}) are expanded as the sine series in the vertical direction, which automatically satisfy the free-slip boundary conditions
\begin{equation}
\psi^{[r]}(x,z,t)=\frac{\partial^{2}\psi^{[r]}(x,z,t)}{\partial z^{2}}=\theta^{[r]}(x,z,t)=0    \label{free-slip_appro}
\end{equation}
at $z=0$ and $z=\pi$, say,
\begin{equation}
\psi^{[r]}(x_{j},z_{k},t)=\psi_{zz}^{[r]}(x_{j},z_{k},t)=\theta^{[r]}(x_{j},z_{k},t)=0    \label{free-slip_discrete}
\end{equation}
at $k=0$ and $k=N_z/2$.

\subsection{Computational efficiency}    \label{app4}

In this section, we illustrate that the above-mentioned CNS algorithm in physical space needs much less amount of calculation than the previous CNS algorithm combined with the Fourier-Galerkin spectral method \citep{lin2017origin}. Table~\ref{CPU_time} illustrates the required CPU times of the CNS algorithm combined with the Fourier-Galerkin spectral method (marked by $T_1$) and the CNS algorithm in physical space described above (marked by $T_2$)  by means of different numbers $np$ of threads in parallel computing. Note that $T_1$ is much larger than $T_2$ in the case of $np=16$, $32$, $64$ or $128$: the corresponding time ratio $T_1/T_2$ is more than $1500$ in all cases. In consequence, the CNS algorithm in physical space described above is much more efficient (about three level of magnitude higher) than the previous CNS algorithm combined with the Fourier-Galerkin spectral method \citep{lin2017origin}, and thus is used in this paper.

\begin{table}
\tabcolsep 0pt
\vspace*{-2pt}
\begin{center}
\def\temptablewidth{1.0\textwidth}
%{\rule{\temptablewidth}{1pt}}
\begin{tabular*}{\temptablewidth}{@{\extracolsep{\fill}}cccc}
\multirow{2}*{$~~np$} & $T_1$ & $T_2$ & \multirow{2}*{$T_1/T_2~~~$} \\
~~~ & (in seconds) & (in seconds) & ~~~~ \\
\hline
$~~16$ & $475348.2$ & $282.0$ & $1686~~~$ \\[3pt]
$~~32$ & $237953.8$ & $138.4$ & $1719~~~$ \\[3pt]
$~~64$ & $123709.2$ & $77.5$ & $1596~~~$ \\[3pt]
$~~128$ & $65149.4$ & $43.2$ & $1508~~~$ \\
\hline
\end{tabular*}
\end{center}
\caption{Required CPU times of the CNS algorithm combined with the Fourier-Galerkin spectral method (marked by $T_1$) and the CNS algorithm in physical space (marked by $T_2$) for solving the governing equations (\ref{RB_psi}) and (\ref{RB_theta}) in the temporal interval $t\in[0, 0.1]$ with the same physical and numerical parameters: the aspect ratio $\Gamma=L/H=2\sqrt{2}$, Prandtl number $Pr=6.8$, Rayleigh number $Ra=10^{7}$, the mode numbers $|n_x|, |n_z|\leq127$ of spatial discretizations, the $10$-th order ($M=10$) of the truncated Taylor series in temporal dimension with a fixed time step $\Delta t=10^{-3}$, and the number $N_s=100$ of significant digits in the multiple precision scheme, where $np$ denotes the number of threads in parallel computing.}    \label{CPU_time}
\end{table}

%%Vancouver style references.
\bibliographystyle{jfm}
\bibliography{refs}

\end{document}